\documentclass[useAMS,usenatbib]{mn2e}

\usepackage{graphicx}
\newcommand {\aplt} {\ {\raise-.5ex\hbox{$\buildrel<\over\sim$}}\ }

\title{Four ultra-short period eclipsing M-dwarf binaries in the WFCAM Transit Survey.}

\author[S.V. Nefs]{S.V. Nefs$^{1}$\thanks{E-mail:
nefs@strw.leidenuniv.nl},J.L. Birkby$^{1}$,I.A.G. Snellen$^{1}$,S.T. Hodgkin$^{2}$,D. J. Pinfield$^{3}$,
B. Sip\H{o}cz$^{3}$,\and G. Kovacs$^{2}$,D. Mislis$^{2}$,R. P. Saglia$^{4,5}$,J. Koppenhoefer$^{4,5}$,P. Cruz$^{7}$,D. Barrado$^{7}$,\and
E. L. Martin$^{6}$,N. Goulding$^{3}$,H. Stoev$^{7}$,J. Zendejas$^{4}$,C. del Burgo$^{8}$,M. Cappetta$^{4}$,\and 
Y.V.Pavlenko$^{9}$\\
$^{1}$Leiden Observatory, Leiden University, Niels Bohrweg 2,2333 CA Leiden, The Netherlands\\
$^{2}$Institute of Astronomy, University of Cambridge, Madingley Road, Cambridge, CB3 0HA, UK\\
$^{3}$Centre for Astrophysics Research, University of Hertfordshire, College Lane, Hatfield, Hertfordshire AL10 9AB, UK\\
$^{4}$Max-Planck Institut fur extraterrestrische Physik, Giessenbachstrasse, D-85741 Garching, Germany\\
$^{5}$Universitatssternwarte Scheinerstrasse 1, D-81679 Munchen, Germany\\
$^{6}$Centro de Astrobiologia (CSIC/INTA), Carretera Ajalvir km 4, 28850 Torrejon de Ardoz, Madrid, Spain\\ 
$^{7}$Departamento de Astrofisica, Centro de Astrobiologia (INTA-CSIC), ESAC Campus, PO Box 78, E-28691 Villanueva de la Canada, Spain\\
$^{8}$UNINOVA-CA3, Campus da Caparica, 2825-149 Caparica, Portugal\\
$^{9}$Main Astronomical Observatory, Academy of Sciences of Ukraine, Golosiiv Woods, Kyiv-127, 03680 Ukraine}
\begin{document}

\date{}

\pagerange{\pageref{firstpage}--\pageref{lastpage}} \pubyear{}

\maketitle

\label{firstpage}

\begin{abstract}
We report on the discovery of four ultra-short period (P$\leq$0.18 days) eclipsing M-dwarf binaries in the WFCAM Transit Survey. Their orbital periods are significantly shorter than of any other known main-sequence binary system, and are all significantly below the sharp period cut-off at $P\sim0.22$ days as seen in binaries of earlier type stars. The shortest-period binary consists of two M4 type stars in a $P=0.112$ day orbit. The binaries are discovered as part of an extensive search for short-period eclipsing systems in over 260,000 stellar lightcurves, including over 10,000 M-dwarfs down to J=18 mag, yielding 25 binaries with $P\leq$0.23 days. In a popular paradigm, the evolution of short period binaries of cool main-sequence stars is driven by loss of angular momentum through magnetised winds. In this scheme, the observed $P\sim$0.22 day period cut-off is explained as being due to timescales that are too long for lower-mass binaries to decay into tighter orbits. Our discovery of low-mass binaries 
with significantly shorter orbits implies that either these timescales have been overestimated for M-dwarfs, e.g. due to a higher effective magnetic activity, or that the mechanism for forming these tight M-dwarf binaries is different from that of earlier type main-sequence stars.
\end{abstract}


\section{Introduction}
The period distribution of close binary star systems (e.g. Devor 2005; Derekas et al. 2007) contains important information on binary formation and evolutionary processes. Observations have revealed that there is a sharp cut-off in the period distribution at $\sim$0.22 days (e.g. Rucinski 1992, Norton et al. 2011), and very few binaries have thus far been discovered with significantly shorter periods (GSC 2314-0530 0.192 days; Dimitrov \& Kjurkchieva 2010 and OGLE BW3 V38 0.198 days; Maceroni \& Montalban 2004). Searching for binaries beyond this cut-off is interesting because their frequency of occurence, and the ratio of detached versus contact binary systems, provide direct constraints on theories that model the formation and migration history of low-mass stars, a mass regime that has been fairly poorly characterised so far. It has even been proposed that very short period low-mass binaries could be the progenitors of stellar mergers which may lead to the observed population of very hot 
Jupiters (Martin et al. 2011), and which could explain events such as Nova Sco 2008, where a contact binary with an orbital period of 1.4 days merged into a single star (Tylenda et al. 2010). 

There are several theories that aim to explain the observed cut-off and the apparent lack of systems beyond it. In the first theory, near-contact binaries are formed from initially well-detached systems that undergo angular momentum loss (AML) via a magnetised wind on Gyr timescales. Evidence for such winds, especially in low-mass, short-period binary systems, is provided by observations of extensive cool spot coverage, rapid rotation in tidally locked orbits, strong H$\alpha$ emission, flares and high activity rates (e.g. Morales et al. 2010, Vida et al. 2009). Stepien (1995; 2006; 2011) estimated the timescale required for the components of a detached binary to both reach Roche-lobe overflow through AML, i.e. to become a contact binary. He proposed that the AML timescale is much longer for low-mass systems, such that M-dwarf binaries (except in the most extreme mass-ratio cases) can not reach Roche-lobe overflow within the age of the Universe. The 0.22 day cut-off therefore corresponds to 
a lower limit in total binary mass of $\sim$1.0-1.2$M_{\odot}$ for contact systems. However, detached low-mass ultra-short period binary systems are also extremely difficult to form by this mechanism. In Section 2, we will further explain and quantify the AML evolution for M-dwarf binaries.

Proposing an alternative theory, Jiang et al. (2011) suggest that an instability in mass transfer, when the primary fills its Roche lobe, is responsible for the observed short-period cut-off. This instability is predicted to occur if the primary star has a significant convection zone, implying that binaries with a primary mass lower than 0.63$M_{\odot}$ would merge too quickly to form stable contact systems. In this scheme, M-dwarf binaries in the contact phase are short-lived and are therefore extremely rare. Unlike the AML model of Stepien (2011), detached low-mass short-period binaries are permitted in the mass transfer instability model because its timescale for angular momentum loss is significantly shorter.

However, third-body interactions in the birth environment may be responsible for accelerating the orbital evolution of low-mass binaries beyond the previous predictions. Here, energy is drained from the binary by either ejection of the lowest-mass companion in a low-N `mini-cluster' (e.g. Reipurth \& Clarke 2001, Goodwin et al. 2004) or through the Kozai mechanism (e.g. Eggleton \& Kiseleva 2001, Fabrycky \& Tremaine 2007). Interestingly, an adaptive optics search for companions around solar-mass contact systems by Rucinski et al. (2007) found that the fraction of short-period binaries in hierarchical triples is at least 59\%($\pm$8\%) (see also Tokovinin et al. 2006), indicating that a significant amount of short-period binaries are found in triple systems.

With these competing scenarios in mind, it is interesting to characterise the low-mass M-dwarf population of ultra short-period binaries. Although M-dwarfs form the most common stellar population in our Galaxy ($\sim$70\% by number; Henry et al. 1997), their intrinsic faintness is a challenge when trying to obtain a sufficiently large M-dwarf sample from a magnitude limited optical survey in order to study their binary characteristics. In this paper we describe a search for the shortest period eclipsing binary systems in the WFCAM Transit Survey (WTS) down to $J$=18. The WTS is an infrared photometric monitoring survey running on the 3.8m United Kingdom Infrared Telescope (UKIRT) with its main goal to find planets transiting M-dwarfs. Because M-dwarf spectral energy distributions (SEDs) peak in the infrared, the WTS is sensitive to redder eclipsing systems, with a significant population of binary systems down to $\sim$M5.

In Section 2 we will expand on our motivation of this work. In Section 3 we describe our observations and data reduction, including the WTS survey itself and low resolution spectroscopic follow-up of some of the candidates. In Section 4 we use a variability statistic to find stars with correlated variability and select eclipsing binary candidates from the WTS data. We subsequently determine orbital periods using box-fitting and Fourier techniques. With simple colour cuts we then pre-select binaries with primary mass $\mathcal{M}_1<0.7M_{\odot}$ (K5). In Section 5 we obtain estimates of the binary effective temperature through broadband SED fitting and spectral template matching. By fitting Fourier series to the data we identify (semi-)detached and contact systems. In Section 6, we compare our results with expectations from the evolution scenarios of low-mass short period binaries. 

\section{Motivation}
\begin{figure*}
\begin{center}$
\begin{array}{lllll}
\includegraphics[width=0.5\textwidth]{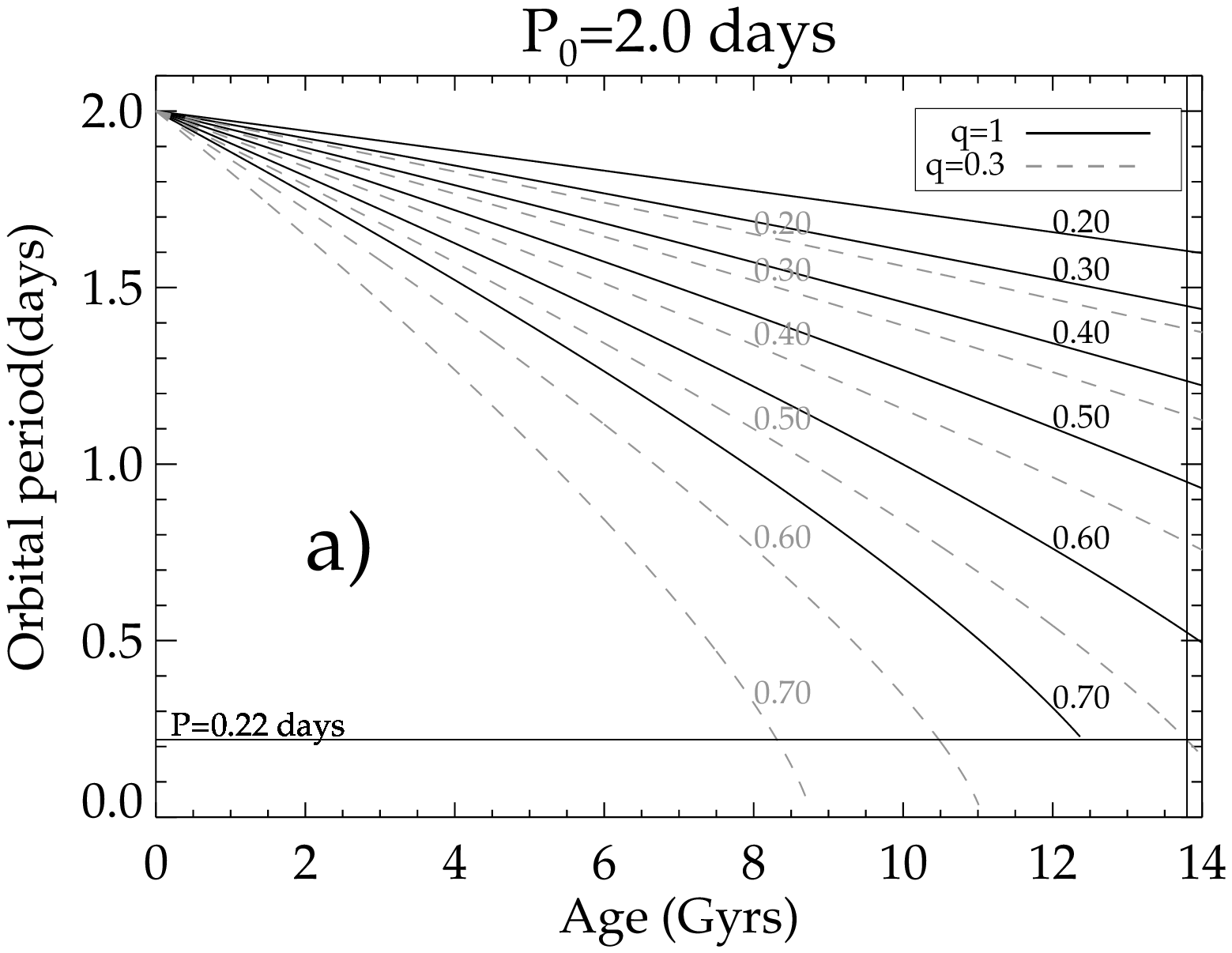}&
\includegraphics[width=0.5\textwidth]{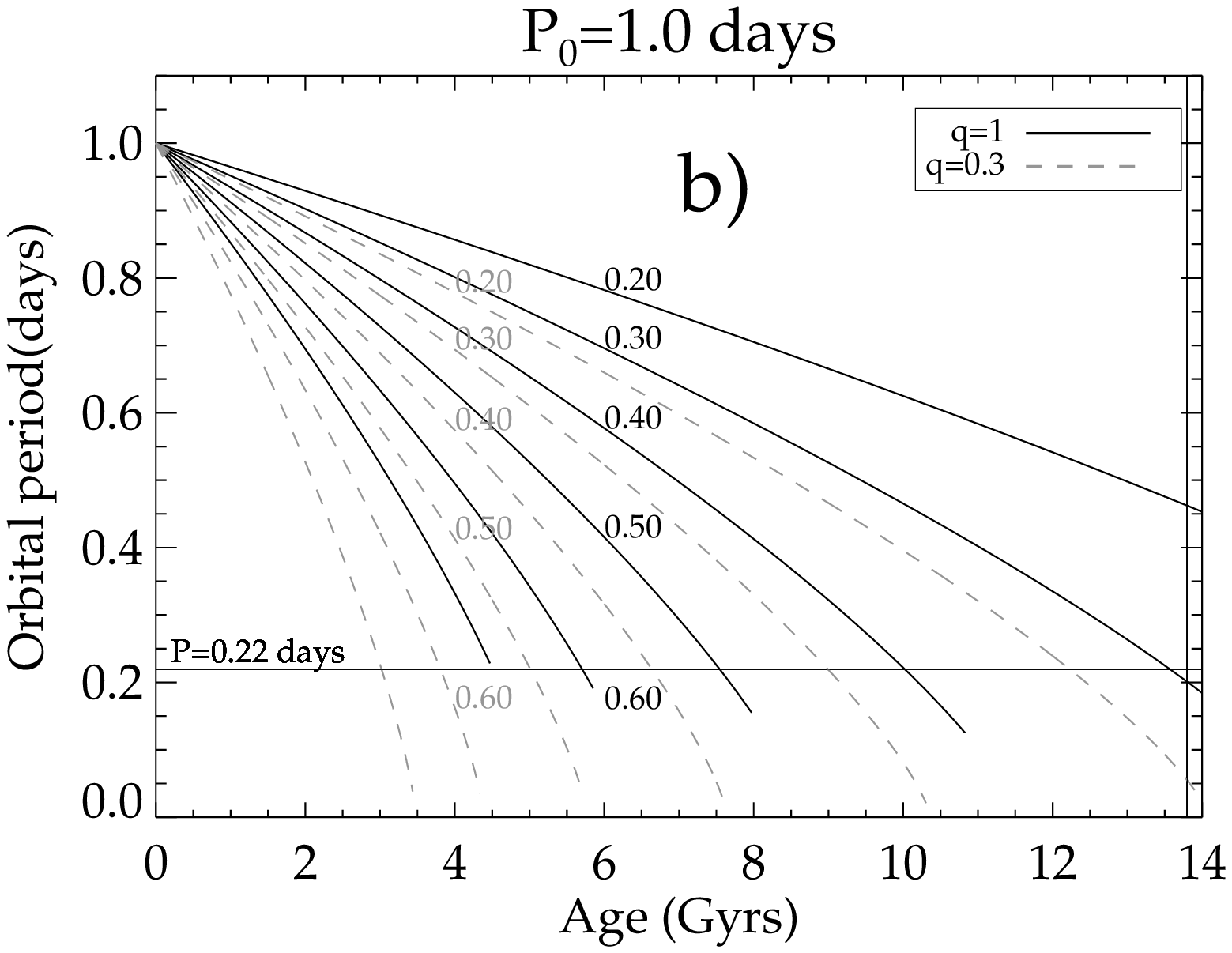}\\
\includegraphics[width=0.5\textwidth]{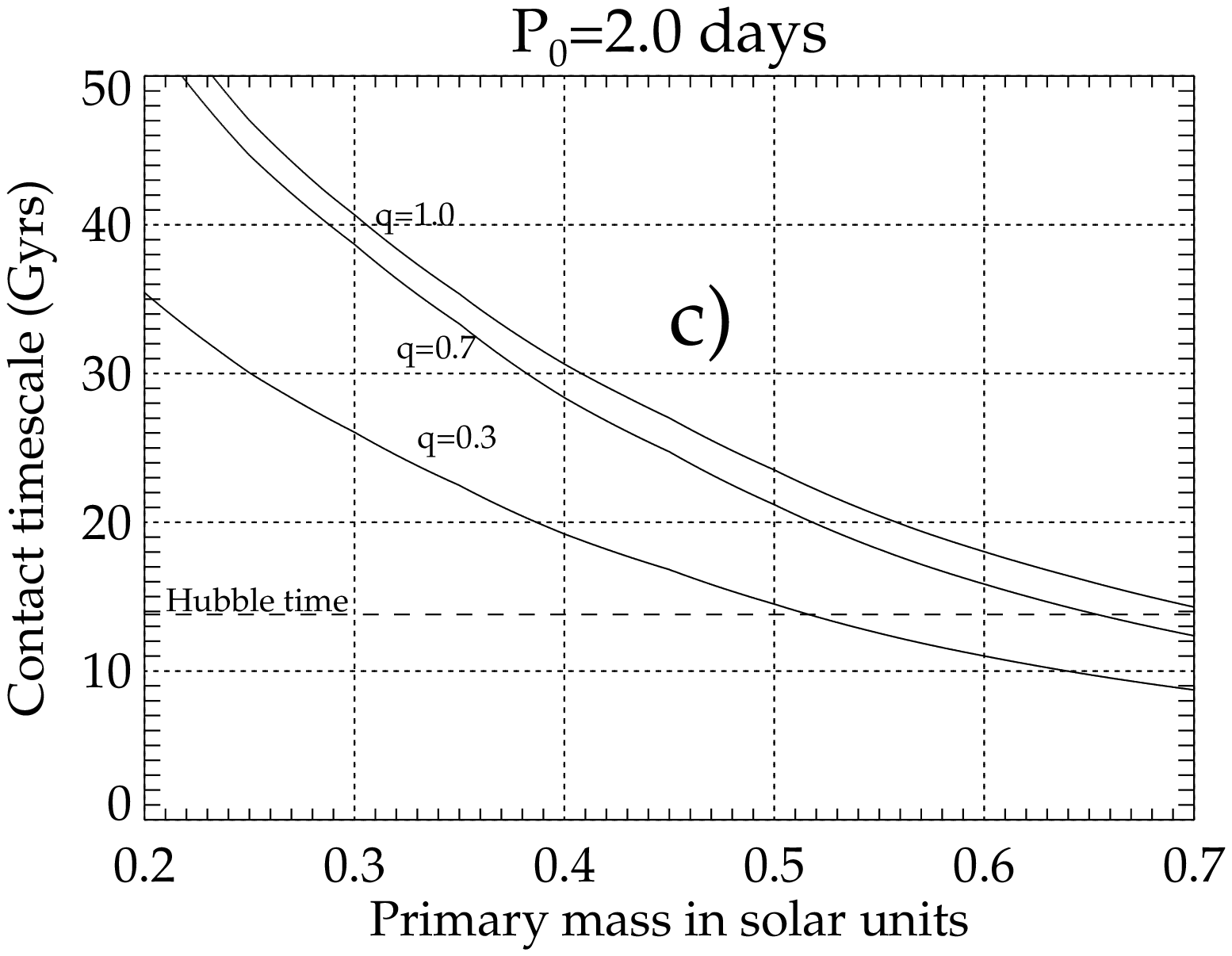}&
\includegraphics[width=0.5\textwidth]{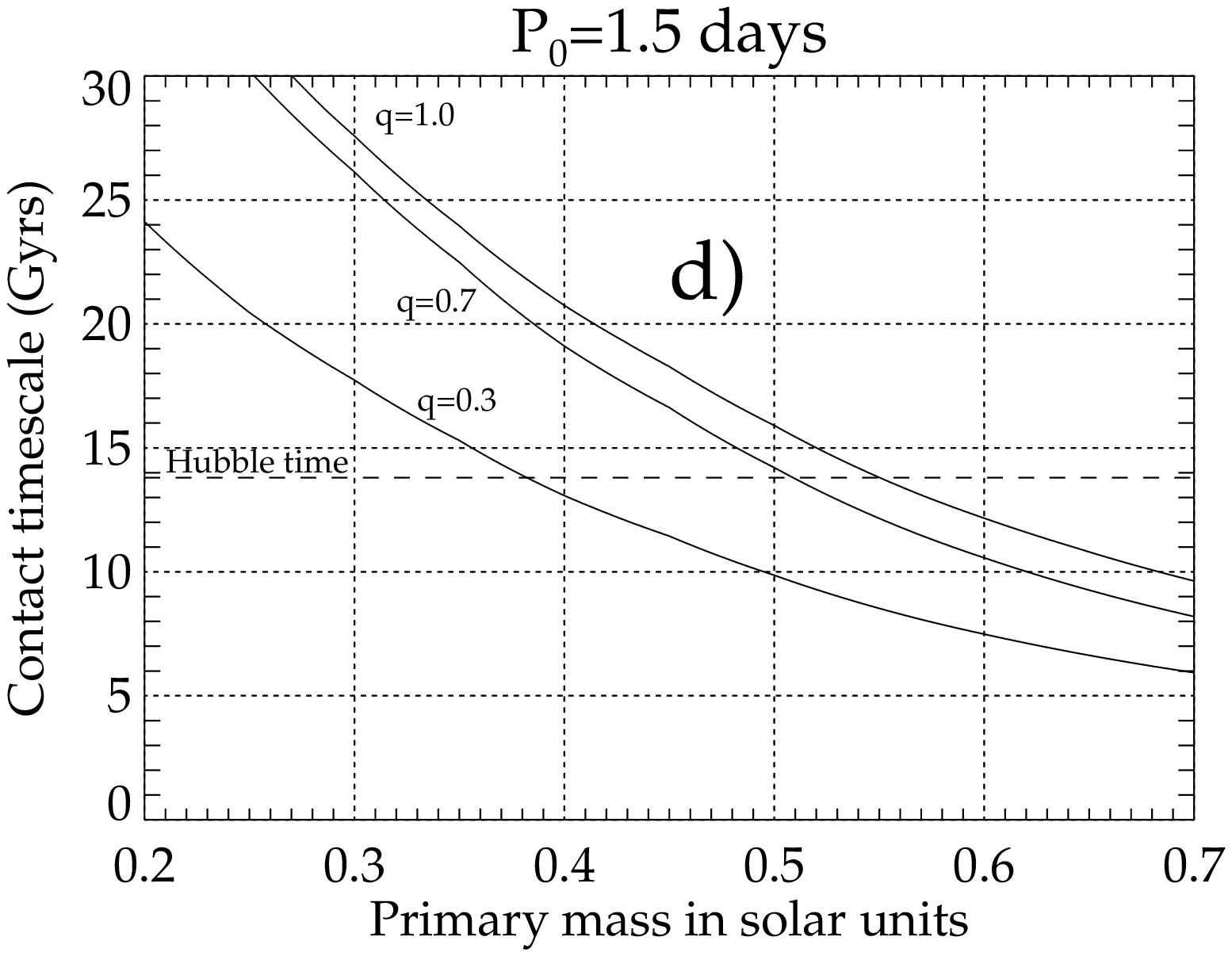}\\
\includegraphics[width=0.5\textwidth]{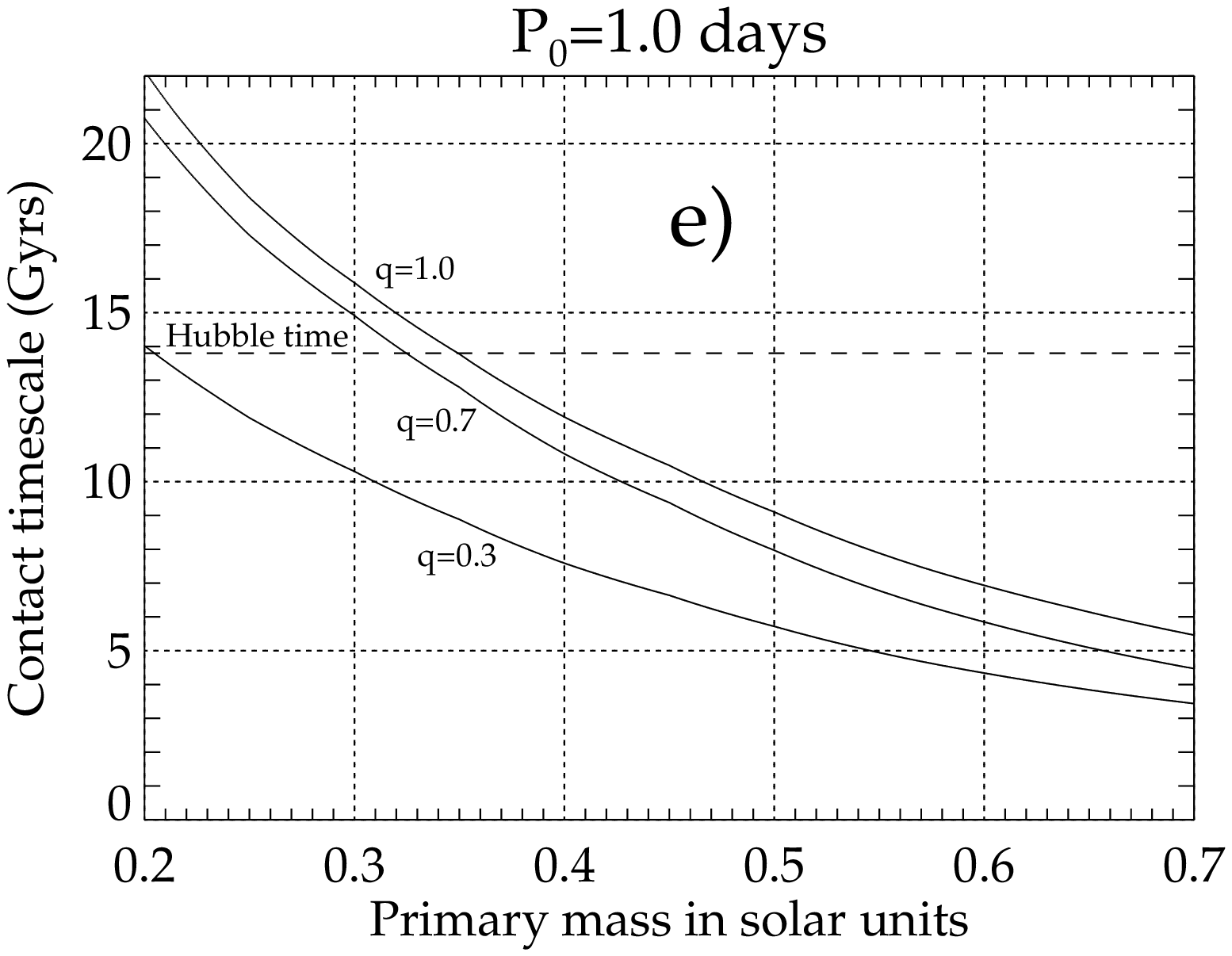}\\
\end{array}$
\end{center}
\caption{\small{{\bf{Panel a)}}: The orbital evolution of M-dwarf binaries, assuming $P_0$=2.0 days and q=1.0, plotted for decreasing primary stellar mass from 0.7 to 0.2$\mathcal{M_{\odot}}$ in steps of 0.1$\mathcal{M_{\odot}}$ (black solid curve). The dashed grey curves are models with mass ratio q=0.3. The models are cut off at the point of Roche lobe overflow, which depends on total mass and the mass-ratio. The horizontal line indicates the 0.22 day period cut-off, whereas the vertical line shows the Hubble time. Binaries with mass less then $\sim$0.65$\mathcal{M}_{\odot}$ could not have reached contact. {\bf{Panel b)}}: same models, but for $P_0$=1.0 day. {\bf{Panels c) through e)}}: the timescales to reach contact as a function of primary mass for $P_0$=2.0, 1.5 and 1.0 days and mass ratios q=1.0, 0.7, and 0.3.}}
\label{BigFig}
\end{figure*}

\subsection{The AML timescale argument}
Single stars with masses$\aplt$1.5$\mathcal{M}_{\odot}$ have an outer convective layer and can be chromospherically very active. This activity is expected to cause the loss of angular momentum through a magnetised wind of material that is forced to co-rotate with the star. The rotation rate of a star is found to correlate with the activity. Young active stars rotate rapidly but their rotation slows down with time leading to a decrease in their activity (e.g. Irwin et al. 2011). In a close binary, with synchronised orbits and spins, the angular momentum loss due to the spin-down of the individual stars causes a decrease of the binary orbital angular momentum, which tightens the orbit, but this then spins up the stars. A popular scenario is that when given enough time, the loss of angular momentum will bring the two stars into contact.

Stepien (1995, 2006, 2011) estimated timescales required for a detached binary to reach Roche-lobe overflow through AML and to evolve into a contact system, concluding that for a solar-type, equal-mass binary with a starting separation that corresponds to a period $P_0$=2.0 days this timescale is $\sim$6.5Gyr. He proposed that the AML timescale is significantly longer for decreasing mass systems and that therefore Roche-lobe overflow will not occur within the Hubble time for binaries with an initial primary mass of less than $\sim$0.7$M_{\odot}$, again assuming an initial orbital period of 2.0 days (an assumption which is further explained below). The 0.22 day period cut would therefore correspond to a lower limit on the total mass of contact systems of 1.0-1.2$M_{\odot}$ (i.e. a binary consisting of two M0 dwarfs). This suggests that evolution timescales are too long for M-dwarf binaries to decay into tight orbits within the age of the Universe. 

One solution to this is to choose a shorter starting period ($P_0$). To calculate the orbital evolution for M-dwarfs, and to investigate this evolution for different starting periods and mass ratios, we equate the change in the total binary orbital angular momentum ($H_{orb}$), which scales with the orbital frequency $\omega$ as $dH_{orb}/dt\sim\omega^{-4/3}\dot{\omega}$, to the change in the sum of spin angular momenta ($H_{spin}$) for the binary components, $dH_{spin}/dt\sim\omega$, following the derivation in Section 3 of Stepien (1995). We assume that mass-loss is small over the Hubble time for M-dwarfs and that the orbit remains circular and synchronised throughout binary evolution. The change in the orbital period as function of time can then be written as:
\begin{equation}
\frac{dP}{dt}=-\mathcal{A}(r_1,\mathcal{M}_1,r_2,\mathcal{M}_2 )P^{-1/3},
\end{equation}
where $\mathcal{A}$ is constant with time and $r_{1,2}$ and $\mathcal{M}_{1,2}$ are the radii and masses of the primary and secondary binary components. This equation has the solution:
\begin{equation}
P(t)=(P_0^{4/3}-\frac{4}{3}\mathcal{A} t)^{3/4},
\end{equation}
where $\rm{P}_0$ is the starting period at formation (in days) and P(t) is the observed orbital period at time t.

In panel a) of Figure \ref{BigFig} we show the orbital evolution of M-dwarfs, assuming $P_0$=2.0 days and q(=$\mathcal{M}_2/\mathcal{M}_1$)=1.0, plotted for decreasing primary stellar mass from 0.7 to 0.2$\mathcal{M_{\odot}}$ in steps of 0.1$\mathcal{M}_{\odot}$. In the same panel, we show the evolution for M-dwarfs with q=0.3 (grey dashed curves). Binary evolution is significantly faster for low-mass ratio systems.  Clearly, unless a low mass-ratio is assumed, there is insufficient AML over the Hubble time for any M-dwarf binary to reach contact. In panel b) of Figure \ref{BigFig} we plot the same binary evolution for M-dwarfs but with $P_0$=1.0 day, showing that a shorter starting period results in faster evolution.

In panels c) through e) of Figure \ref{BigFig} we show the time required to reach contact as a function of primary mass assuming $P_0$=2.0, 1.5, and 1.0 days for mass ratios q=1.0, 0.7, and 0.3. Clearly, in the $P_0$=1.0 day model, evolution is much faster and q=1 systems with primaries more massive than $\sim$0.35$\mathcal{M}_{\odot}$ could be brought into contact within the Hubble time. This means that the upper limit on the total mass of contact systems would shift to masses corresponding to later-type M-dwarfs, assuming that the formation mechanism for such binaries is the same as that of earlier-type stars. Also, even for a 0.5+0.5$\mathcal{M}_{\odot}$ binary the evolution timescale is $\sim$8 Gyrs in the fastest model, which would mean that such binaries would be some of the oldest objects in our current universe. 

\begin{figure}
\centering
\includegraphics[width=0.5\textwidth]{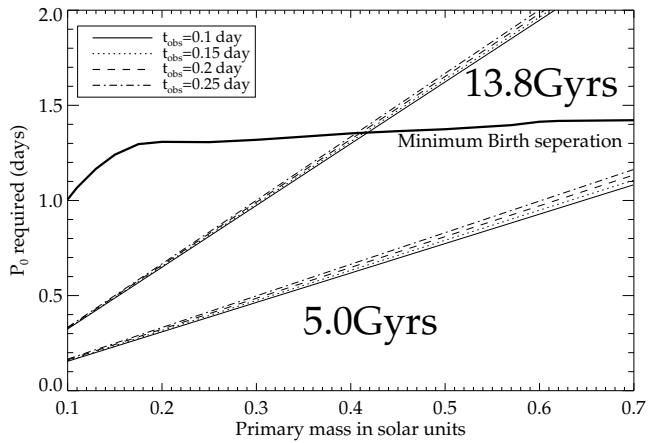}
\caption{\small{The starting period $P_0$ required to bring an equal-mass binary of given primary mass into a currently observed orbit of 0.1,0.15,0.2, and 0.25 days (the four lines, observed period increasing from bottom to top) for t=13.8 Gyr (upper set of four lines) and t=5.0 Gyr (lower set of four lines). The thick black solid curve indicates the estimated minimum possible separation period at birth for a binary using 1.0 Myr Baraffe models.}}
\label{P0plot}
\end{figure}

We have calculated what $P_0$ is minimally required to bring a given M-dwarf binary into a given orbital period today, within a given time-frame. We show this as a function of primary mass, derived using Equation 2, in Figure \ref{P0plot} for equal-mass binaries, assuming observed orbits of 0.10, 0.15, 0.20 and 0.25 days. For example, a minimum starting period of $P_0$=1.0 day is required for a 0.3+0.3$\mathcal{M}_{\odot}$ binary to reach $P$=0.2 days within the Hubble time. For a typical 5.0 Gyr thin disk binary, the required $P_0$ is 0.4 days for the same system. Clearly, to constrain AML theory, it is vital to assess from observations what is the actual frequency of (ultra)short-period M-dwarf binaries.

Although shorter starting periods could at least partly explain the existance of ultra-short period M-dwarf systems within the AML framework, binaries with these short starting periods should then be found abundantly in young cluster environments. However, pre-main sequence binary progenitors with $P_0\leq$2.5 days appear to be very rare, with 4 currently observed. HD15555 (a G5+K1 in a 1.7 day orbit), MML53 (a G2+K2 in a 2.1 day orbit), V4046Sgr (K5+K7;2.42 days) and 155913-2233 (K5+K5;2.42 days) are the only known (Melo et al. 2000, Hebb et al. 2010). Also, contact binaries of any spectral type are extremely rare in open clusters younger than $\sim$4 Gyr, only TXCnC (0.38 day orbit; Zhang et al. 2009) is linked to the intermediate-age open cluster Preasepe ($\sim$600 Myr). Cargile et al. (2008) report the discovery of a 4.7 day eclipsing pre-main sequence M-dwarf with masses 0.39 and 0.40$\mathcal{M}_{\odot}$, which suggests that shorter-period pre-main sequence M-dwarfs are not subject to 
observational bias. 

The minimum possible orbital separation of a close binary during the fragmentation of a protostellar cloud is an important constraint on the assumed $P_0$ as it prevents explaining ultra-short period binaries as simply forming very close together. This is because T-Tauri stars of subsolar mass have radii of $\sim$2-3$\rm{R}_{\odot}$ (Baraffe et al. 1998). A pile-up of detached binaries with periods of 2-3 days is also expected because the Kozai mechanism in a triple system can effectively produce such binaries within 50 Myrs (e.g. Fabrycky and Tremaine 2007), but is less efficient for shorter periods. We approximately estimate the shortest possible $P_0$ assuming pre-main sequence stellar radii using Baraffe (1998) 1.0 Myr models (the thick black solid curve in Figure \ref{P0plot}). Note that for the lowest mass M-dwarfs no $P_0$ can yield very short-period binaries. This makes it difficult to justify any starting period of $P_0\aplt$1.5 days.

\begin{figure}
\centering
\includegraphics[width=.5\textwidth]{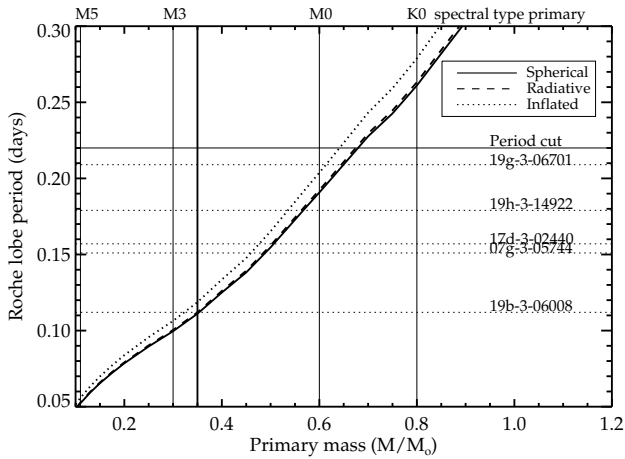}\\
\caption{\small{The estimated binary orbital period (Roche lobe period) for low-mass q=1 contact binary systems, plotted as function of primary mass for three different models, 
all assuming the Baraffe et al. (1998) 1Gyr mass-radius relation: i) a spherical model without distorted radius (solid line); ii) a fully radiative n=3 polytrope model (dashed line;no distortion) and iii) a fully convective stellar atmosphere with distortion due to tidal effects and rotation (Knigge et al. 2011; dotted line). The thick vertical line indicates the full convection limit ($\sim$0.35$\mathcal{M}_{\odot}$), whereas we also indicate approximate K0, M0,M3 and M5 type stars. Also shown are the 0.22 day period cut (horizontal solid line) and the five shortest period ($P\leq$0.21 days) eclipsing binaries from the WTS red binary sample (presented in Section 6).}}
\label{orbit}
\end{figure}

\subsection{Periods at contact}
We estimated the typical orbital periods we would expect for M-dwarf contact binary systems in order to focus our search for these systems. To do this, we determined their critical Roche lobe size $r_{1,2}$ (e.g. Eggleton 1983). For example, for an equal-mass binary this is approximately given by $r_1=r_2\sim0.379a$, where $a$ is the binary separation. Using the main-sequence mass-radius relation for low mass stars at 1Gyr of age (Baraffe et al. 1998) and Keplers law we calculate corresponding critical Roche lobe periods $P_{Roche}$. For illustration this $P_{Roche}$ is plotted as a function of primary mass in Figure \ref{orbit} (solid curve). For a binary of two M0 stars (assuming a primary mass of 0.6$M_{\odot}$; Baraffe \& Chabrier 1996), $P_{Roche}$ is $\sim$0.19-0.21 days, for a twin M5 system $\sim$0.05-0.07 days and $\sim$0.35 days for a solar type binary. The Roche period is also a function of $q$; for a 0.6$\mathcal{M}_{\odot}$ system with q=0.8, $P_{Roche}$ is $\sim$0.13 days and $\sim$0.07 
days for a q=0.5 system. This indicates that binaries with lower q have even shorter $P_{Roche}$. The dotted curve in Figure \ref{orbit} is computed from the expression for the radius of a tidally and rotationally deformed Roche-lobe filling star (Knigge et al. 2011, Section 5.2.2). For such a star $P_{Roche}$ is higher because the equilibrium radius of each star is now affected by the deformation. Note that there is still considerable uncertainty in the mass,radius,$\rm{T}_{eff}$, spectral type calibration (e.g. Baraffe \& Chabrier 1996), and magnetic activity may be a contributing factor to radius inflation in a tight low mass binary system (e.g. Rozyczka et al. 2009), which in turn increases the length of $P_{Roche}$.

Therefore, to critically test the predictions of AML and other paradigms, it is essential to determine from observations the abundance of M-dwarf binaries that are near contact, i.e. binary stars which are expected to have ultra-short orbital periods on the order of only a few hours. Because of the intrinsic faintness of M-dwarfs and the strong line broadening induced by rapid stellar rotation it is difficult to find such binaries using data from current radial velocity surveys. We therefore present the results of a campaign utilising photometric measurements obtained by the WFCAM Transit Survey, which has both the cadence and sensitivity to detect and characterise ultra-short period binaries in the uncharted M-dwarf regime.

\section{Observations and data reduction}
\subsection{WTS J band time-series photometry}
The WFCAM Transit Survey (WTS) is an ongoing photometric monitoring campaign operating on the 3.8m United Kingdom Infrared Telescope (UKIRT) at Mauna Kea, Hawaii, since August 2007. The Survey uses the Wide-Field Camera (WFCAM), which has four 2048$\times$2048 18$\mu$m HgCdTe Rockwell Hawaii-II, infrared imaging arrays which each cover 13.65'$\times$13.65' (0.4"/pixel), and are separated by 94\% of a chip width\footnote{http://casu.ast.cam.ac.uk/surveys-projects/wfcam} (Casali et al. 2007). Observations for the WTS are obtained in the J-band (1.25$\mu$m), near the peak of the spectral energy distribution (SED) of a typical M-dwarf. 

The WTS is primarily designed to find planets transiting red dwarf stars and to detect low mass eclipsing binaries, by observing $\sim$6000 early to mid M-dwarfs with $J\leq$16 (in this paper we consider stars down to J=18). The WTS observing strategy uses the queue-schedule mode of UKIRT and can operate in mediocre seeing and thin cloud cover. Four target fields, 1.5 square degree each and passing within 15 degrees of zenith, were selected to give year-round visibility. The fields were selected as regions of the sky where the ratio of dwarfs to giants was maximized, reddening relatively low (E(B-V) between 0.057 and 0.234) and overcrowding reduced by observing close to but outside the galactic plane ($b>5^{\circ}$). 
\begin{table}
\tiny{
   \begin{tabular*}{0.5\textwidth}
   {@{\extracolsep{\fill}}llllll}
   \hline
Name	&(RA)         &(DEC)     &\#stars             &\#epochs     &\#cand.\\
\hline
03hr field	&03$^h$39$^m$	    &+39$^d$14$^m$   &10827(36306)	  &392	&74\\
07hr field	&07$^h$05$^m$	    &+12$^d$56$^m$   &16623(56070)	  &626	&140\\		
17hr field	&17$^h$14$^m$	    &+03$^d$44$^m$   &9621(39879)	  &709	&68\\		
19hr field	&19$^h$35$^m$	    &+36$^d$29$^m$   &34452(130320)	  &1154	&375\\				
\hline
     \end{tabular*}
}
\caption{The main properties of the four WTS survey fields. Indicated are the approximate centres of the fields (right ascension and declination), the total number of stars with $J\leq$16 ($J\leq$18 in brackets), the number of epochs as of September 7th 2011, and the number of binary candidates per field.}
\label{fields}
\end{table}
In Table \ref{fields} we summarise the main properties of the four survey fields (the 03hr field, 07hr field, 17hr field and 19hr field). A field is observed by dithering the four detectors of WFCAM (a paw print) through 8 pointings, where each pointing is labeled with a single letter from \textit{a} to \textit{h} (e.g. '19a'), which are then tiled to give uniform coverage across the field. In itself, each single pointing consists of a 9-point jitter pattern with 10 second exposures at each jitter position. In this way, the near-infrared lightcurves have an average cadence (including overheads) of 15 min. Typically, WTS observations are taken only at the beginning of a night, just after twilight in $>$1" seeing.

Data reduction of the raw 2D J-band images is performed using the Cambridge Astronomical Survey Unit (CASU) pipeline, which is based on the INT wide-field survey pipeline (Irwin \& Lewis
2001). Astrometric and photometric calibration is obtained using 2MASS. Source detection is done on stacks of the 20 best exposures (the `master frame') and aperture photometry is performed as described in Irwin et al. (2007). The brightest stars (saturation occurs around J=12-12.5) have a precision of $\sim$3 mmag per data point, whereas 1\% photometry is still achieved for J=16 ($\sim$7\% for J=18). 

The short-period eclipsing binary candidates described in this paper were obtained from WTS lightcurves reduced on 7 September 2011 of all four survey fields. The 19hr field had the most extensive coverage ($\sim$1100 data points). For each field, we also obtained single deep exposures in WFCAM $ZYJHK$. These will be used together with $griz$ photometry from the Sloan Digitized Sky Survey (SDSS) where available to create SEDs and estimate the effective temperature of the candidates. 

\subsection{INT broad-band photometry}

The 17hr field in the WTS is not covered by SDSS, therefore we obtained optical photometry of this field to use in our SED fitting and effective temperature estimates. We used the Wide Field Camera (WFC) on the 2.5m Isaac Newton Telescope (INT) in La Palma on the nights of 8,10 July 2009 and 26 July 2010 to obtain g, r and i photometry of this field. The smaller field-of-view of WFC compared to WFCAM required 14 pointings to cover the entire field. We used the following exposure times for each pointing: $g$ 2$\times$250 seconds, $r$ 500 seconds, and $i$ 100 seconds, in order to match the typical depth achieved by SDSS. The data were reduced using custom-made IDL procedures to bias-subtract, flat-field (and de-fringe in the case of the i-band, where the fringe-frame was created from a number of deep exposures of the moonless sky). We used the IDL procedures $find.pro$ to identify the sources and $aper.pro$ to extract the photometry, calibrating to SDSS magnitudes using INT observations of the 19hr field 
in the same filters.

\subsection{Low resolution spectroscopy}
As an independent check on the effective temperatures derived for our binary candidates in Section 5.2, from broad-band colour fitting of their SEDs, we obtained low resolution spectral observations of three of our targets. Optical spectra of 19h-3-14922 (P=0.1798 days, (r-i)=1.65) and 19g-3-06701 (P=0.2090 days, (r-i)=1.01), were taken with the 4.2m WHT on the night of August 17 2010. We used the red arm and the R158 grism of the ISIS dual beam spectrograph, giving a dispersion of $\sim$1.8\AA/pixel using a slit width of 1.0" matching the seeing. In addition, a spectrum of 07e-2-03887 (P=0.2734 days, (r-i)=1.0) was obtained with ACAM ($\sim$3.3\AA/pixel) on the night of 08 September 2011. The spectra were reduced using a combination of IRAF (the \textit{apall} package) and custom IDL routines. In IDL, the spectra were trimmed to encompass the length of the slit, bias-subtracted, flat field corrected and median-filtered to remove cosmic rays. Background subtraction and optimal extraction of the 1D 
spectra was performed in IRAF. We used frames taken with the CuNe+CuAr standard lamp to determine the wavelength solution. For flux-calibration, we obtained reference spectra of standard stars. 

\section{Sample selection}
\subsection{Variability statistic}
The WTS 3.0 release consists of $\sim$262,000 stellar lightcurves with $J\leq18$. First, we select only those lightcurves that have less than 20\% bad datapoints and clip the bad data before further processing. We define as bad data a J-band measurement that either has infinite value (saturation, near-edge detections) or a photometric error larger than 1.0 magnitudes. We subsequently use the Stetson $J_S$-statistic (Stetson 1996) as a quantative estimator for photometric variability, analogous to Pepper \& Burke (2006). 
\begin{figure}
\includegraphics[width=.5\textwidth]{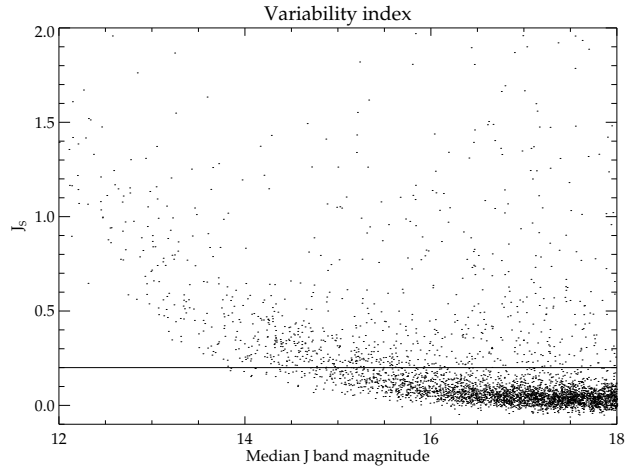}
\caption{\small{The Stetson $J_S$ variability statistic for one detector of
the 19h WTS paw, plotted as a function of WFCAM J band magnitude. We use a cut
$J_S>0.2$ to select candidate eclipsing binaries. At the brightest magnitudes,
outliers are dominated by systematic effects. These are removed through
visual inspection.}}
\label{jstat}
\end{figure}
The Stetson statistic weights photometric observations that are correlated in time using subsets of data that are separated by less than 0.03 days. For a nonvariable star showing only random (Gaussian) noise, this statistic will be around 0, whereas it will be positive for stars with correlated (physical) variability. As an example in Figure \ref{jstat} we show the $J_S$ variability index for one WFCAM detector in WTS paw 19h. To select candidate variable stars, we apply a conservative cut of $J_{S,cut}>$0.2. We determine this cut by examining the length of our candidate lists for random WTS paws for values of $J_{S,cut}$ from 0 to 0.5. We subsequently checked each candidate lightcurve and obtain a cutout thumbnail image from the stacked WFCAM J band master frame to eliminate the most obvious false-positives through visual inspection. These false-positives are caused by blending effects from nearby stars, extended wings and bleed trails from bright stars or detector defects. This step eliminates 80-90\% of 
the candidates and typically 20 to 40 objects per individual detector pass the inspection stage. 
\begin{figure}
\includegraphics[width=.5\textwidth]{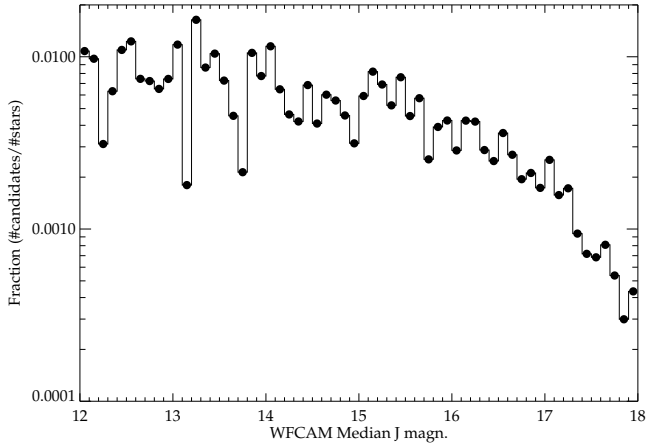}
\caption{\small{The ratio of binary star candidates over the number of available good quality WFCAM lightcurves as a function of WFCAM J magnitude. Each black dot indicates this fraction determined in 0.1 magnitude bins. The plot is truncated at J=12 because of saturation effects for brighter stars.}}
\label{detections}
\end{figure}
In total, we find 656 variables with $J\leq$18 in this way. 375 candidates are from the 19hr field (which contributes 49.6\% of the total number of lightcurves
with $J\leq$18), 68 from the 17hr field (15.2\%), 140 from the 07hr field (21.4\%) and finally 74 candidates are from the 03hr field (13.8\%). In Figure \ref{detections} we show a plot of the candidate fraction as a function of J band magnitude. At $J$=16, roughly 0.5\% of the high quality lightcurves is classified as a candidate, but for fainter magnitudes this fraction quickly decreases, reflecting the poorer photometry. For this reason, we have not investigated any lightcurves with $J\geq$18. 
\subsection{Orbital period determination}
To determine periods for all binary candidates we adopt a simple box-fitting algorithm (Kovacs et al. 2002), such as used by Collier-Cameron et al. (2006). We limit our search to $P\geq$0.1 days because our frequency spectra are very noisy at the shortest periods, which we suspect is due to strong aliasing caused by the peculiar form of the WTS window function. Note that shorter period systems could still be detected as an alias of the real period. We phase-fold the lightcurves over 20,000 periods between 0.1 and 1.0 days and for every period we vary the primary eclipse epoch $T_0$ and box width, and calculated the error-weighted signal compared to white noise (S/N). We then refined our analysis by selecting the period and epoch with the highest significance and re-iterate the box-fitting procedure with a smaller range of parameters, centered around the frequency peak. In this way, we generally achieve uncertainties in period of better than $\sim10^{-5}$ days and better than $\sim10^{-4}$ days on $T_0$. 
In this procedure we exclude periods within 10 minutes of 0.33, 0.5 and 1.0 days. In the case the algorithm confuses the secondary with the primary eclipse, this is picked up through visual inspection. In these cases the lightcurves are folded on several period aliases. As a crosscheck on the performance of the box-fitting algorithm, we also compute the Lomb-Scargle power spectrum (Scargle 1982) between 1 hr and 2 days (with a period sampling of 0.0001 days) for each candidate. Since the Lomb-Scargle method uses Fourier analysis, it is mostly sensitive to W Uma variables, pulsators and rotators, and less sensitive to the narrow eclipses that are expected for (semi-)detached systems. In the end we adopt the method that provides the lowest rms in the phase-folded lightcurve.

\subsection{Selection of the final sample}
We applied a first-pass colour cut using SDSS photometry (Kowalski et al. 2009) to select late K and M-dwarf short-period eclipsing binaries, which is verified using the fitting of the broadband SEDs in Section 5.2. SDSS photometry is available for three WTS survey fields, the 03hr field, 07hr field and 19hr field. For the 17hr field, which does not have SDSS coverage, we use $gri$ photometry obtained on the INT. We made cuts at (r-i)$\geq$0.50 and (i-z)$\geq$0.25 and selected orbital periods with $P\leq$0.3 days. This defines our `red' sample, the principal focus of this paper. In addition we use the well-calibrated SDSS colour selection criteria from Sesar et al. (2007) to exclude RR Lyrae. As there may be overlap in estimated effective temperature between late K-type stars and early M stars we also kept binaries which have $P\leq$0.23 days ($\sim$20,000 seconds, analogous to the sample of Norton et al. 2011) that are bluer than our cut, but are excluded as pulsators using the Sesar et al. (2007) 
colour criteria. This is our `blue' sample. We have removed objects from our red and blue samples that have sinusoidal phase-folded light curves with a large scatter in their amplitude, which would indicate variability caused by star spots (the amplitude depends on the spot-size which is expected to vary on timescales shorter than the 4 year observing span of the WTS). Note that this could remove a small fraction of genuine grazing equal-mass contact systems at our faintest magnitudes where our photometry is less precise.

\begin{figure}
\includegraphics[width=0.5\textwidth]{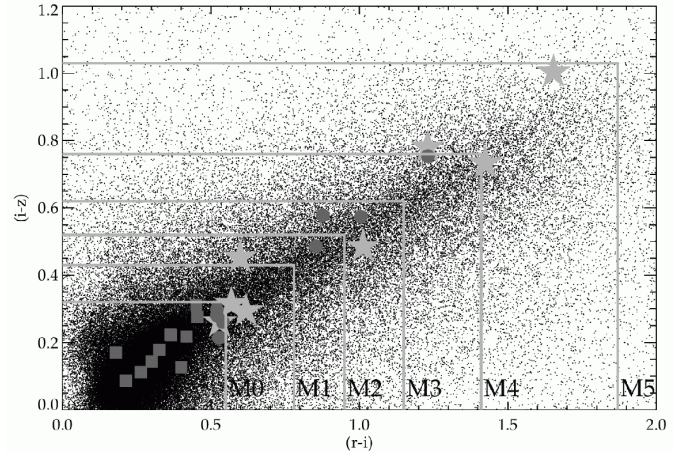}
\caption{\small{SDSS colour-colour diagram for objects in the 19hr field, showing (r-i)
versus (i-z). The light grey solid lines indicate the median average M-dwarf spectral types from Kowalski et al. (2009). Note that these averages generally have 1$\sigma$ uncertainty of 0.05 to 0.15 in (r-i) and (i-z). Black dots are objects in the WTS 19hr field (for plotting purposes we only show one field), which have SDSS DR7 photometry. Dark grey squares indicate our 'blue' short period EB sample ($P\leq$0.23 days), dark grey filled dots indicate our red sample, where the light grey stars in the figure are red sample binaries (with P$\leq0.23$ days).}}
\label{colors}
\end{figure}

In Figure \ref{colors} we show our red and blue samples in a (r-i) versus (i-z) SDSS colour-colour diagram. The light grey lines indicate the median M dwarf stellar types, as presented in Kowalski et al. (2009). The colours of these spectral types generally have a 1$\sigma$ uncertainty of 0.05-0.15. Dark grey filled dots and light grey stars indicate objects from our red sample, whereas dark grey filled squares represent the blue sample. The lightcurves and the optical colours of all of our candidates do not show features which would suggest they are M-dwarf white dwarf systems (Covey et al. 2005).

\subsubsection{Reddening estimates}
We have investigated the reddening effect for the 19hr field using a model for interstellar extinction presented by Amores \& Lepine (2005). At J$\aplt$17, assuming no reddening, the WTS is distance-limited to $\sim$1800 pc for the earliest M-dwarfs. At 1.8 kpc, the reddening model gives an extinction of $A_V$=0.37 mag (E(B-V))=0.12 mag) in the direction of the 19hr field, and using relations presented in Table 6 of Schlegel et al. (1998), we converted between filters to find $A_g$=0.43 mag, $A_K$=0.042 mag, E(r-i)=0.076, E(i-z)=0.069 and E(J-H)=0.037. The reddening effect along the line-of-sight to the field thus appears to be small, and similar results are obtained for the other three WTS fields. This makes it unlikely that any of our short-period red binary candidates are highly reddened pulsators. Also note that the pulsation periods of giant stars are inversely proportional to the square root of the mean stellar density, meaning that the low densities of giant stars result in significantly longer 
pulsation periods than we detect in our candidates. Rodriguez-Lopez et al. (2012) suggested that M-dwarfs may be pulsating, however it is not clear if such M-dwarf pulsations are stable and whether they could be detected in M-dwarfs with Gyr ages. They have not been confirmed observationally to date. Note that any pulsation would also have to be stable on the order of four years (the span of the WTS observing campaign).
\begin{figure*}
\includegraphics[width=1.0\textwidth]{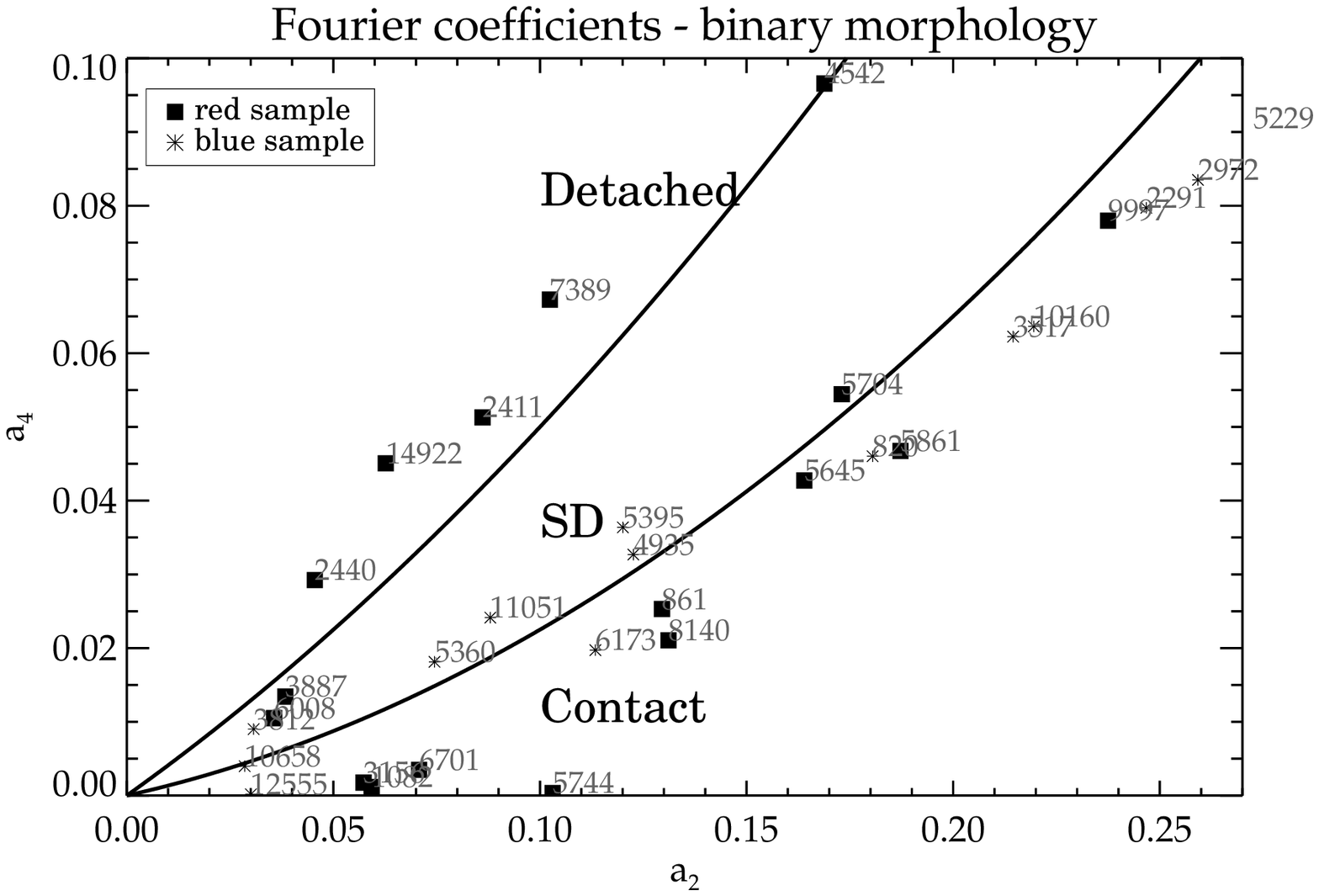}
\begin{center}$
\begin{array}{ll}
\includegraphics[width=0.5\textwidth]{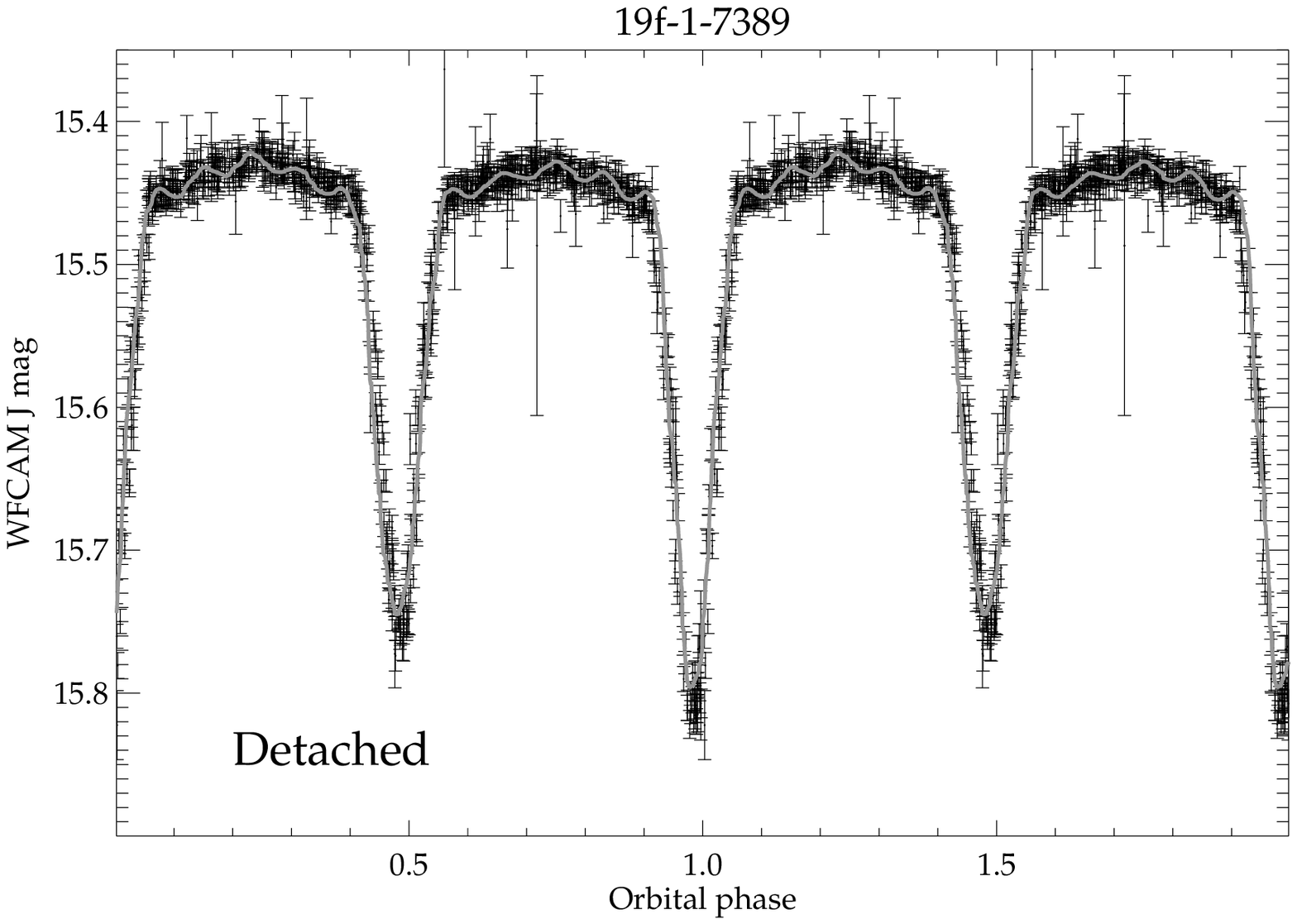}&
\includegraphics[width=0.5\textwidth]{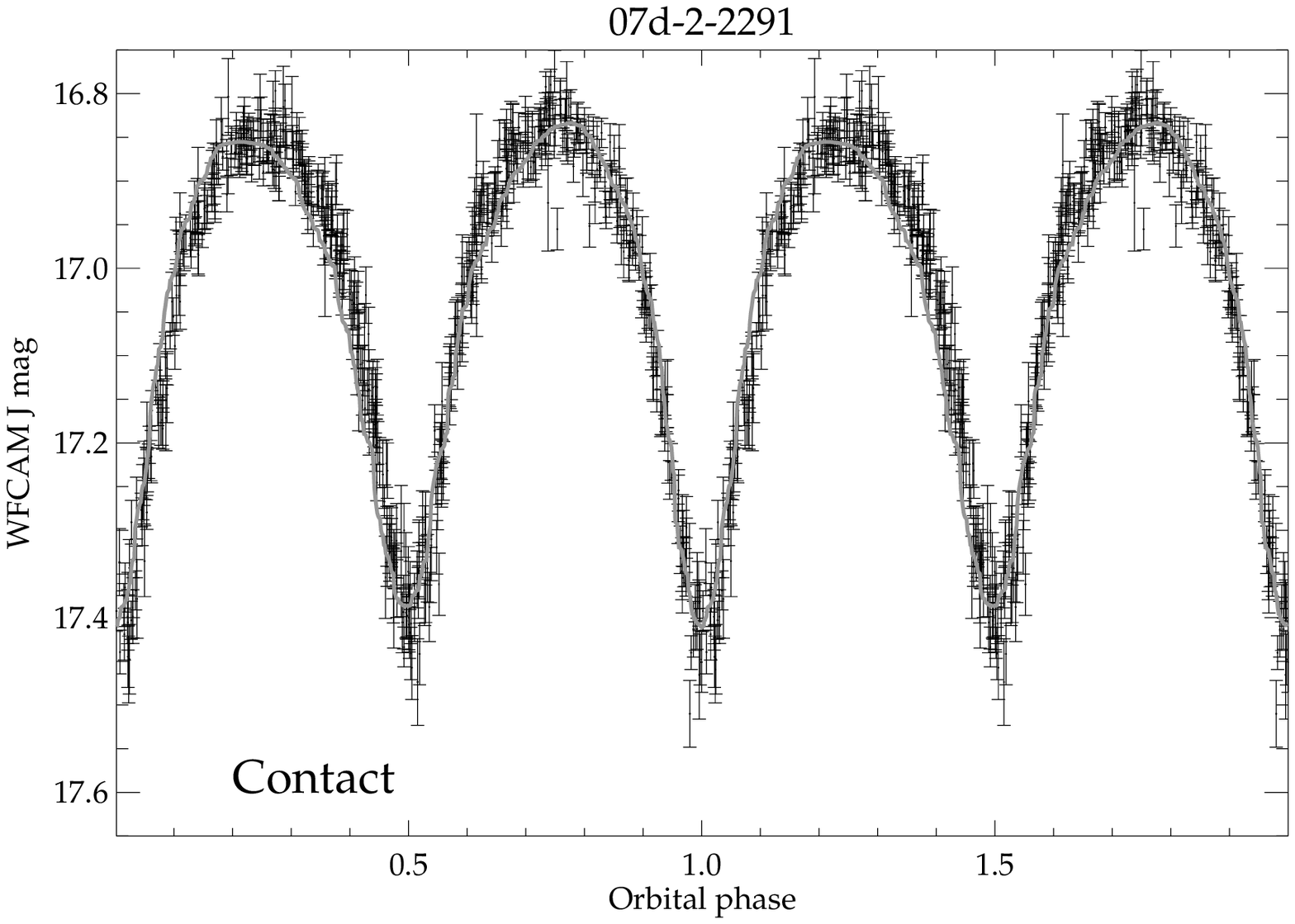}\\
\end{array}$
\end{center}
\caption{\small{{\bf{Top panel}}: Classification of our binary sample in the $a_2,a_4$ Fourier coefficient plane, where the numbers identify the candidates of Table 2, and filled squares refer to the red sample whereas asterisks indicate the blue sample. The lower solid line denotes the envelope for contact, which is relating the two coefficients through $a4=a_2(0.125+a_2)$. Sources below the envelope are classified as contact binaries. 'SD' indicates semi-detached binaries. {\bf{Lower panel}}: two representative binaries classified as detached and contact systems. The grey solid line is the best-fit Fourier decomposition (Section 5).}}
\label{a2a4}
\end{figure*}

\section{Characterisation of the eclipsing binary systems}
\subsection{Binary classification}
For a crude initial classification of a binary in the three categories \textit{detached}, \textit{semi-detached} or \textit{(over)contact}, we follow Rucinski (1993) and decompose the phased lightcurve $m(\Phi)$ into Fourier components with coefficients $a_i$ and $b_i$:
\begin{equation}
m(\Phi)=m_0+\sum_{i=1}^{10}[a_icos(2\pi i\Phi)+b_isin(2\pi i\Phi)],
\end{equation}
where $m_0$ is the mean lightcurve magnitude. The number of Fourier components is limited to 10, because for higher $i$ the quality of the fit does not improve for the typical uncertainties reached. For the fitting of the Fourier coefficients we use the IDL routines \textit{fourfit} and \textit{fourfunc} from Marc Buie's IDL archive\footnote{Available at http://www.boulder.swri.edu/$\sim$buie/idl/}. Based on a grid of coefficients derived from fits to model eclipsing binary lightcurves the $a_2$ and $a_4$ coefficients allow a distinction to be made between the three categories by plotting the sources in the ($a_2,a_4$) plane (Figure \ref{a2a4}). The coefficient $a_2$ is related to the depth of the eclipses, which depends on the inclination of the orbit and the mass ratio. For example, we find $a_2\sim$0.036 for 19b-3-06008, a relatively shallow semi-detached binary, whereas deep eclipse systems such as 07c-4-05645,19e-4-00861 and 19c-2-08140 have $a_2\sim0.13-0.16$. The coefficients $a_1$ and $a_3$ are 
related to the ratio of primary versus secondary eclipse depths and $b_1$ to the difference in the lightcurve maxima at phase 0.25 and 0.75 (star spots or mass transfer can bias this value). Pulsating variables can be distinguished from genuine eclipsing systems, in addition to the colour cuts previously described, by using the $b_2$ and $b_4$ coefficients (e.g. Pojmanski 2002). Our candidates have $b_2$ and $b_4$ consistent with non-pulsating stars.
\subsection{The effective temperature}
We estimate the effective temperature $\rm{T}_{eff}$ of the sources in our sample using an 8-band weighted SED fit to combined SDSS griz and WFCAM YJHK photometry. We $\chi^2$ fit the observed SED to a grid regularly spaced by 50 K in $\rm{T}_{eff}$ with theoretical broad-band colours based on Baraffe et al. (1998) 1Gyr, log(g)=5.0, solar metallicity models. In this process, we re-iterate the fitting procedure after removing the band that gives the largest outlier from the initial $\chi^2$ fit. For three sources we obtain an independent check on our results using low resolution spectra from the WHT (Section 2.2). For the spectroscopy data we $\chi^2$ fit NexTGen model spectra (Allard et al. 1997) with steps in effective temperature $\Delta \rm{T}_{eff}$=100K, solar metallicity and surface gravity log(g)=5.0 to our observed spectra resulting in uncertainties in spectral type of about one subclass (corresponding to $\sim$100-150K). We mask out the strong telluric ozone bandhead around 7600 $\rm{\AA} $ and the 
$\textrm{H}_{\alpha}$ emission line at 6563 \AA. 
\begin{figure}
\includegraphics[width=0.5\textwidth]{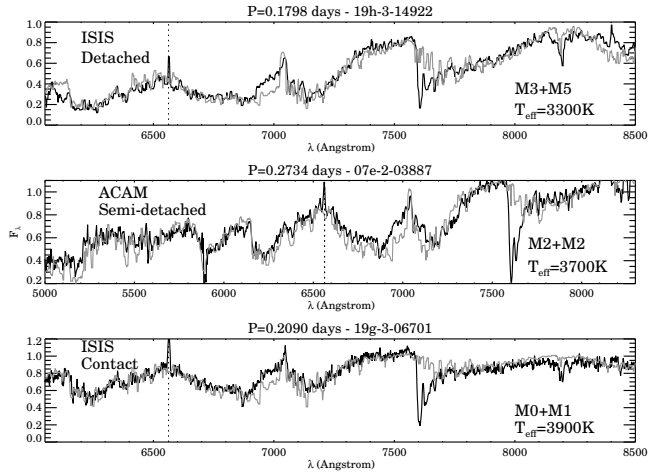}
\caption{\small{Two low resolution WHT ISIS red arm spectra and one ACAM spectrum of our short period binaries. The spectra are flux calibrated and normalised. {\bf{Top panel}}: 19h-3-14922, a detached system of $\sim$M3 and M5 stars. {\bf{Middle panel}}: 07e-2-03887, a (semi-)detached system of M2 stars. {\bf{Lower panel}}: 19g-3-06701 a contact system of $\sim$M0 and M1 stars. The grey curves are best-fit NextGen model spectra for solar metallicity, log(g)=5.0 and age 1.0 Gyrs. The dotted vertical line indicates the position of the $\textrm{H}_{\alpha}$ line at 6563 \AA, which is related to emission from the stellar chromosperes induced by spun up rotation in the binary.}}
\label{shortspec}
\end{figure}

In Figure \ref{shortspec} we show our three low resolution spectra (solid black lines) with overplotted best-fit model spectra (grey lines). In this figure we indicate approximate spectral types for the binary components by crude estimates based on the simplified assumption that the stars radiate as blackbodies. We use the relative depths of the primary and secondary eclipses to obtain the ratio of surface brightness and together with the ratio of radii, which we estimate from the square of the primary and secondary eclipse depths, we solve for the individual temperatures $\rm{T_{1,2}}$, analogous to Coughlin et al. (2011). For the conversion to spectral type we use the tabulated model spectral types as a function of $\rm{T}_{eff}$ from Baraffe \& Chabrier (1996). Note that the models consistently underpredict the flux around 6900 \AA, which may be related to missing opacities of TiO (Titanium-Oxide) in the code we use. We find that the $\rm{T}_{eff}$ derived from the SED fitting is generally lower than the 
corresponding temperature from the spectral template matching, typically by one to two subclasses, corresponding to 100-300 K, which we attribute to missing opacities in the optical. However, given that we estimate typical errors of 200-300K on the SED result, the two methods are still in agreement. In Table 2 we show $\rm{T}_{eff}$ for each binary candidate.

All three binaries show significant $\rm{H}_{\alpha}$ emission, which indicates chromospheric activity (the absense of the Lithium absorption line appears to rule out accretion from young stars). Using the IDL routine $\textit{feature}$ we estimate the equivalent width (EW) of the $\rm{H}_{\alpha}$. The calculated EW values ( -7.9 \AA$ $ for 19h-3-14922, -3.8 \AA$ $ for 07e-2-03887 and -5.7 \AA$ $ for 19g-3-06701) are consistent with or slightly higher than those corresponding to active single stars with similar spectral types (Riaz et al. 2006). The three systems have equivalent widths between 30-70\% of the empirical accretor/non-accretor division for their estimated spectral types (Barrado y Navascues \& Martin 2003). The presence of the Na (Sodium) absorption doublet around 8200 \AA$ $ and the absence of a strong Ca (Calcium) triplet in the near-IR indicate that the three candidates are high gravity objects, i.e. M-dwarfs and not giants.  

\begin{figure*}
\begin{center}$
\begin{array}{llllllll}
\includegraphics[width=0.5\textwidth]{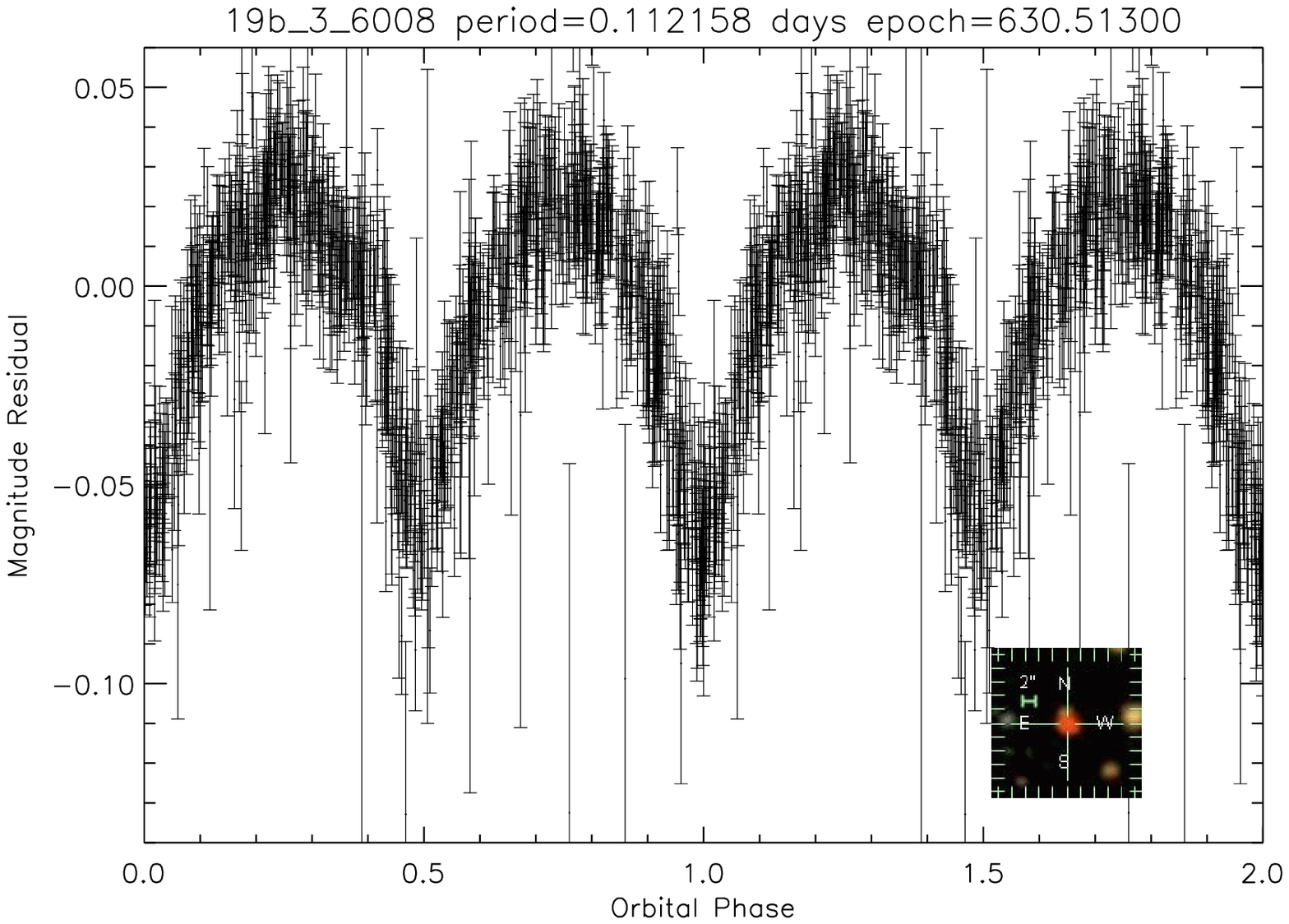} &
\includegraphics[width=0.5\textwidth]{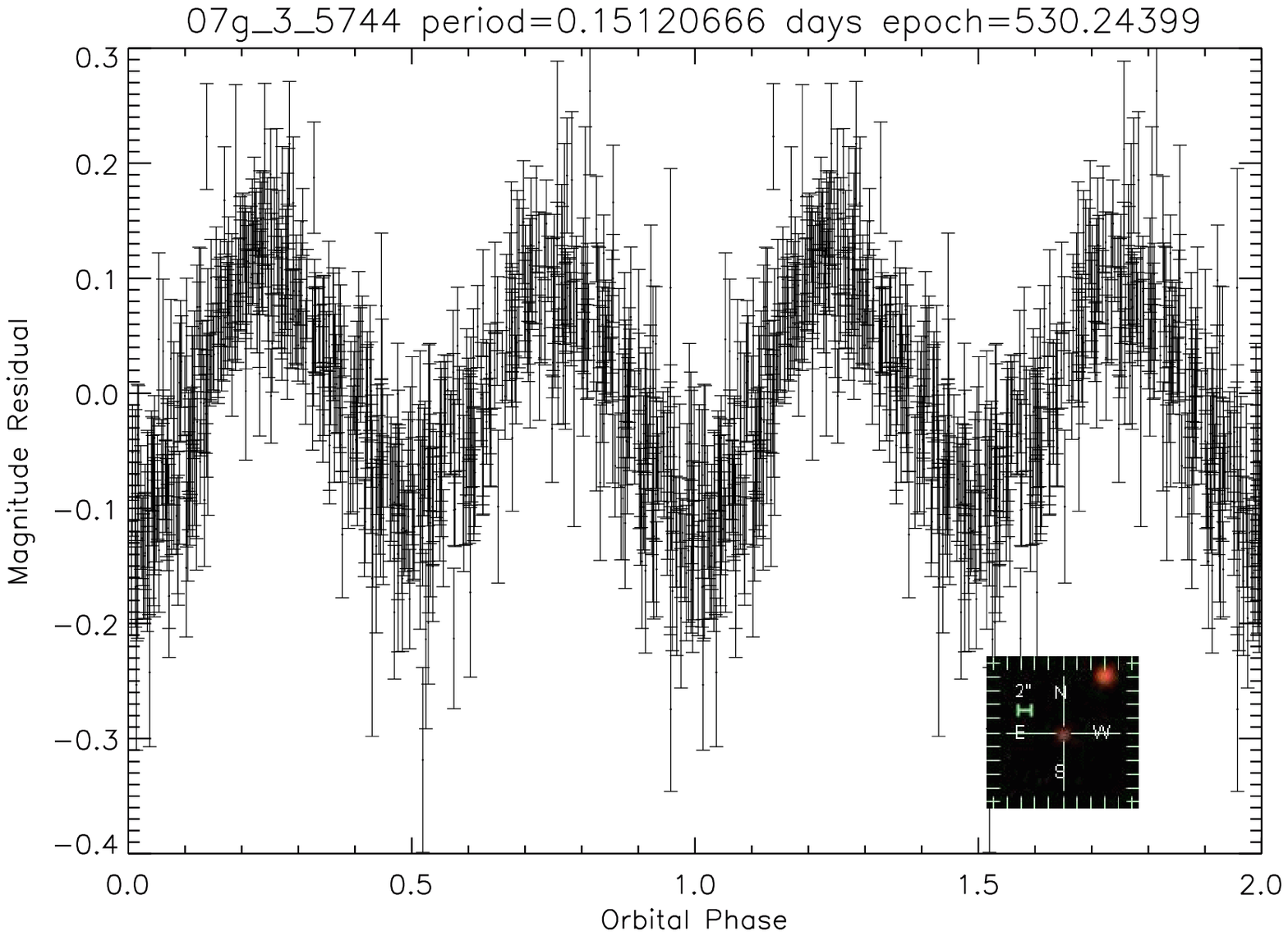} \\
\includegraphics[width=0.5\textwidth]{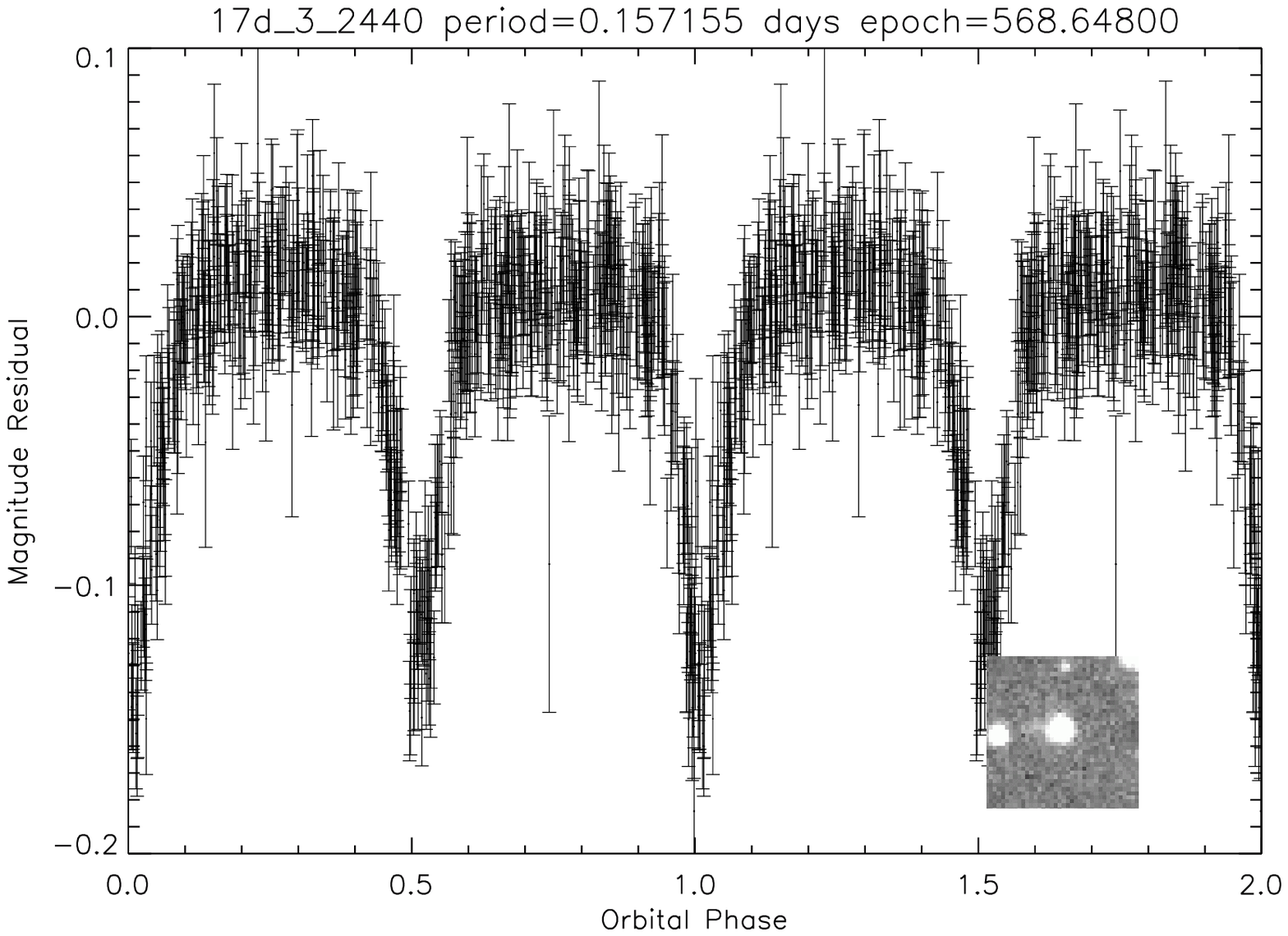} &
\includegraphics[width=0.5\textwidth]{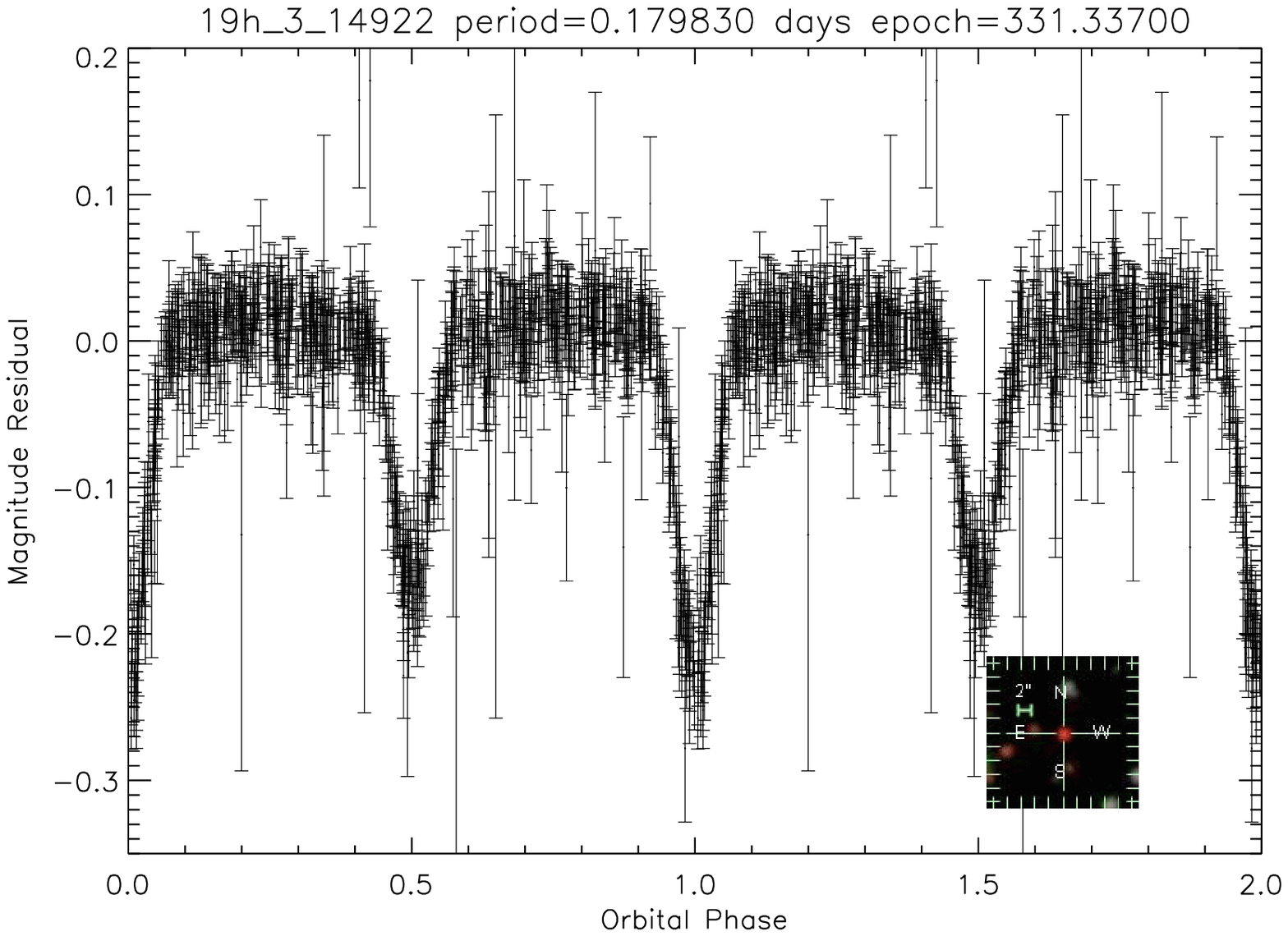} \\
\end{array}$
\end{center}
\caption{The (median subtracted) WFCAM J band lightcurves of the four candidate ultra-short period eclipsing M-dwarfs binaries, phase-folded to the best-fit primary eclipse epoch and orbital period. The boxes in the lower right corner of each plot show a three colour SDSS composite image of each source. For 17d-3-02440, which does not have SDSS coverage, we show the WFCAM J band image. A colour version of the thumbnails is available in the online paper.}
\label{BlowUp}
\end{figure*}

\begin{figure*}
\begin{center}$
\begin{array}{ll}
\includegraphics[width=.5\textwidth]{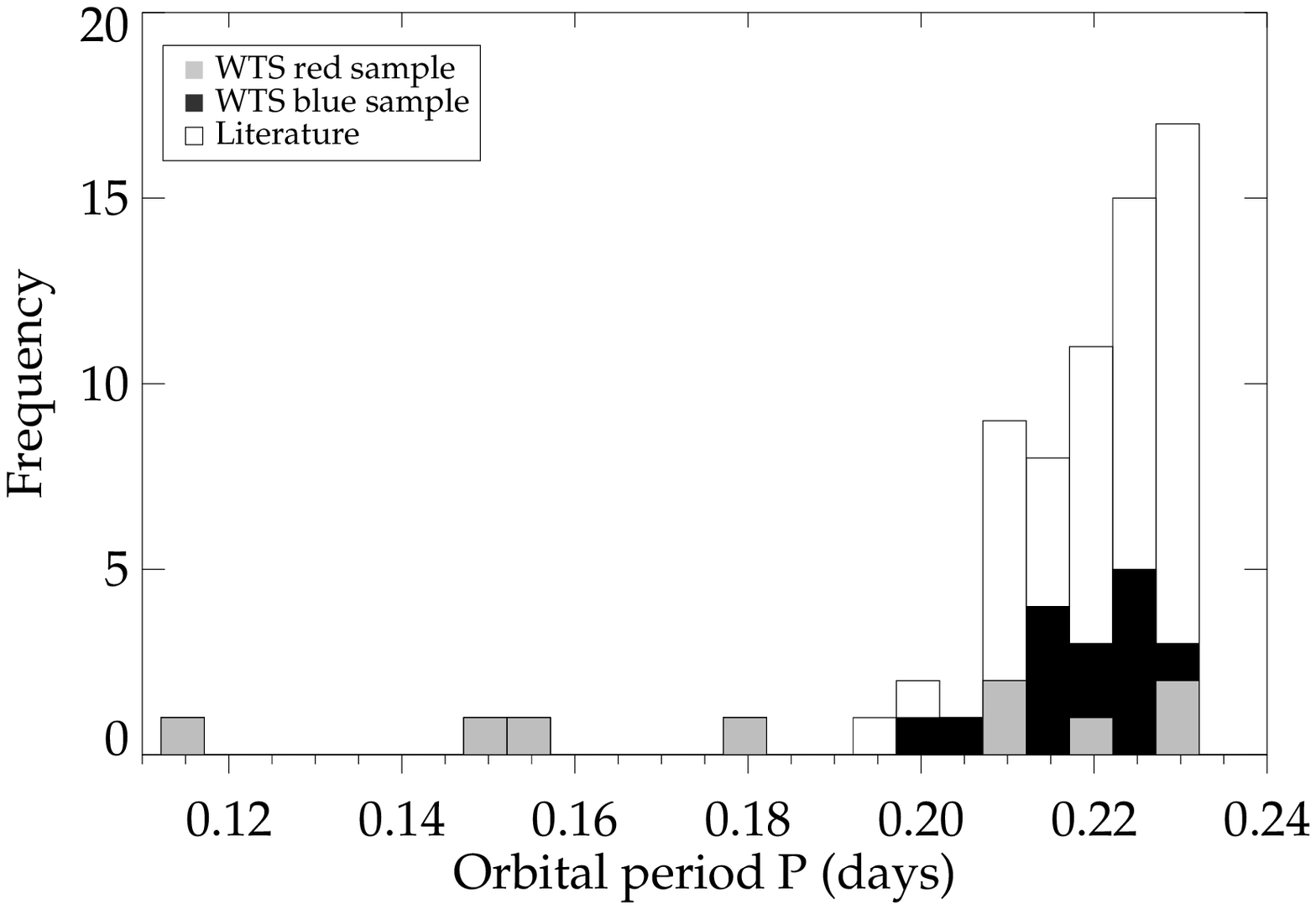}&
\includegraphics[width=.5\textwidth]{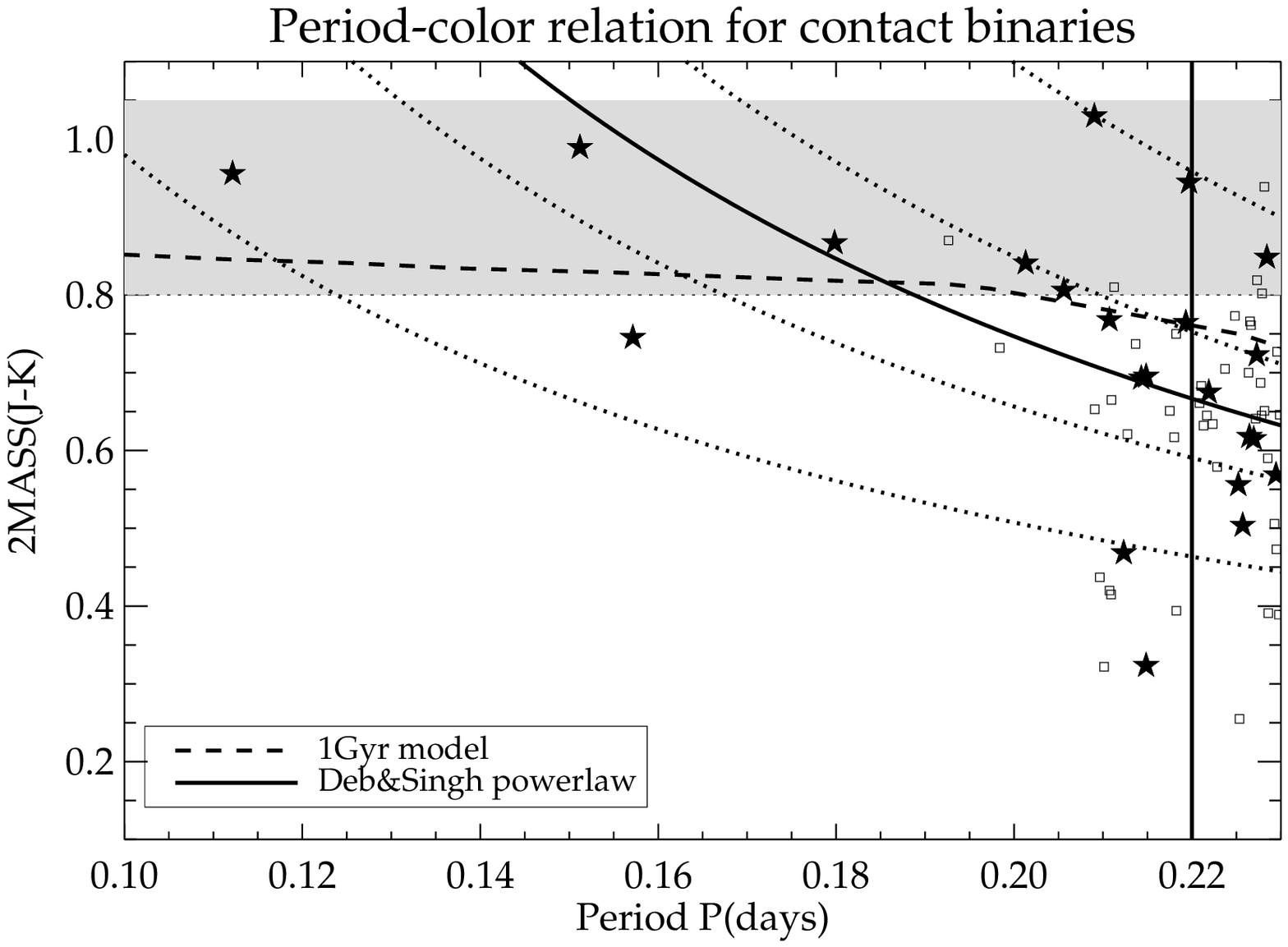}\\
\end{array}$
\end{center}
\caption{\small{ {\bf{Left panel}}: Period histogram of the WTS short period binaries (grey and black filled bars), combined with sources from the literature (white bars). {\bf{Right panel}}: The black stars indicate our WTS eclipsing binary sample with $P\leq$0.23 days, whereas open boxes are sources from the literature. The period-colour (J-K) power-law relationship for contact binaries from Deb \& Singh (2010) is indicated by the solid black curve. The dotted curves show the 1 and 3$\sigma$ boundaries on the power-law index of the correlation. The dashed curve shows the theoretically predicted colours for periods at Roche lobe overflow using Baraffe 1Gyr single star models. The grey band indicates the range of (J-K) colours for a sample of nearby active M-dwarfs classified as M4 (Riaz et al. 2006).}}
\label{periodcolor}
\end{figure*}

\section{Results \& Discussion} 
The final sample consists of 25 eclipsing binary systems with orbital periods shorter than 0.23 days (20,000 seconds), of which 9 have optical to near-infrared colours or optical spectra consistent with M-dwarfs. Four of these 9 have periods shorter than any known main-sequence binary. In Figure \ref{BlowUp} we show the WFCAM J band lightcurves of these four binaries phase-folded to the best-fit epoch and period determined from the box-fitting algorithm. Five binaries from the red sample with periods shorter than 0.23 days have Fourier coefficients ($a_2$,$a_4$) typical of a contact system (`C'), two binaries are identified as semi-detached systems (`SD'), whereas two other systems are detached (`D). From the blue sample, 9 out of 14 binaries ($\sim64\%$) are identified as contact systems.  We have summarised the binary characteristics for both the red sample and the blue sample in Table \ref{proptable}, including the Fourier classification. For all of the 25 binaries, lightcurves are shown and 
coordinates are given (in Table 3) in the Appendix.

The shortest period system, 19b-3-06008, is a binary consisting of two $\sim$M4 stars ($\rm{T}_{eff}\sim$3150 K) with P$\sim$0.1121 days. This binary appears to be close to contact. Assuming orbital synchronisation, the estimated masses and radii for this binary correspond to a fast rotational velocity between 15-25\% of the breakup speed (Herbst et al. 2001), but slow enough to keep the system stable. A second system, 07g-3-05744, has a period of $\sim$0.151 days, shows slightly unequal eclipses and a $\rm{T}_{eff}$ of 3400 K, suggesting this is a binary with spectral types of $\sim$M2-3. This system is identified as a contact system by the Fourier decomposition. Two other systems, 17d-3-02440 ($\sim$M1.5) and 19h-3-14922 (WHT ISIS spectrum indicating $\sim$M3 and M5 stars) with periods $\sim$0.1572 and $\sim$0.1798 days, are detached binaries with unequal ratio of primary versus secondary eclipse depth. Figure \ref{orbit} illustrates why these binaries are classified as detached systems despite their 
ultra-short periods, whereas 07d-2-02291 (a blue sample K type binary with P$\sim$0.2013 days) is a contact system; the critical Roche lobe period for contact depends on primary mass. Nevertheless, significant out-of-eclipse variability in the lightcurves of the four ultra-short period binaries indicates there is significant tidal interaction between the primary and secondary in every system.

\subsection{Comparison with previous studies}
This study probes for the first time the low-mass regime of ultra-short period eclipsing binaries in an extensive way. Recently, Norton et al. (2011) presented 43 eclipsing binary candidates identified in the SuperWASP Survey with periods shorter than 0.23 days, but only three objects in their sample have (J-K)$\geq$0.8 ($\sim$M0 spectral type). Their shortest period system is GSC2314-0530 (P$\sim$0.1926 days), which was modelled by Dimitrov \& Kjurkchieva (2010) as a semi-detached M-dwarf binary with components of 0.51$M_{\odot}$ and 0.26$M_{\odot}$ (corresponding to spectral types of $\sim$M0 and M4). Two other binaries in their sample have $P\leq$0.21 days (0.20908 and 0.20964 days). Maceroni \& Montalban (2004) found an almost twin M3 system (OGLE BW3 V38; semi-detached, P=0.1984 days) with strong chromospheric emission and masses of 0.44 and 0.41$M_{\odot}$. Other surveys found only binaries with P$>$0.21 days. Miller et al. (2010) found seven candidate contact binaries with periods between 0.2109 
and 0.23 days in a 0.25 square degree Galactic plane survey with the ESO 2.2m telescope, and Weldrake et al. (2004) found a P=0.2155 day system in the globular cluster 47 Tuc, whereas Pribulla et al. (2009) report a P=0.211249 day system. The left panel of Figure \ref{periodcolor} shows a histogram of all literature binaries plus our new systems with P$\leq$0.23 days (where grey and black filled bars indicate those from the WTS). The WTS sample follows the literature distribution, but suggests that rather than a sharpe cut-off at 0.22 days, the population extends into a tail of significantly shorter periods for lower-mass systems.

\subsection{The period-colour relation}
Our observations suggest a correlation between spectral type and the shortest possible binary period. This is in line with the period-colour relation as found by Eggen (1967) for contact binaries, implying that bluer systems have longer orbital periods. This is because for main-sequence stars the effective temperature (colour) is linked with stellar radius, and as a consequence orbital period for a contact binary. The period-colour relation can be used to identify genuine contact systems and weed out pulsating stars with comparable periods (i.e. $\delta$ Sct and $\gamma$Dor) and has been used for distance estimates (e.g. Rucinski 2007).  

Recently the same relation was presented for infrared colours by Deb \& Singh (2010) using 2MASS (J-K) photometry, which is relatively free from extinction effects. We have collected (J-K) colours from the literature of $P\leq0.23$ days binary systems and combined them with our WTS results in a period-near-infrared colour diagram in Figure \ref{periodcolor} (right panel). In this figure the stars indicate WTS objects, whereas the open boxes are literature sources. We have converted the WFCAM infrared photometry to 2MASS colours using the equations in Hodgkin et al. (2009). We extrapolated the best-fit powerlaw period-colour relation of Deb \& Singh (black solid line) to the ultrashort periods of our WTS binaries. The dotted curves show the 1 and 3$\sigma$ boundaries of this power-law index. There is significant scatter in the period versus colour sample, yet, $\sim$65\% of all systems with $P\leq$0.23 days fall within the 1$\sigma$ limits of the correlation, indicating that the majority of our systems 
are consistent with (near-)contact binaries rather than pulsators. However, we believe this scatter is intrinsic (the 2MASS J and K bands are obtained at the same epoch) and could be due to varying degrees of contact or age differences within the sample. Our shortest period binary system (19b-3-06008, P$\sim$0.1121 days) is an outlier, indicating that a simple powerlaw is not sufficient at ultra-short periods. This is because for mid- and late M-dwarfs (J-K) colour is nearly constant with spectral type and as a result the period-colour relation breaks down. We used Baraffe (1998) models to convert the maximum binary mass for a certain period to the corresponding infrared colours for a single star with the same $\rm{T}_{eff}$ as the binary. We plot this for equal-mass binaries as a dashed line in the right-hand panel of Figure \ref{periodcolor}. It is consistent  with the Deb \& Singh relation down to $\sim$0.20 days within their 1$\sigma$ uncertainties, but is a better fit to the data at shorter periods. 
Riaz et al. (2008) performed a spectroscopic study of 1080 nearby active M-dwarfs and obtained 2MASS colours. The grey shaded region indicates (J-K) colours for their M-dwarfs classified as M4, showing that the infrared colours for the ultra-short period binaries are consistent with observations of active single mid-type M-dwarfs.
\begin{table*}
\tiny{
   \begin{tabular*}{1.0\textwidth}
   {@{\extracolsep{\fill}}|llllllllllllllll|}
   \hline
Red sample  & & & & & & & & & & & & & & &\\ 
($P\leq0.23$ days) & & & & & & & & & & & & & & &\\ 
9 objects & & & & & & & & & & & & & & &\\ 
Name        &period       &epoch &$\frac{d_2}{d_1}$     &u        &g        &r
      &i       &z       &J &H &K  &(r-i)   &(i-z) &State &$T_{eff}$\\      
\hline
19b-3-06008 &0.11215791  &630.51306  &0.83  &24.43  &20.74  &19.26  &17.83 &17.10  &15.75 &15.16 &14.80 &1.43   &0.73  &SD &3150\\
07g-3-05744 &0.15120666  &530.20618  &0.93  &23.16  &22.26  &21.53  &20.30 &19.52  &17.91 &17.48 &16.96 &1.23   &0.78  &C &3400\\
17d-3-02440 &0.15715549  &568.64622  &0.97  &-      &21.08  &19.81  &18.76 &-      &16.91 &16.38 &16.21 &1.05   &-     &D &3600\\
19h-3-14922 &0.17983010  &331.33704  &0.86  &24.83  &23.06  &21.16  &19.51 &18.50  &16.87 &16.23 &16.05 &1.65   &1.00  &D &3300\\
19g-3-06701 &0.20903132  &320.24438  &0.77  &24.52  &20.62  &19.07  &18.05 &17.57  &16.20 &15.61 &15.22 &1.01   &0.49  &C &3700\\
07c-4-05645 &0.21072274  &578.31655  &0.91  &21.85  &19.18  &17.85  &17.28 &16.97  &15.62 &14.94 &14.89 &0.57   &0.32  &C &4150\\
19e-3-05704 &0.21968029  &323.13207  &0.91  &24.81  &21.92  &19.45  &18.85 &18.40  &17.19 &16.51 &16.30 &0.60   &0.45  &SD &3950\\
19c-2-08140 &0.22728102  &321.37419  &0.97  &21.18  &18.54  &17.43  &16.90 &16.64  &15.59 &15.00 &14.86 &0.53   &0.26  &C &4300\\ 
19e-4-00861 &0.22842771  &332.28102  &0.80  &23.57  &20.35  &18.95  &18.33 &18.04  &16.74 &16.02 &15.95 &0.62   &0.29  &C &4100\\
\hline
Red sample & & & & & & & & & & & & & & &\\
(P$\leq$0.30 days) & & & & & & & & & & & & & & &\\
8 objects & & & & & & & & & & & & & & &\\
\hline
19c-1-09997 &0.24660068  &335.22912  &0.91  &23.90  &19.73  &18.46  &17.92 &17.66  &16.83  &16.37 &16.35 &0.54  &0.26 &C &4350\\
03f-1-01082 &0.25016408  &518.31103  &0.99  &23.67  &21.11  &19.51  &18.63 &18.05  &16.55  &16.34 &15.63 &0.88  &0.58 &C &3400\\
03b-3-02411 &0.26251989  &1050.5354  &0.61  &23.71  &21.88  &20.15  &18.92 &18.17  &16.76  &16.21 &16.03 &1.23  &0.75 &D &3250\\
19f-1-07389 &0.26986911  &324.49300  &0.89  &21.19  &18.72  &17.48  &16.95 &16.73  &15.43  &14.95 &14.64 &0.53  &0.21 &D &4250\\
07h-4-03156 &0.27032161  &551.23573  &0.71  &20.28  &17.91  &16.71  &16.19 &15.89  &14.75  &14.10 &13.91 &0.52  &0.30 &C &4200\\
19d-4-05861 &0.27219697  &336.29275  &0.70  &21.96  &19.73  &18.83  &18.31 &18.03  &17.14  &16.41 &16.49 &0.52  &0.28 &C &4650\\
07e-2-03887 &0.27335648  &533.21993  &0.95  &21.69  &19.51  &18.10  &17.10 &16.52  &15.25  &14.68 &14.39 &1.01  &0.57 &SD &3700\\
19a-4-04542 &0.29367660  &324.31748  &0.74  &22.53  &20.22  &18.73  &17.88 &17.39  &16.08  &15.40 &15.64 &0.85  &0.49 &D &3900\\

\hline
Blue sample & & & & & & & & & & & & & & &\\
(P$\leq$0.23 days) & & & & & & & & & & & & & & &\\
14 objects & & & & & & & & & & & & & & &\\
\hline
07d-2-02291 &0.20130714  &594.33319  &1.00  &21.20  &19.35  &18.45  &18.13 &17.95  &16.80  &16.01 &16.00 &0.33   &0.18 &C &4650\\
07f-1-05360 &0.20561088  &783.57709  &0.90  &22.96  &20.52  &19.42  &19.01 &18.79  &17.66  &17.13 &16.90 &0.42   &0.22 &SD &4450\\
07b-4-06173 &0.21233759  &785.57965  &1.00  &20.89  &18.74  &17.79  &17.39 &17.26  &16.04  &15.58 &15.61 &0.40   &0.13 &C &4750\\
07a-1-03517 &0.21430452  &553.24191  &0.81  &19.63  &17.64  &16.82  &16.45 &16.23  &15.05  &14.60 &14.36 &0.37   &0.22 &C &4800\\
07c-3-05395 &0.21484877  &528.30800  &0.85  &21.30  &19.10  &18.20  &17.83 &17.61  &16.59  &16.21 &16.30 &0.36   &0.22 &SD &4850\\
07g-3-04935 &0.21484957  &876.36140  &0.96  &21.30  &19.10  &18.20  &17.83 &17.61  &16.63  &16.23 &15.97 &0.36   &0.22 &SD &4800\\
03e-4-02972 &0.21931536  &1155.4332  &0.92  &22.07  &19.73  &18.79  &18.36 &18.42  &17.06  &16.35 &16.36 &0.42   &-0.06 &C &4650\\
03e-4-03812 &0.22189564  &534.24769  &0.90  &19.27  &18.01  &17.53  &17.31 &17.23  &16.26  &16.06 &15.56 &0.22   &0.09 &SD &5600\\
19h-1-10160 &0.22338883  &374.31584  &0.85  &21.09  &21.12  &18.03  &17.58 &17.27  &15.77  &- &- &0.45   &0.31 &C &-\\
19c-3-12555 &0.22521823  &324.36446  &0.18  &18.77  &17.40  &16.76  &16.49 &16.38  &15.30  &14.99 &14.74 &0.27   &0.11 &C &5250\\
19b-2-05229 &0.22570901  &336.39181  &0.97  &21.78  &19.39  &18.38  &17.92 &17.65  &16.68  &16.24 &16.22 &0.46   &0.28 &C &4650\\
19a-2-10658 &0.22645761  &334.36200  &0.89  &18.72  &17.15  &16.54  &16.36 &16.19  &15.17  &14.79 &14.56 &0.18   &0.17 &C &5300\\
07g-3-00820 &0.22695102  &528.30249  &0.87  &17.10  &15.32  &14.60  &14.30 &14.15  &13.23  &12.76 &12.63 &0.30   &0.14 &C &5100\\
19g-4-11051 &0.22945028  &317.38529  &0.97  &20.26  &18.28  &17.50  &17.16 &16.96  &16.14  &15.63 &15.61 &0.35   &0.20 &SD &4900\\
  \hline  
     \end{tabular*}
}
\caption{Source property table for our red short period sample, $P\leq$0.30 days and (r-i)$\geq$0.5 and (i-z)$\geq$0.25, and our blue sample with $P\leq$0.23 days. The epoch column shows MJD-245000. The ratio $d_2/d_1$ indicates the secondary eclipse depth over the primary eclipse depth in the phasefolded J band lightcurve, determined by median averaging the data in a
phase range of $\pm$0.01 around phase 1.0 and 0.5. Quoted magnitudes are SDSS DR7 ugriz and WFCAM JHK, rounded off to two decimals. Our full SEDs have 8 band grizYJHK photometry. The state column indicates the estimate from Fourier fitting (Section 4.1) of the morphological state of the binary ('D' for detached, 'SD' for semi-detached and 'C' for contact.).}
\label{proptable}
\end{table*}

\subsection{Constraints to binary evolution scenarios}
Our results imply that for M-dwarf binaries the orbital periods can extend significantly beyond the proposed period cut-off of $\sim$0.22 days. For our (near-)contact M-dwarf system 19b-3-06008, the fitted $T_{eff}\sim$3150 K, which corresponds to a mass of 0.15-0.25$\mathcal{M}_{\odot}$, implies that there is insufficient angular momentum loss over the Hubble time within the AML framework to explain the observed orbital period with starting periods of $P_0$=2.0-1.5 days. For the $P_0$=1.0 day model and a mass-ratio of $\sim$0.3 the contact timescale is of the order of the Hubble time. However, the near-equal ratio of secondary to primary eclipse depths suggests that this binary is rather near-equal-mass, unless the secondary is extremely bloated in radius with respect to its mass. Also, a mass-ratio of 0.3 and a primary mass of 0.2$\mathcal{M}_{\odot}$ would bring the secondary mass uncomfortably close to the hydrogen burning limit, whereas an intrinsically dark and low-mass secondary such as a brown 
dwarf would be unlikely to produce the observed secondary eclipse depth and the out-of-eclipse variations. Again, for the other three ultra-short period binaries only the $P_0$=1.0 day model can reproduce the observed orbits for the estimated primary mass, under the requirement that we assume $\mathcal{M}_{1}>0.35\mathcal{M}_{\odot}$ and/or an extreme mass ratio, the latter which may be the case for 19h-3-14922 (where we estimate that q could be as low as 0.5). Given our estimates of the minimum birth separation for M-dwarfs (Figure \ref{P0plot}), it seems unlikely that these binaries were formed at such short periods.

Another possibility is that the evolution of M-dwarf binaries is faster than expected. M-dwarfs are known to be active and flaring stars. West et al. (2011) find an activity fraction of 40-80\% for M4-M9 dwarfs, which implies that scaling solar type stars to M-dwarfs is not trivial. It is possible that because of the (near-)convective nature of mid-to-late type M-dwarfs that the ultra-short periods of our binaries, when synchronised with the stellar rotation, cause the magnetic field lines to be significantly twisted (the $\alpha^2$ dynamo; Radler et al. 1990). In other words, the topology of the magnetic fields could be significantly different from that of solar-type stars, which may directly affect the overall activity and as a consequence the rate at which angular momentum is lost. 

Alternatively, the formation mechanism for M-dwarfs may be different from that of earlier type main sequence stars. It is possible that during the pre-main sequence phase an excess amount of angular momentum is removed which accelerates the orbital evolution. Possible sources of such enhanced evolution could be dynamical interactions with other stars, accretion of the surrounding material and/or interaction with a circumbinary disk (Pringle 1991; Artymowicz et al. 1991; Bate \& Bonnell 1997). The hydrodynamical calculations of binary formation by fragmentation of Bate et al. (2002) indicate that proto-binaries potentially form with large separations ($\geq$ 10 AU) and go through a phase of accretion and orbital evolution towards tighter orbits ($\sim$ 1 AU). Less is known about the later phases of orbital evolution. In this paper, we have demonstrated that periods of $\sim$1.0 day would be required at early age to explain the observed ultra-short periods for our M-dwarf binaries, if magnetic braking is 
the dominant mechanism by which low-mass stars evolve over Gyrs timescales.

Recall that Jiang et al. (2011) concluded that contact M-dwarf binaries (primary mass $\leq$0.63$M_{\odot}$) are short-lived and thus detections of them during this `special' phase of their evolution would be rare. This is because they predict an instability in mass-transfer at the point at which the primary fills its Roche lobe which merges the system almost instantaneously. In combination with the findings by Ge et al. (2010), who show that the instability of mass transfer occurs promptly upon Roche lobe filling of the primary (a semi-detached binary) for stars with a significant convection zone, it seems that the existance of our new near-contact M-dwarf systems, 19b-3-06008 and 07g-3-05744, among $\sim$10,000 M-dwarf sources in the WTS Survey, is unusual. However, detached short-period low-mass binaries are allowed in this model, because the angular momentum loss timescale is significantly shorter than in the Stepien AML model for M-dwarfs. Our detached ultra-short period systems, 17d-3-02440 and 
19h-3-14922, thus appear to fit within the Jiang et al. model predictions.

\section{Conclusion}
In this paper we have presented the results of an extensive search for short-period low-mass eclipsing binary systems in the near-infrared J-band lightcurves of the WFCAM Transit Survey. Probing over 10,000 M-dwarfs down to J=18 we report on the discovery of four ultra-short period (P$\leq$0.18 days) M-dwarf binaries. Their periods are significantly shorter than of any other known main-sequence binary system. All four are below the observed sharp period cut-off at $P\sim0.22$ days as seen in binaries of earlier type stars. The shortest-period binary consists of two M4 type stars in a $P=0.112$ day orbit. In total, we find 25 binaries with orbital periods shorter than 0.23 days of which 9 are identified as likely M-dwarf systems through their broad-band colours or spectra. Of these 9 systems, 5 are indicated as potential contact systems. These detections pose a direct challenge to popular theories that explain the evolution of short period binaries by loss of angular momentum through magnetised winds, or 
by unstable mass-transfer, which predict timescales that are either too long for their formation, or timescales that are too short to observe them in the contact phase. Our discovery of binaries with significantly shorter orbital periods than the previously observed cut-off implies that either these timescales have been overestimated for M-dwarfs, e.g. due to a higher effective magnetic activity, or that the mechanism for forming these tight M-dwarf binaries is different from that of earlier type main-sequence stars.
\section*{Acknowledgments}
{\small{We thank the anonymous referee, whose comments led to a significant improvement of the manuscript. SVN, JLB, IAGS, SH, DJP, RPS, JK, DB, ELM, CdB and YVP have received support from RoPACS during this research, and BS, GK, DM, PC, NG, HS, JZ and MC are supported by RoPACS, a Marie Curie Initial Training Network funded by the European Commission’s Seventh Framework Programme. CdB acknowledges the Fundacao para a Ciencia e a Tecnologia (FCT) through the project Pest-OE/EEI/UI0066/2011. }}

\section*{Appendix}
\begin{figure*}
\begin{center}$
\begin{array}{llllllll}
\includegraphics[width=0.5\textwidth]{ResultsRevised06008.ps} &
\includegraphics[width=0.5\textwidth]{ResultsRevised05744.ps} \\
\includegraphics[width=0.5\textwidth]{ResultsRevised02440.ps} &
\includegraphics[width=0.5\textwidth]{ResultsRevised14922.ps} \\
\includegraphics[width=0.5\textwidth]{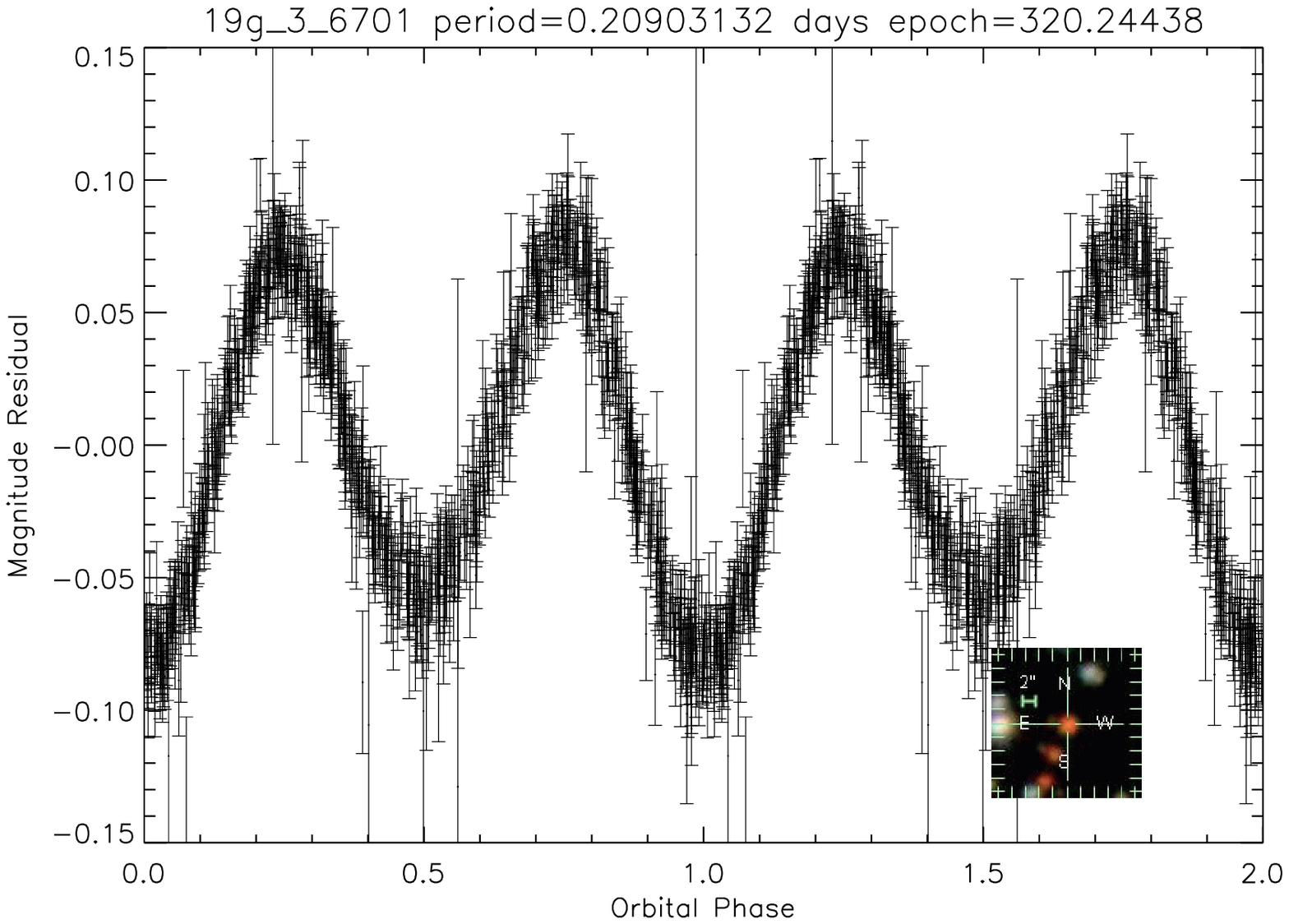} &
\includegraphics[width=0.5\textwidth]{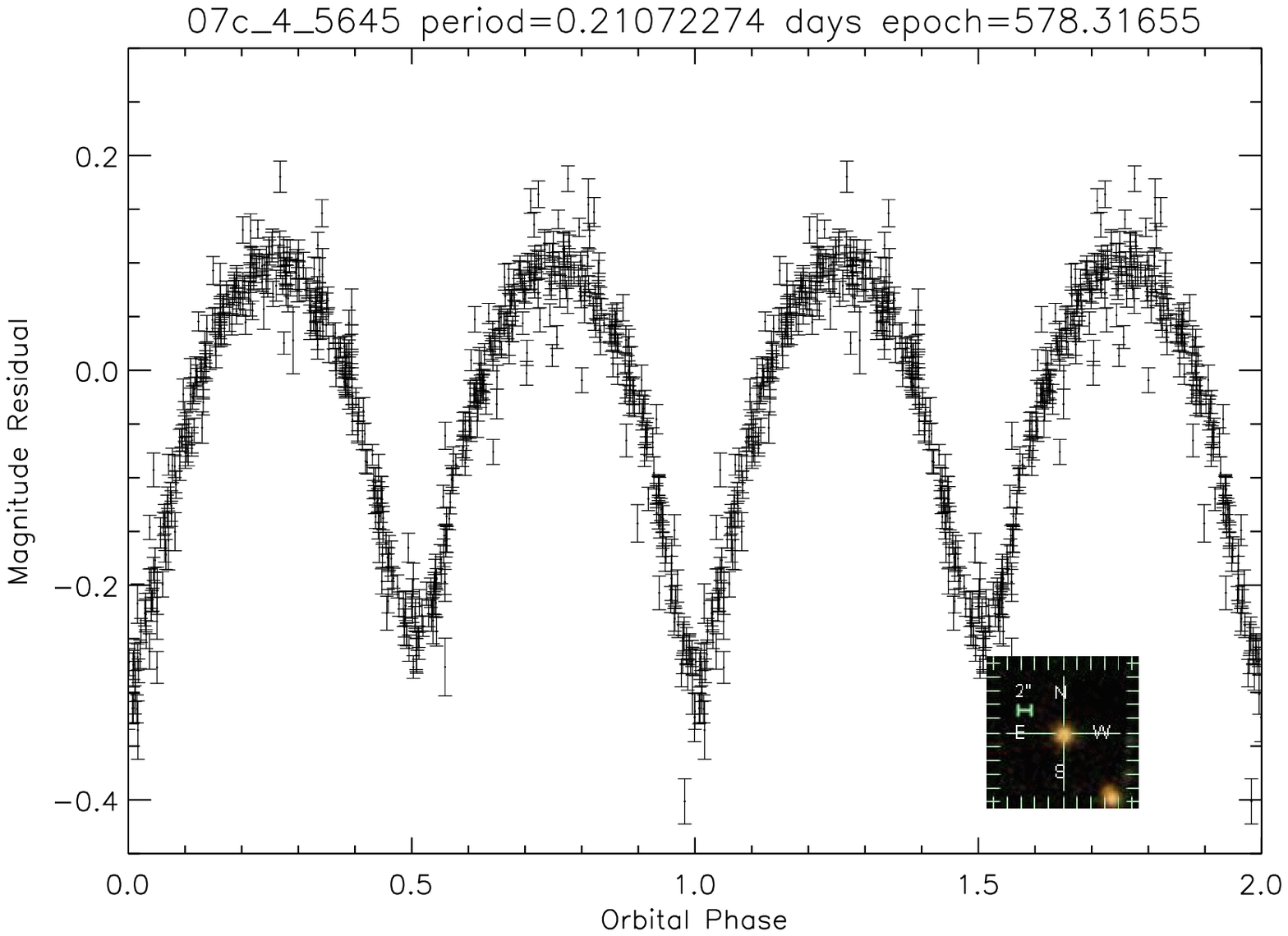} \\
\end{array}$
\end{center}
\caption{The red sample of candidate eclipsing binaries, which satisfies the colour constraints (r-i)$\geq$0.5 and
(i-z)$\geq$0.25, ordered by increasing period $P\leq$0.23 days. We plot two full binary orbits. The boxes in the lower right corner
of each plot show a three colour SDSS composite image of each source. For 17d-3-02440, which does not have SDSS coverage, we show the
WFCAM J band image. A colour version of the thumbnails is available in the online paper.}
\end{figure*}

\begin{figure*}
\begin{center}$
\begin{array}{ll}
\includegraphics[width=0.5\textwidth]{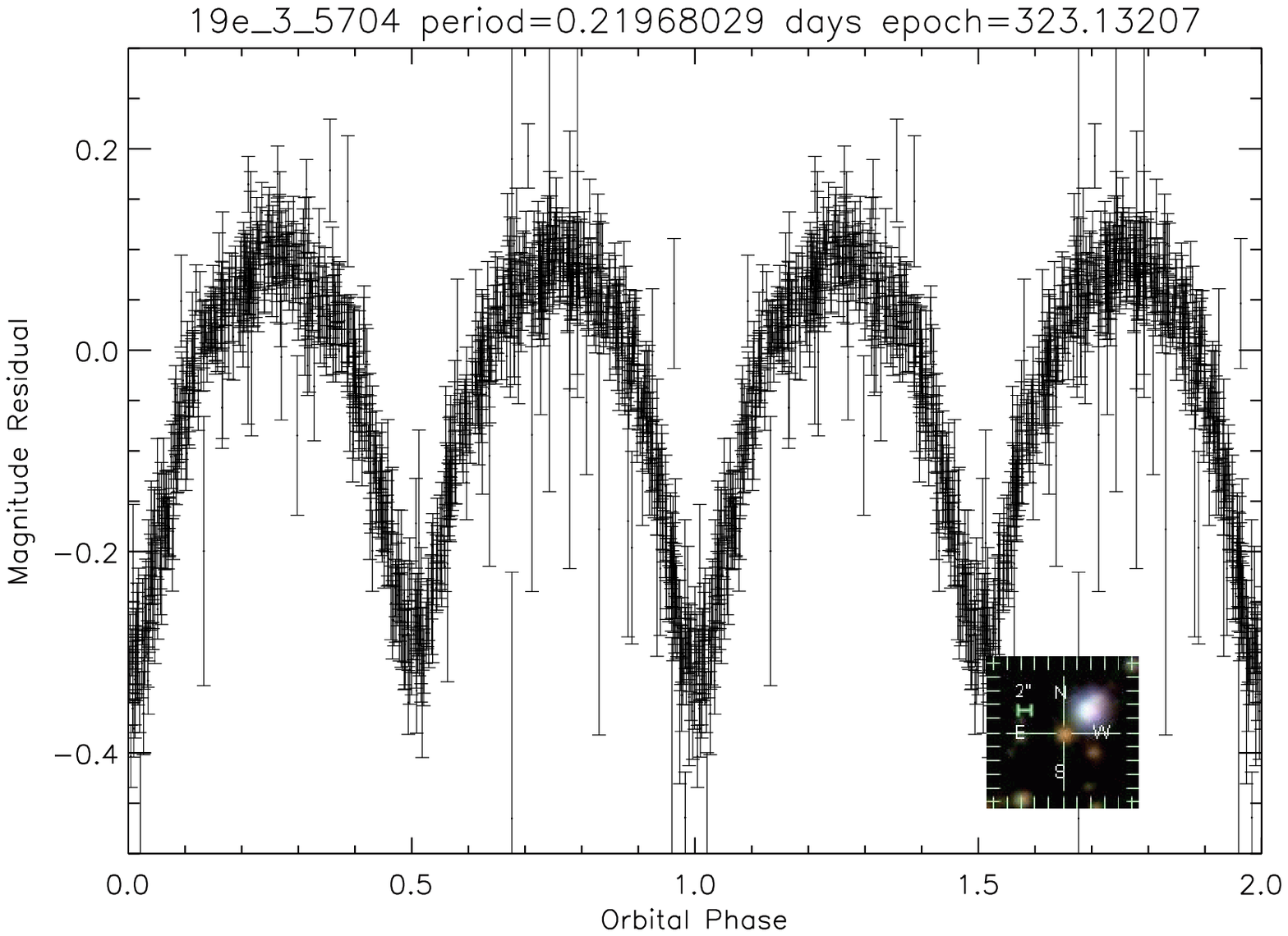} &
\includegraphics[width=0.5\textwidth]{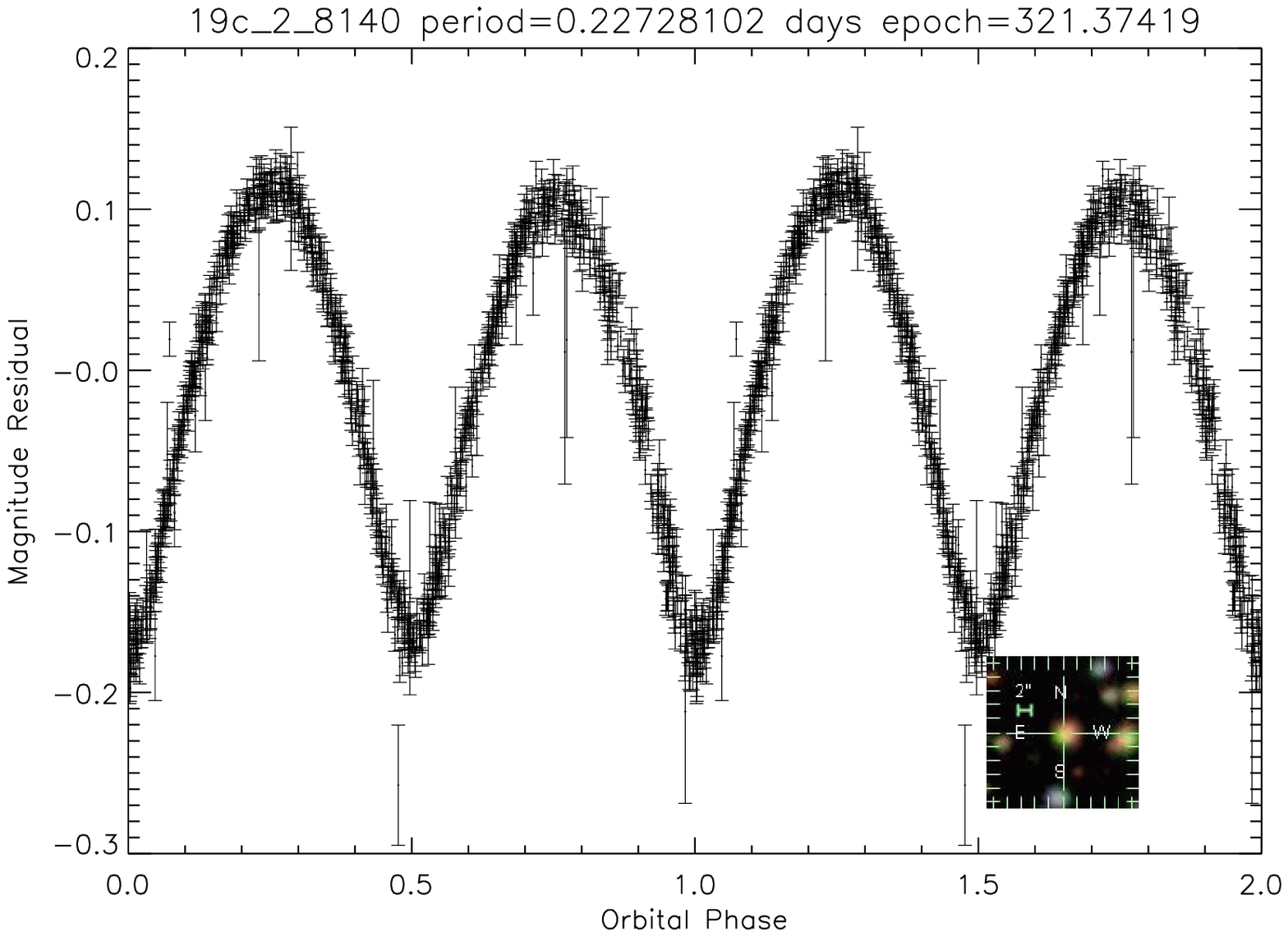} \\
\includegraphics[width=0.5\textwidth]{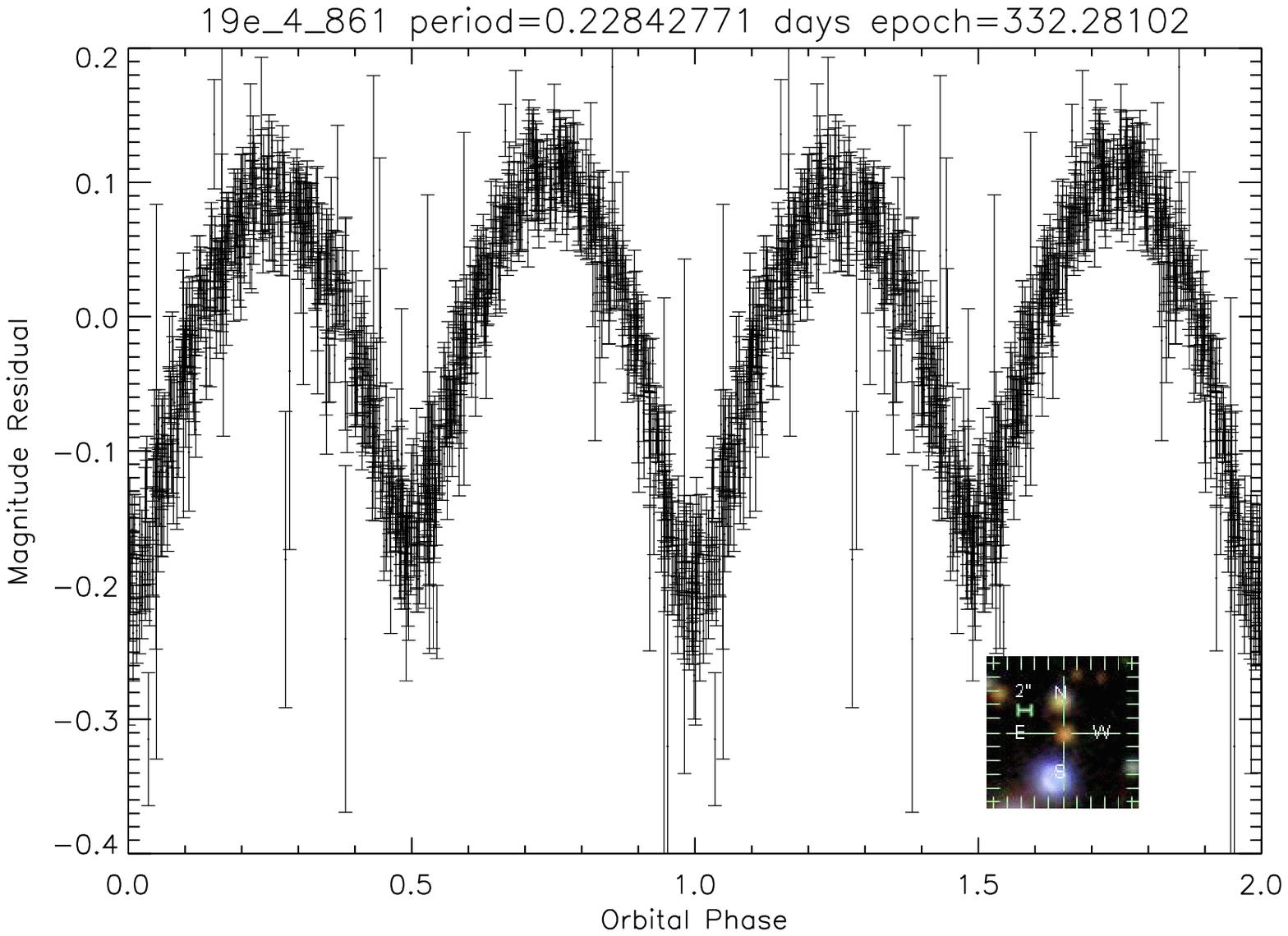} &
\end{array}$
\end{center}
\caption{Red sample with $P\leq$0.23 days continued.}
\end{figure*}

\begin{figure*}
\begin{center}$
\begin{array}{llllll}
\includegraphics[width=0.5\textwidth]{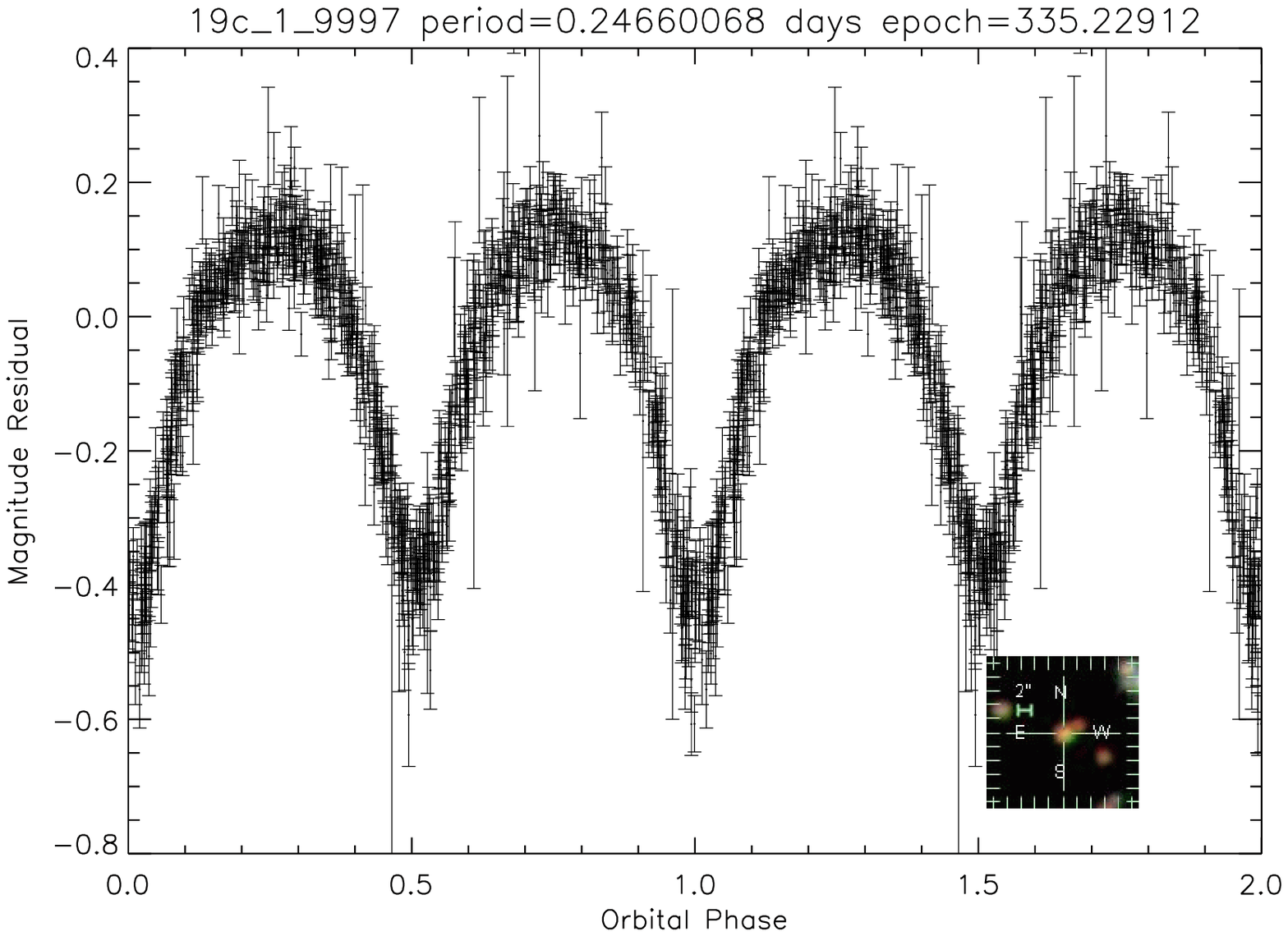} &
\includegraphics[width=0.5\textwidth]{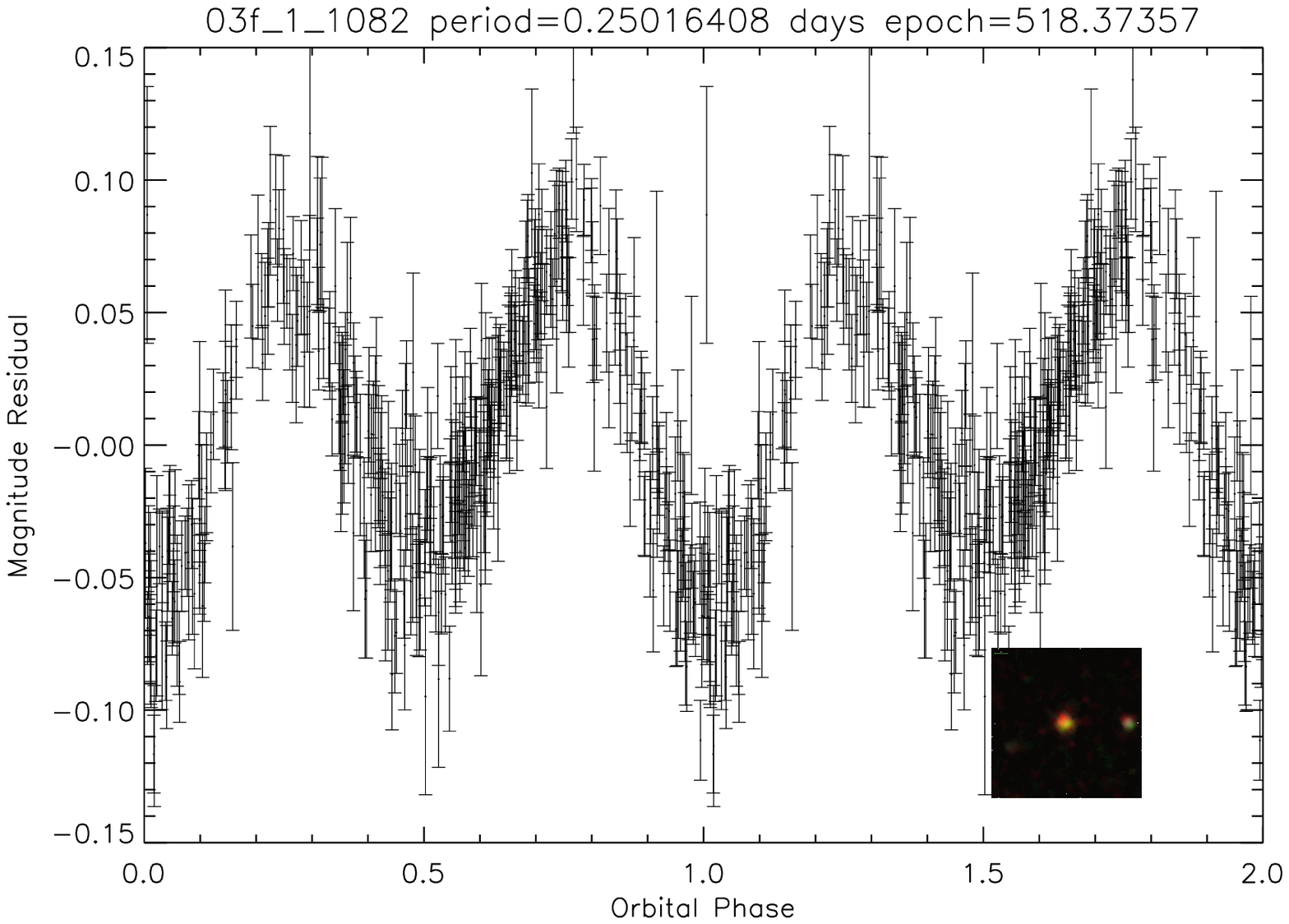} \\
\includegraphics[width=0.5\textwidth]{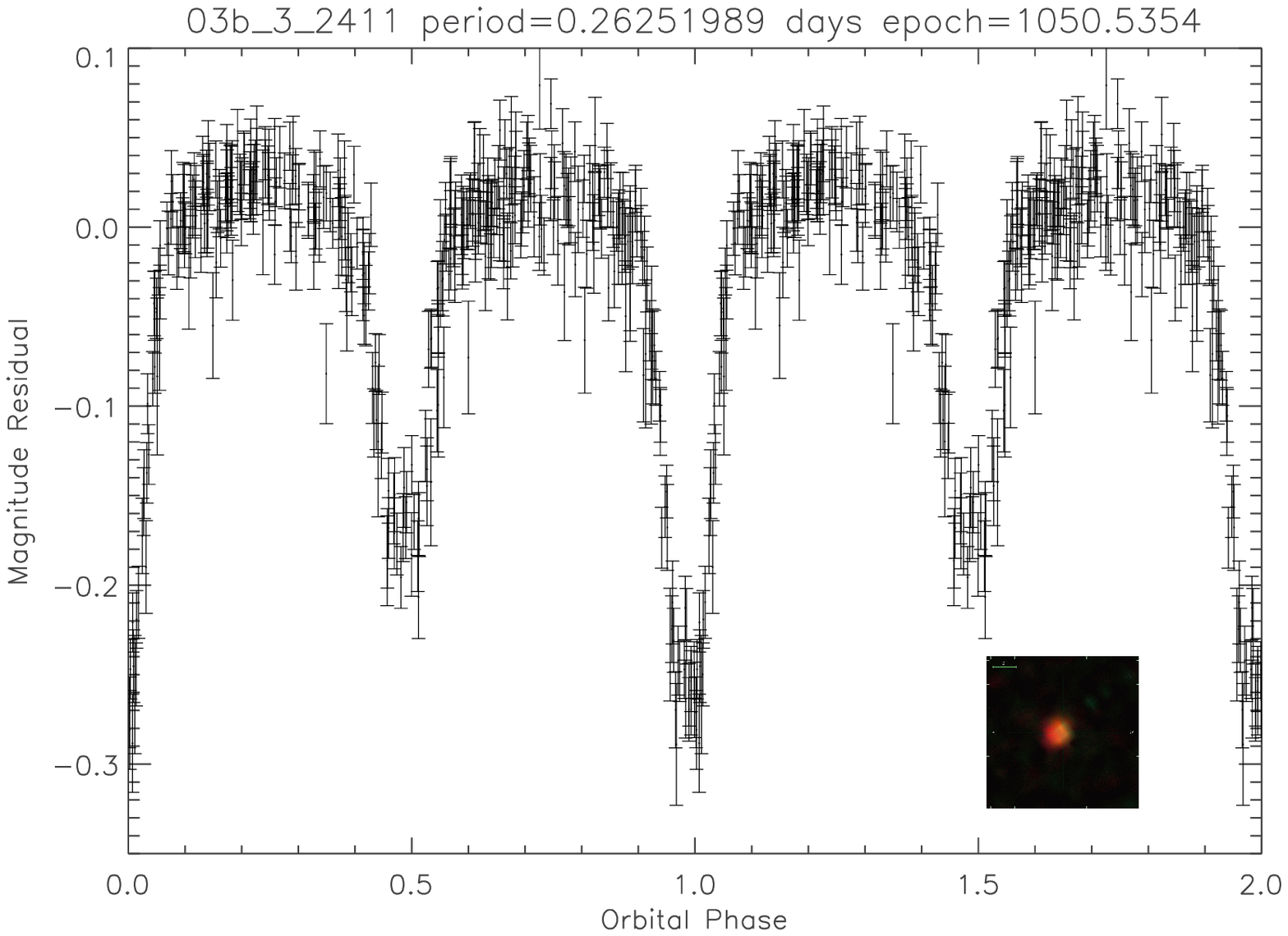} &
\includegraphics[width=0.5\textwidth]{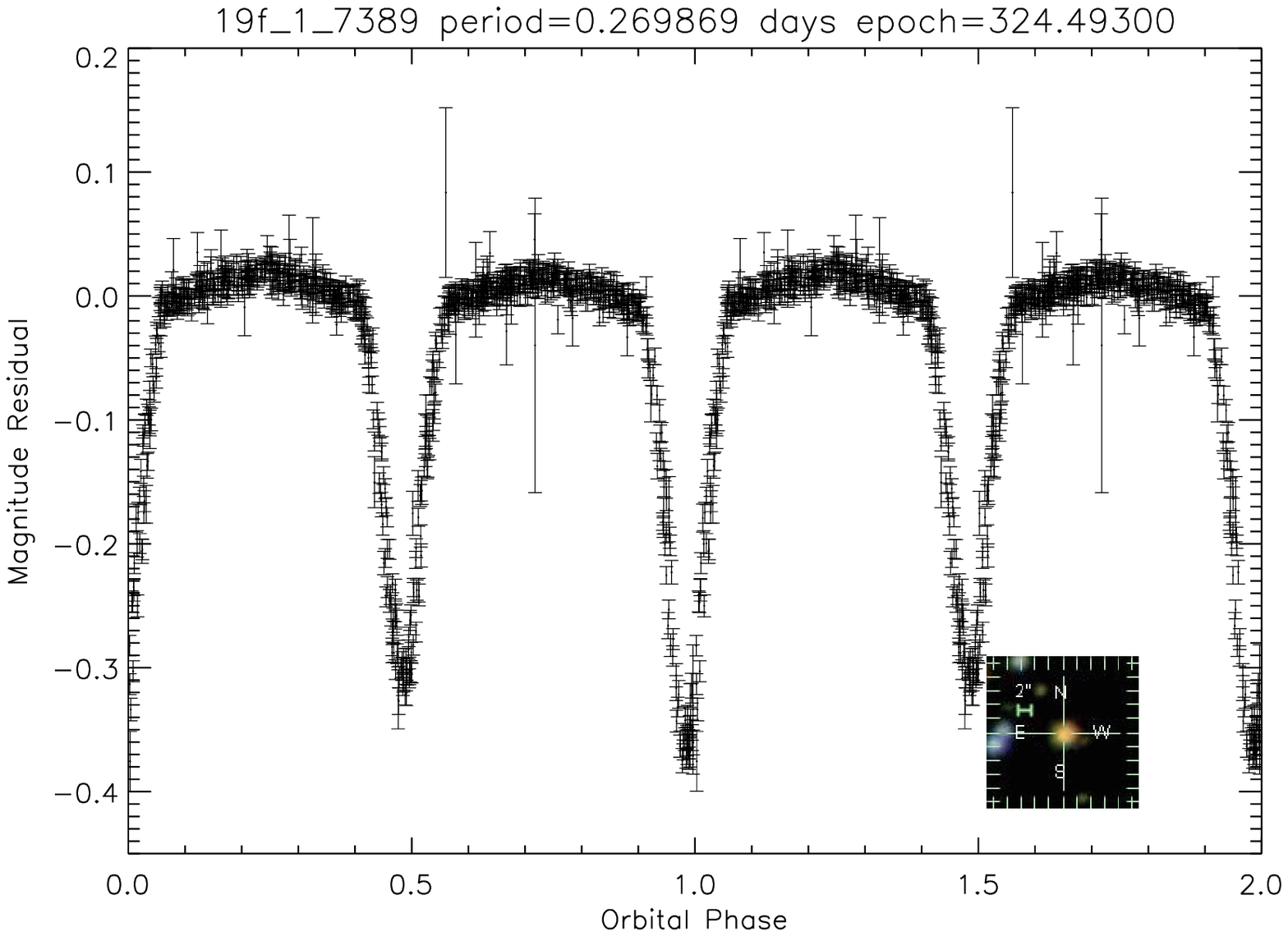} \\
\includegraphics[width=0.5\textwidth]{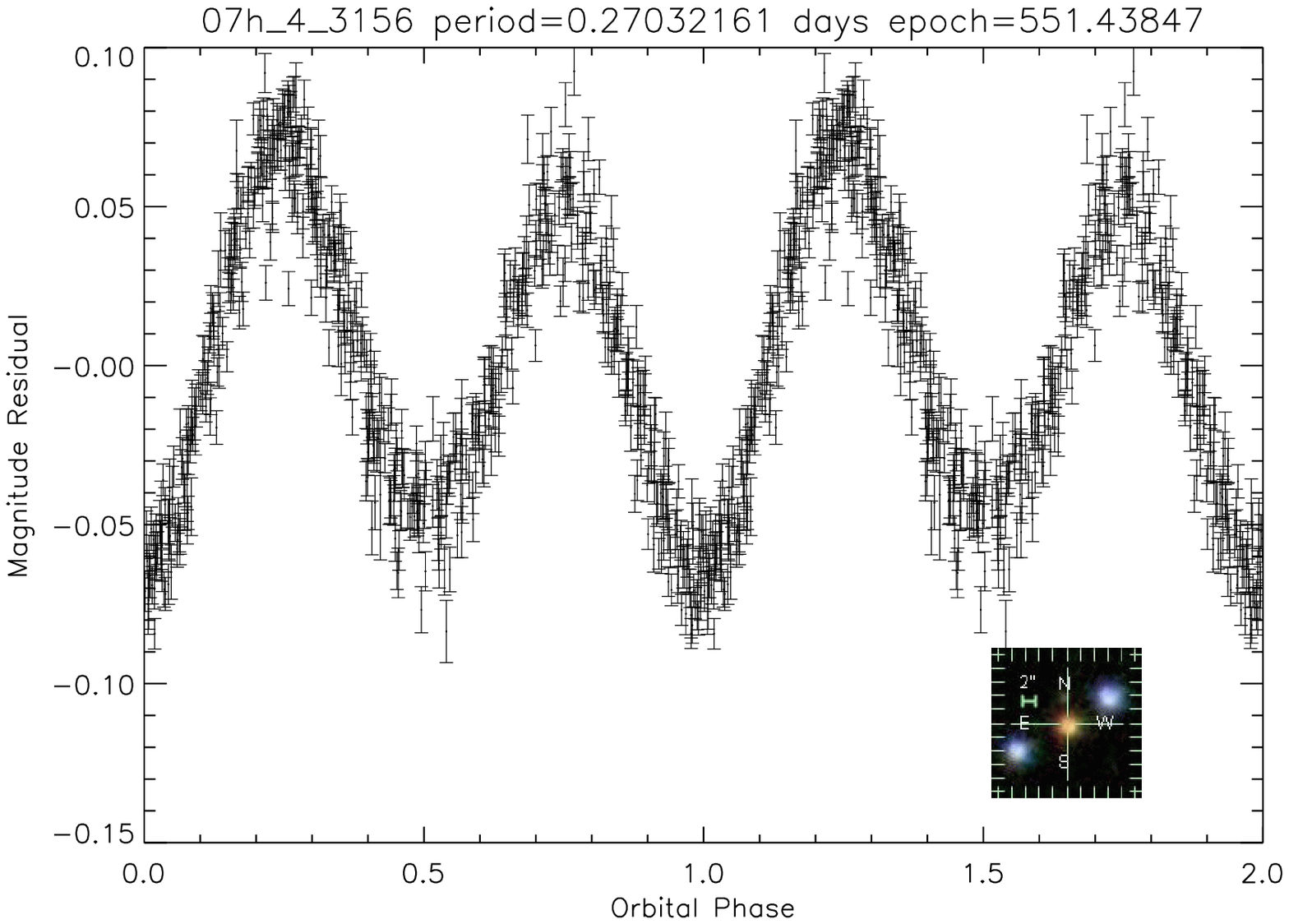} &
\includegraphics[width=0.5\textwidth]{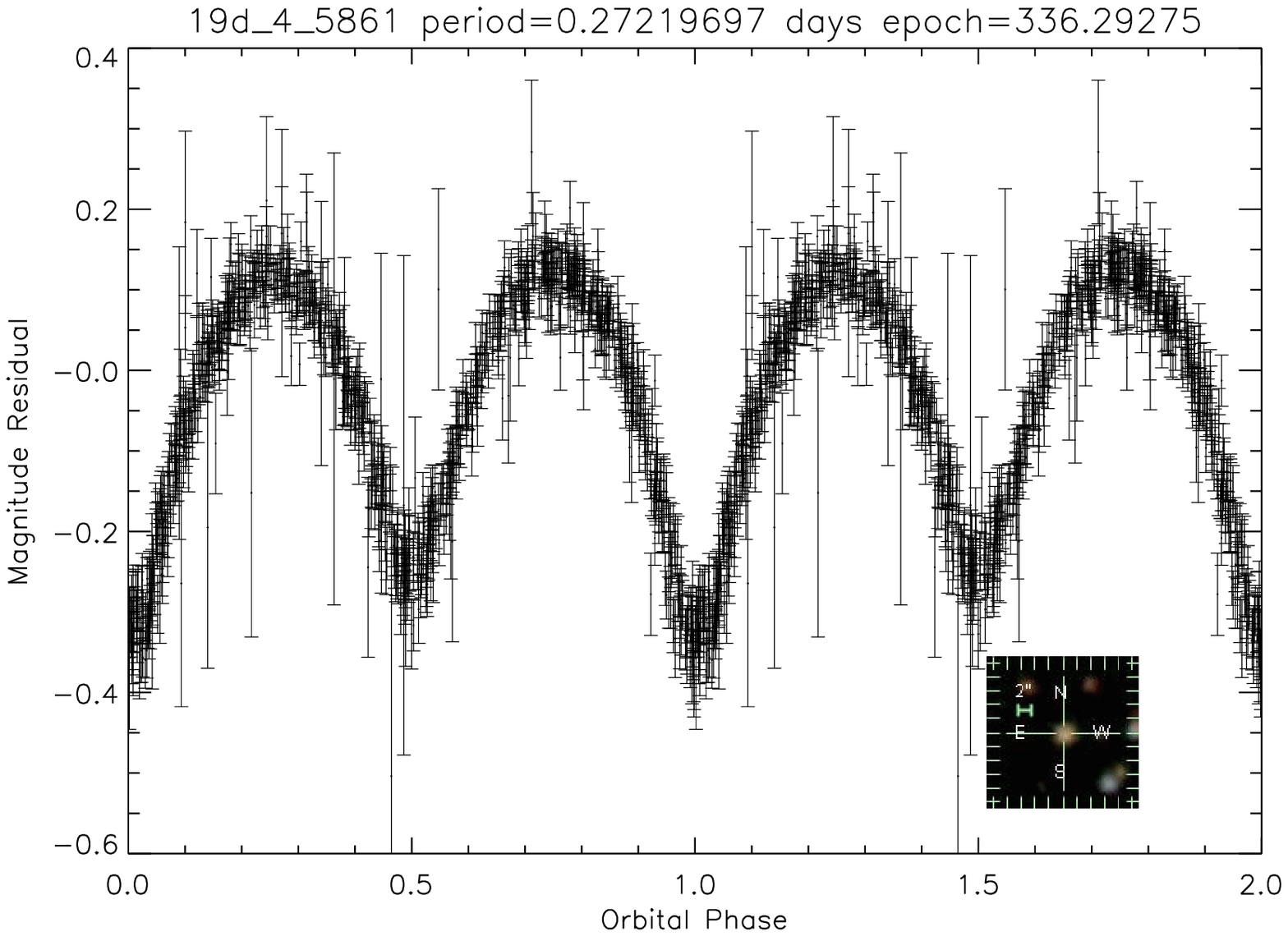} \\
\end{array}$
\end{center}
\caption{Red sample with 0.23$<P\leq$0.3 days.}
\end{figure*}

\begin{figure*}
\begin{center}$
\begin{array}{ll}
\includegraphics[width=0.5\textwidth]{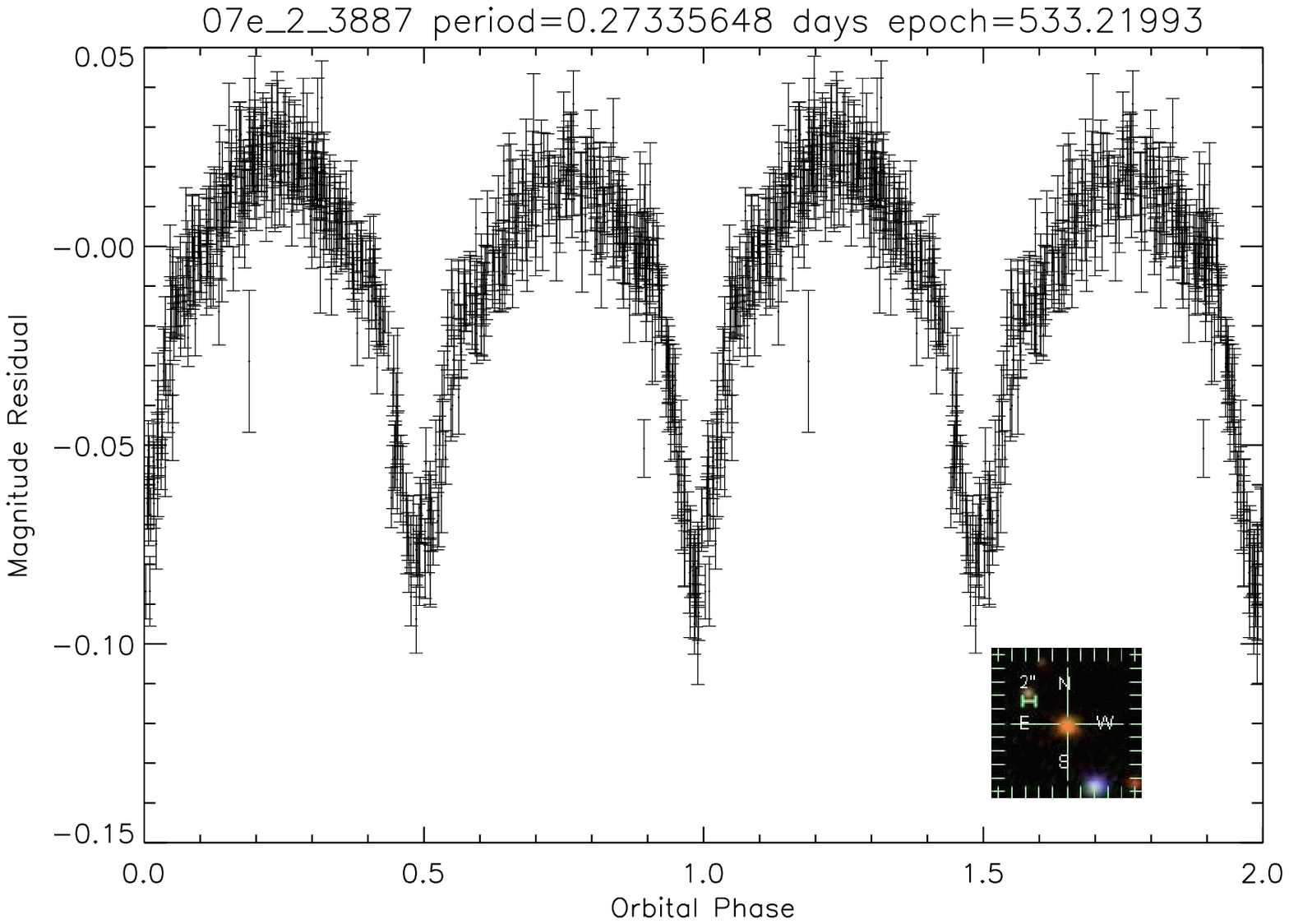} &
\includegraphics[width=0.5\textwidth]{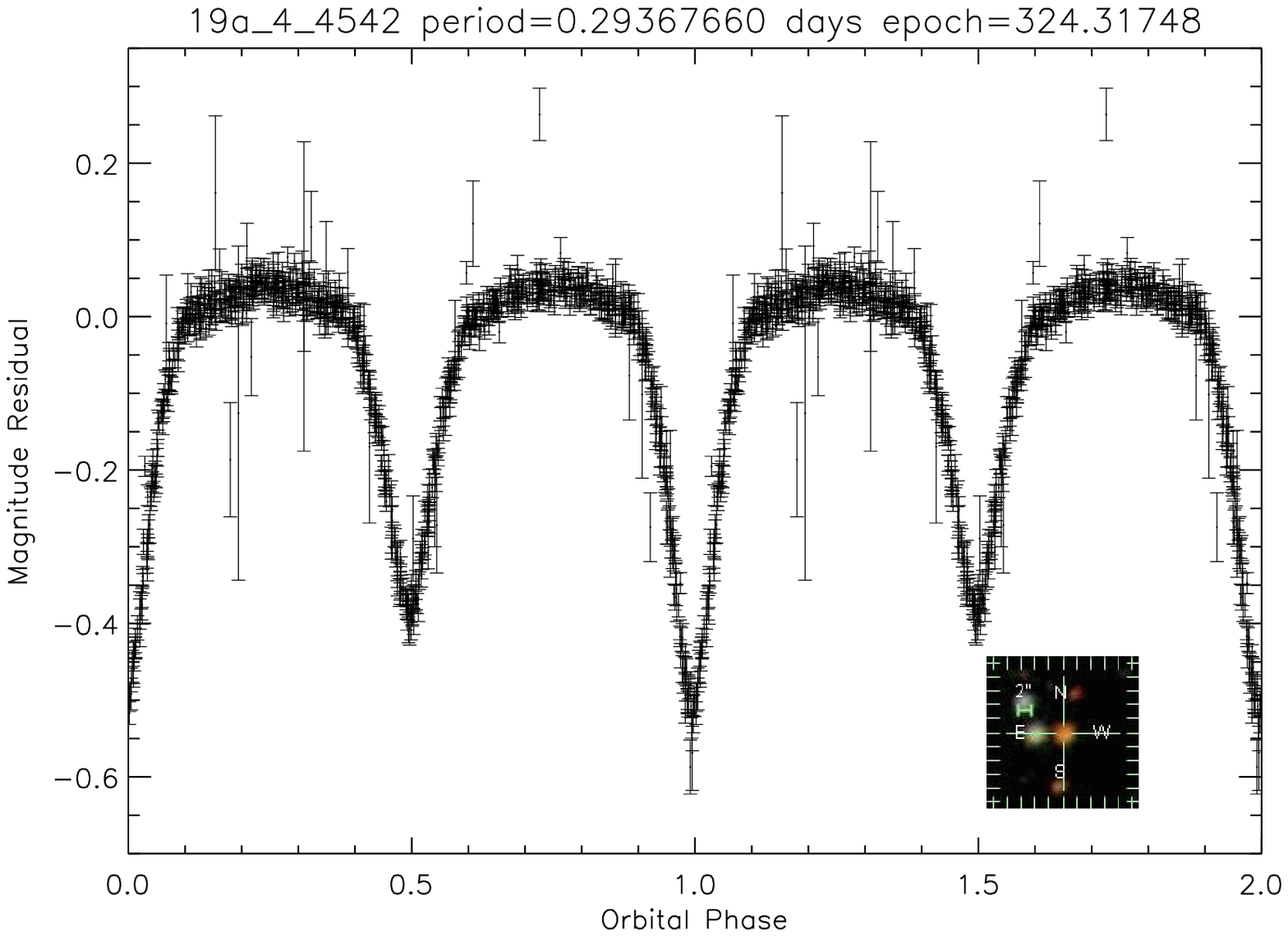} \\
\end{array}$
\end{center}
\caption{Red sample with 0.23$<P\leq$0.3 days continued.}
\end{figure*}

\begin{figure*}
\begin{center}$
\begin{array}{llllll}
\includegraphics[width=0.5\textwidth]{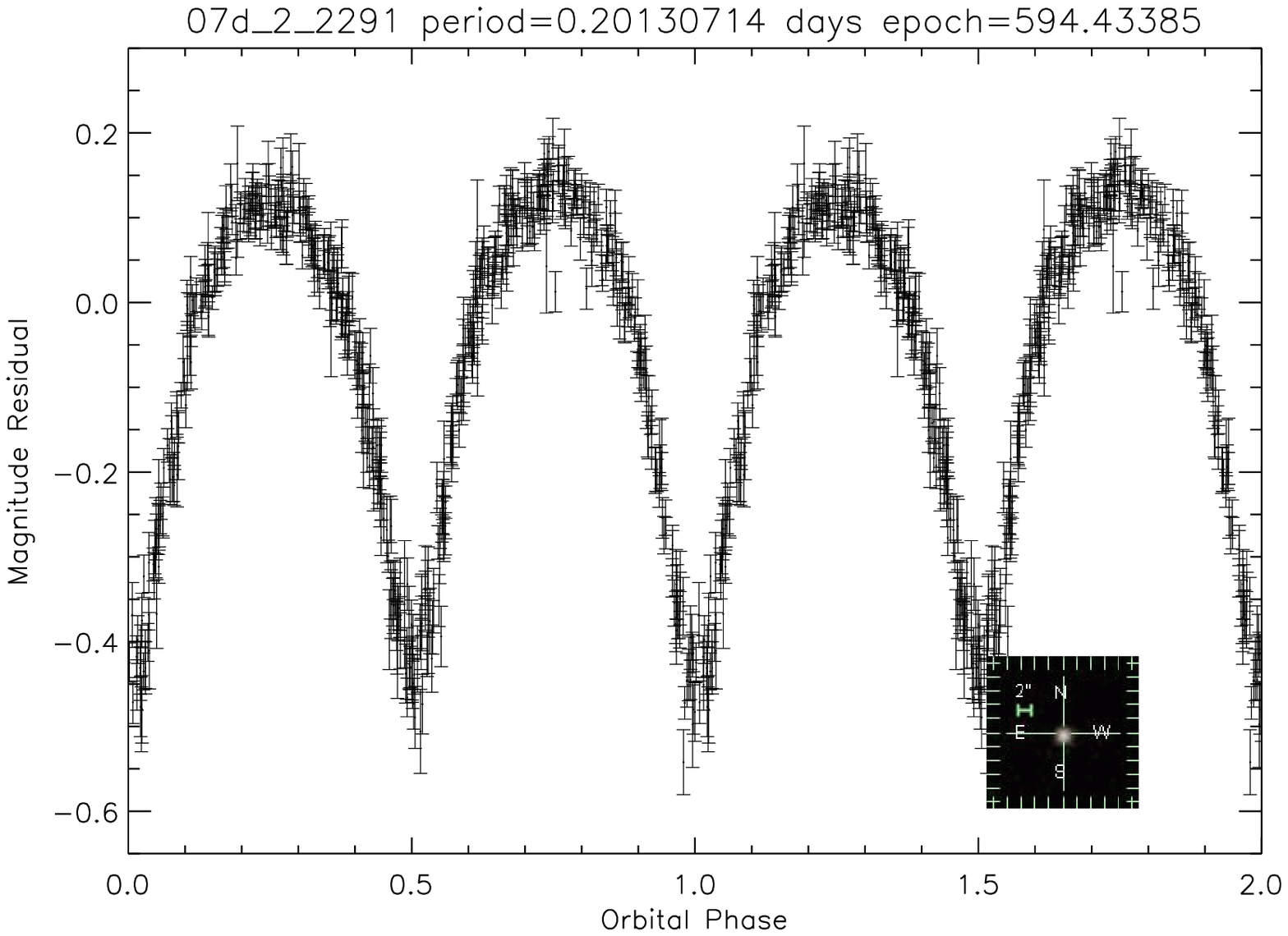} &
\includegraphics[width=0.5\textwidth]{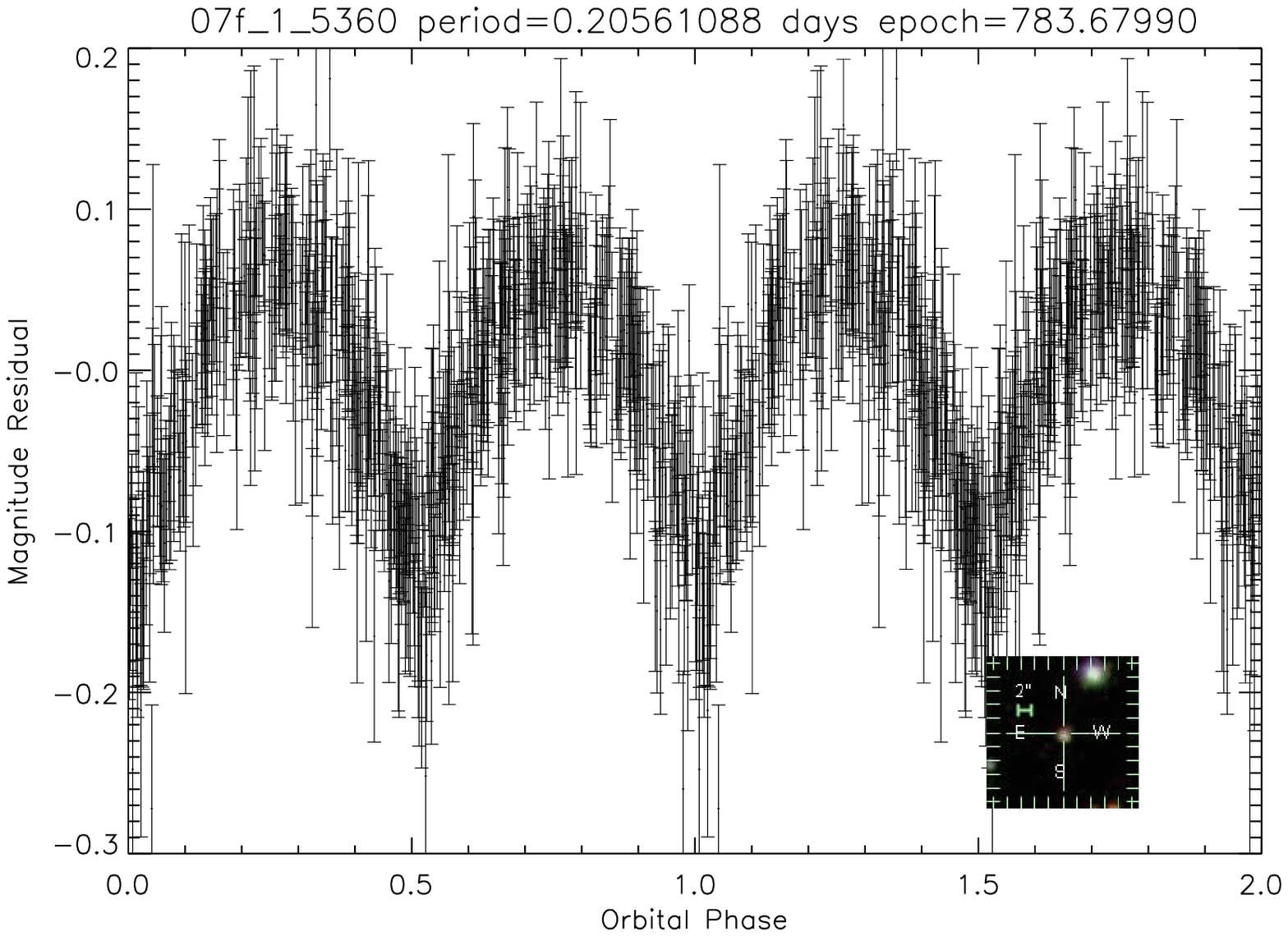} \\
\includegraphics[width=0.5\textwidth]{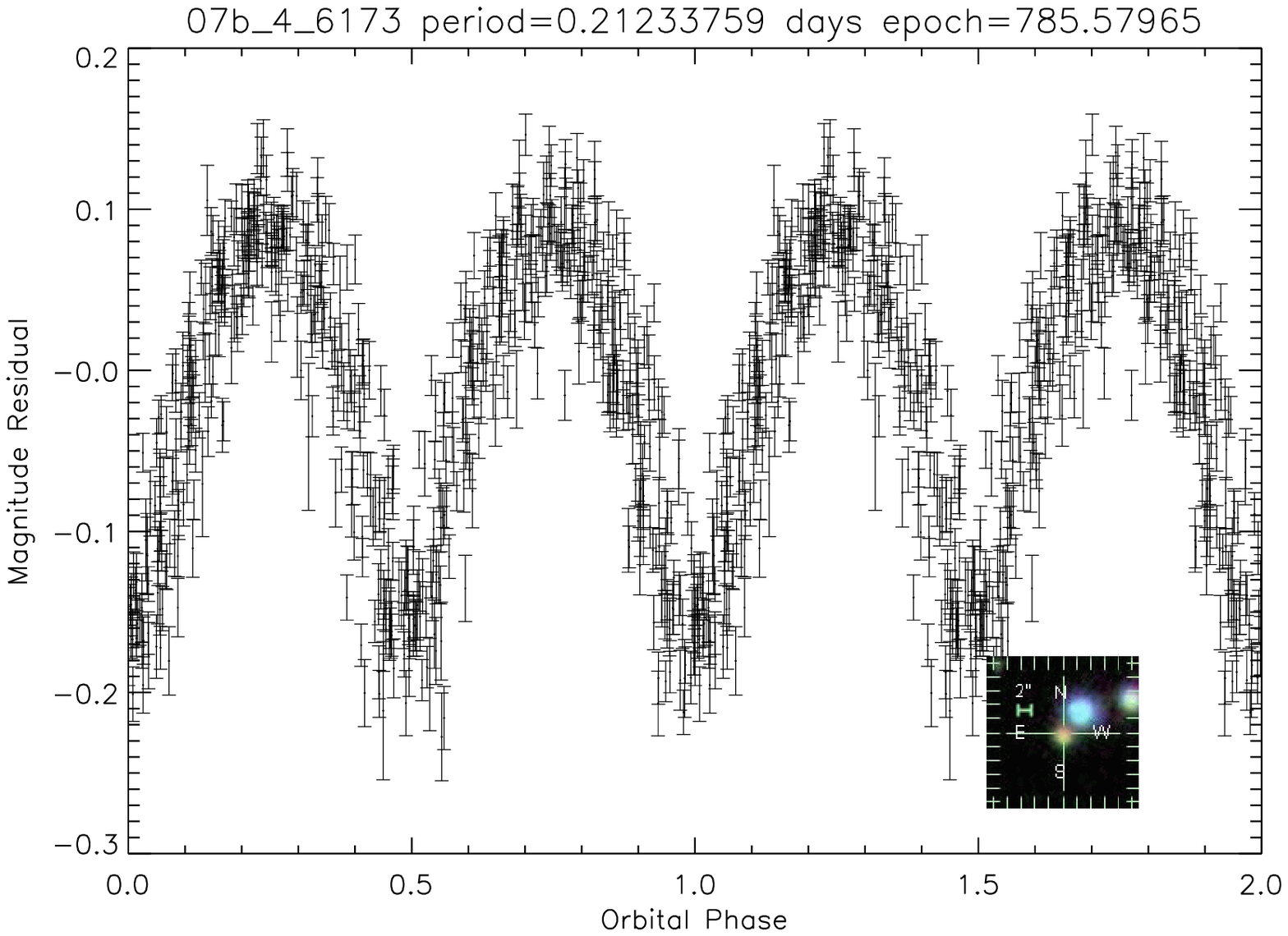} &
\includegraphics[width=0.5\textwidth]{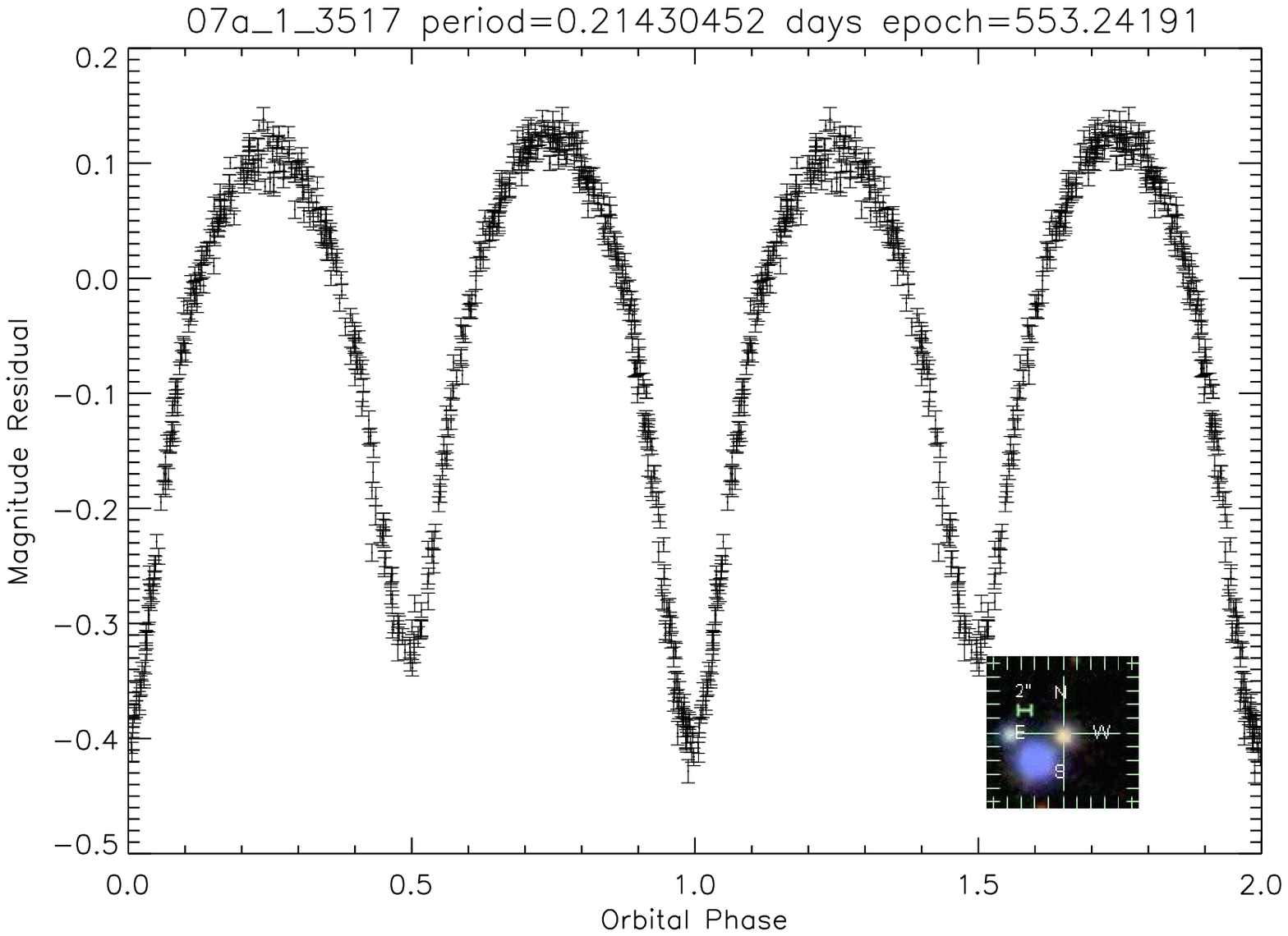} \\
\includegraphics[width=0.5\textwidth]{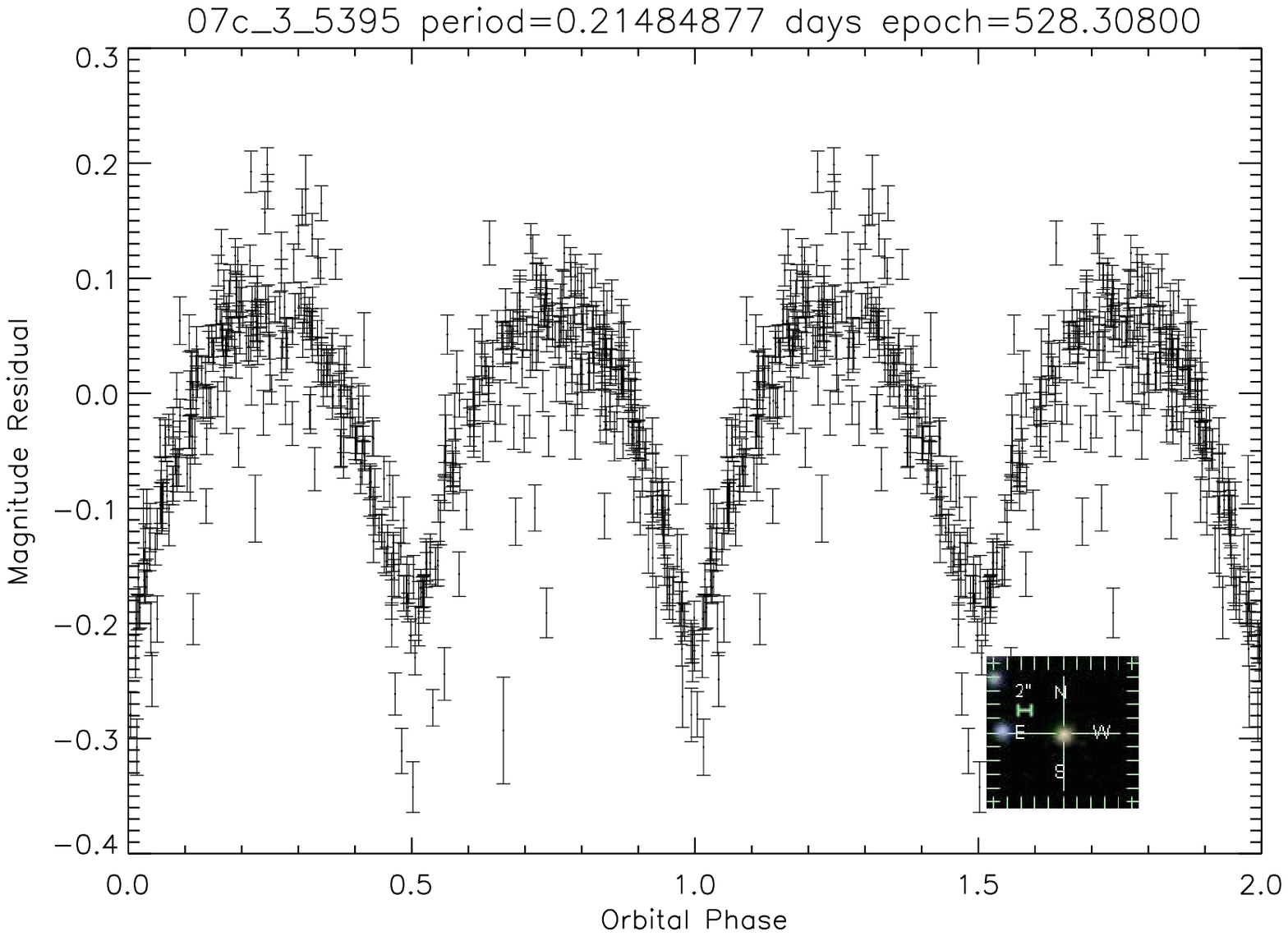} &
\includegraphics[width=0.5\textwidth]{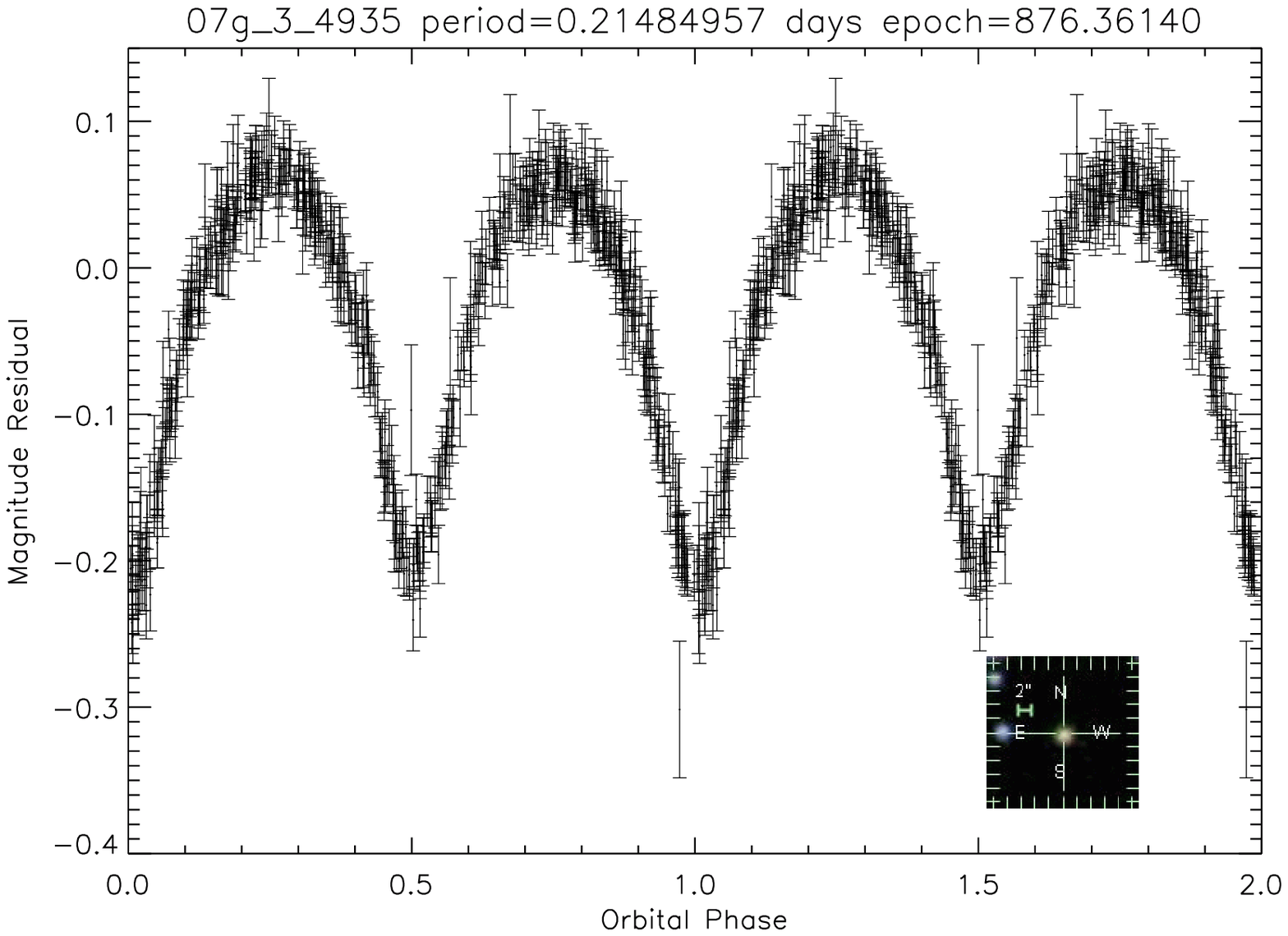} \\
\end{array}$
\end{center}
\caption{Blue sample with $P\leq$0.23 days.}
\end{figure*}

\begin{figure*}
\begin{center}$
\begin{array}{llllll}
\includegraphics[width=0.5\textwidth]{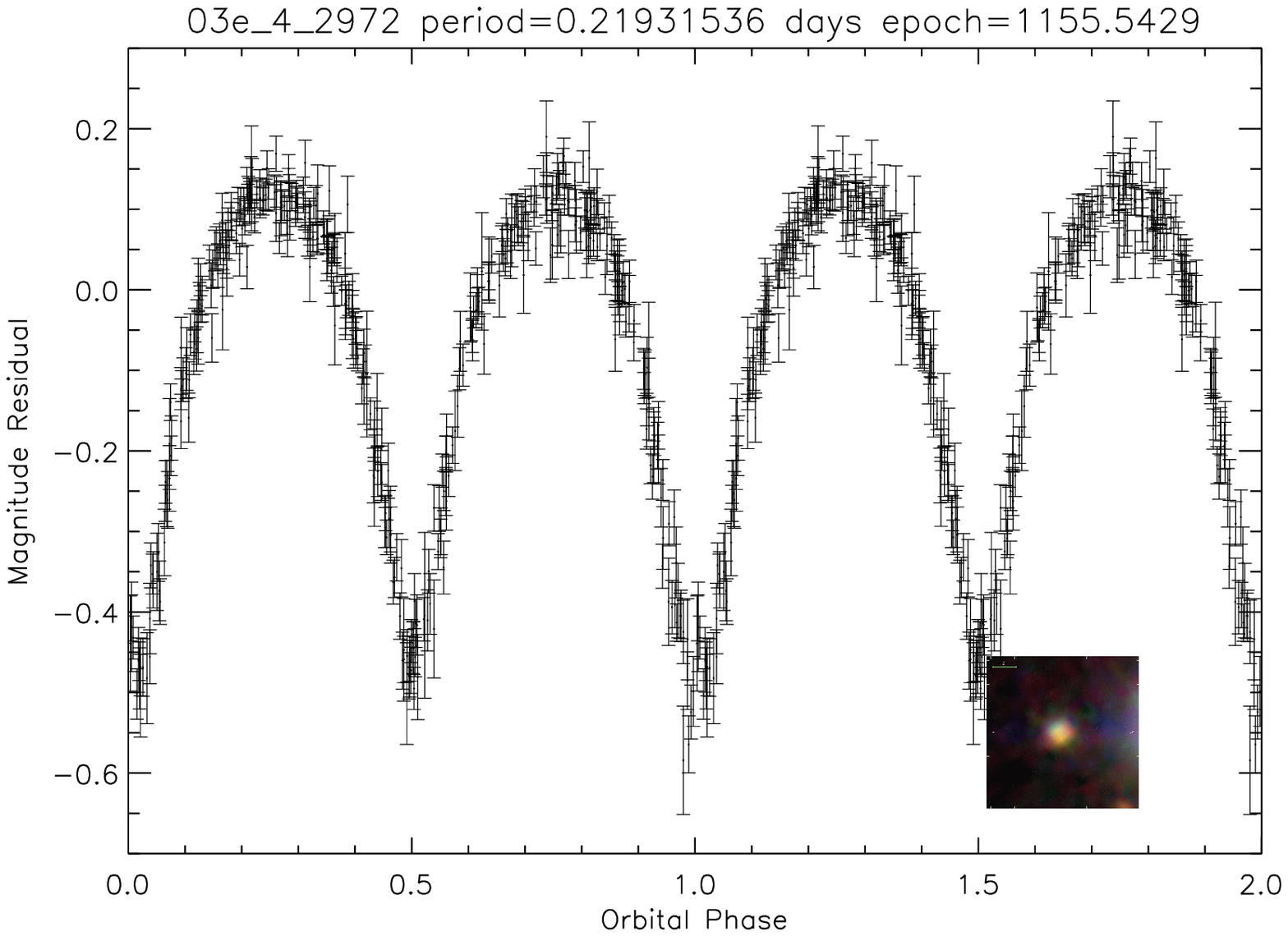} &
\includegraphics[width=0.5\textwidth]{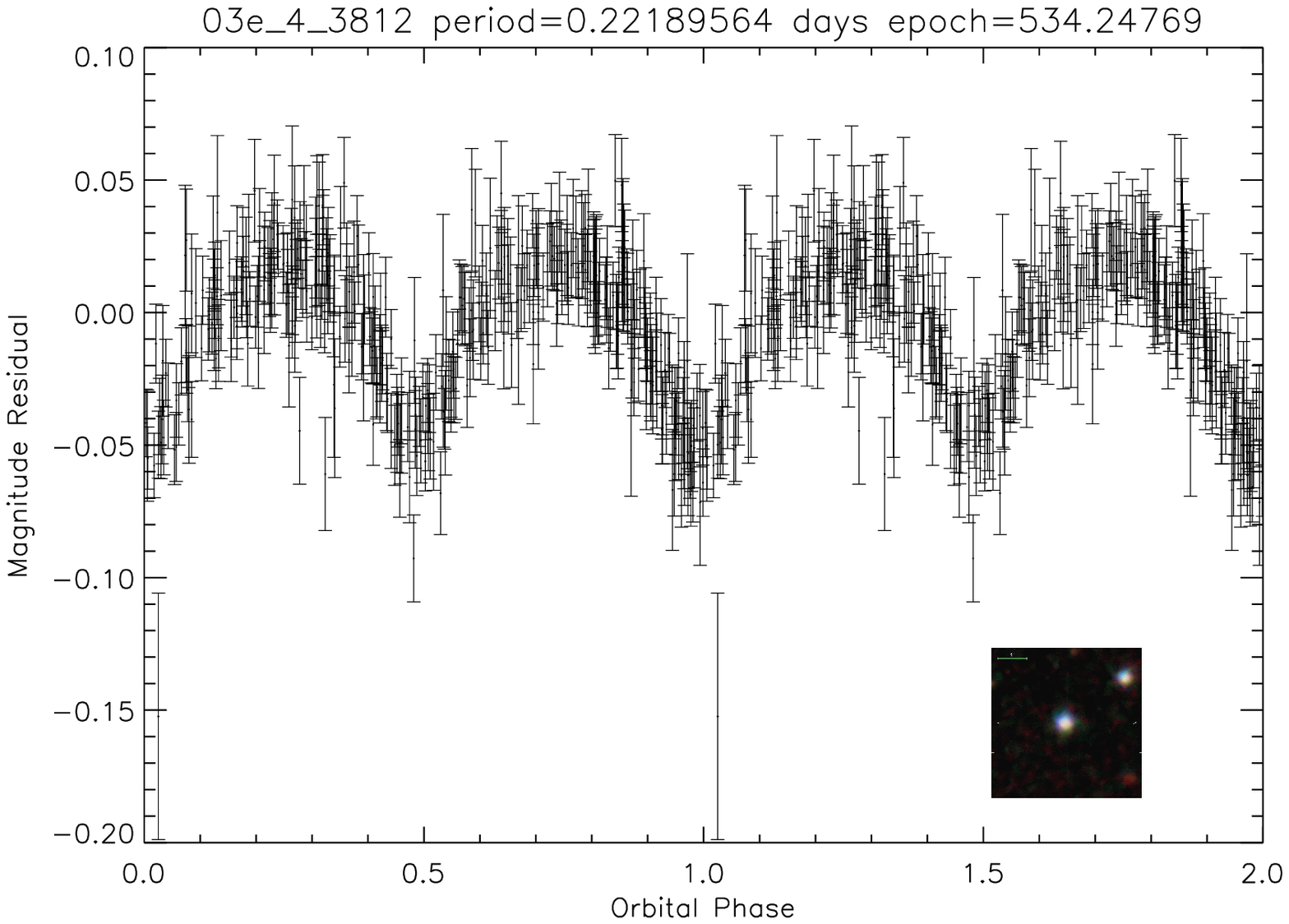} \\
\includegraphics[width=0.5\textwidth]{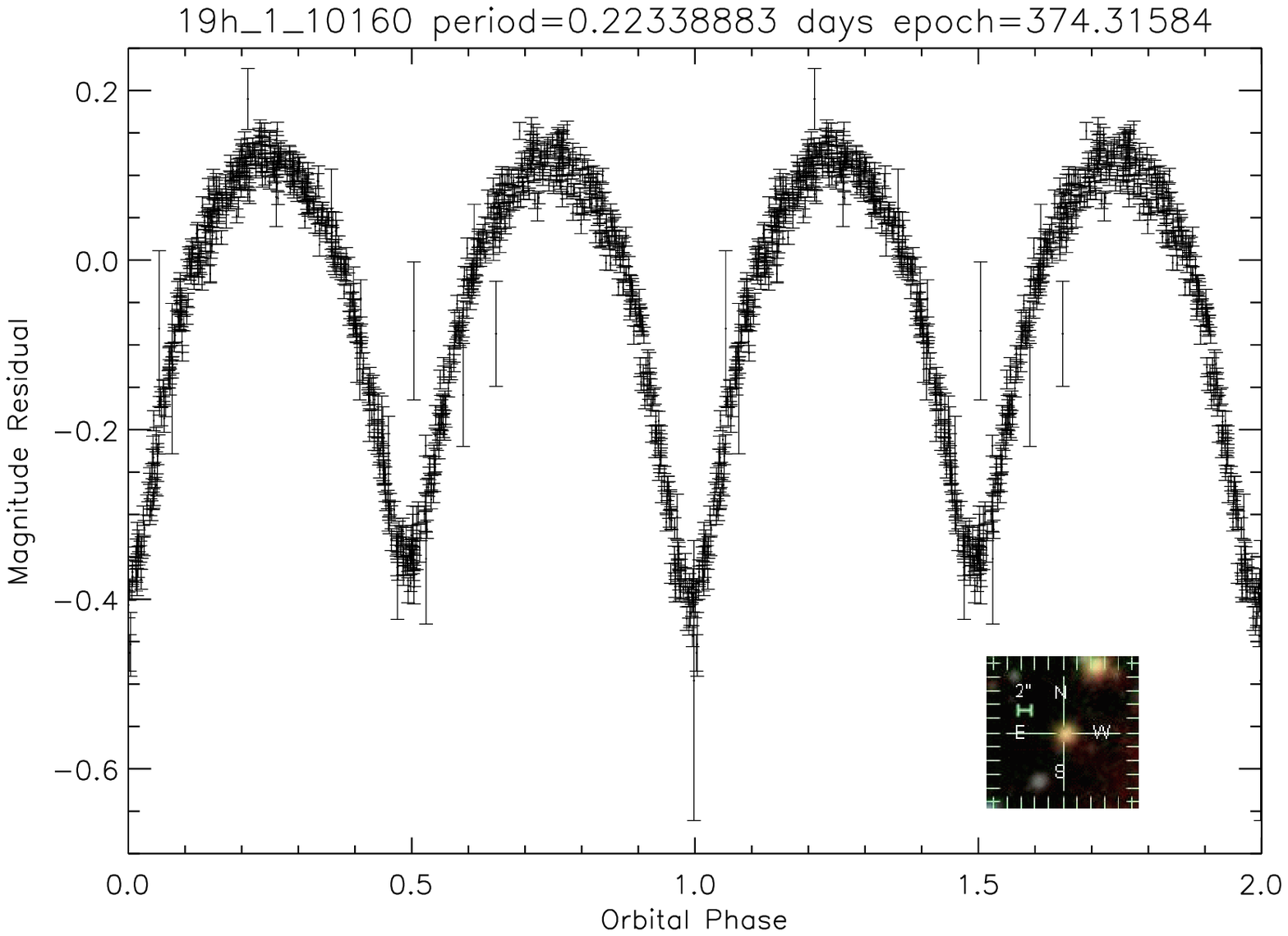} &
\includegraphics[width=0.5\textwidth]{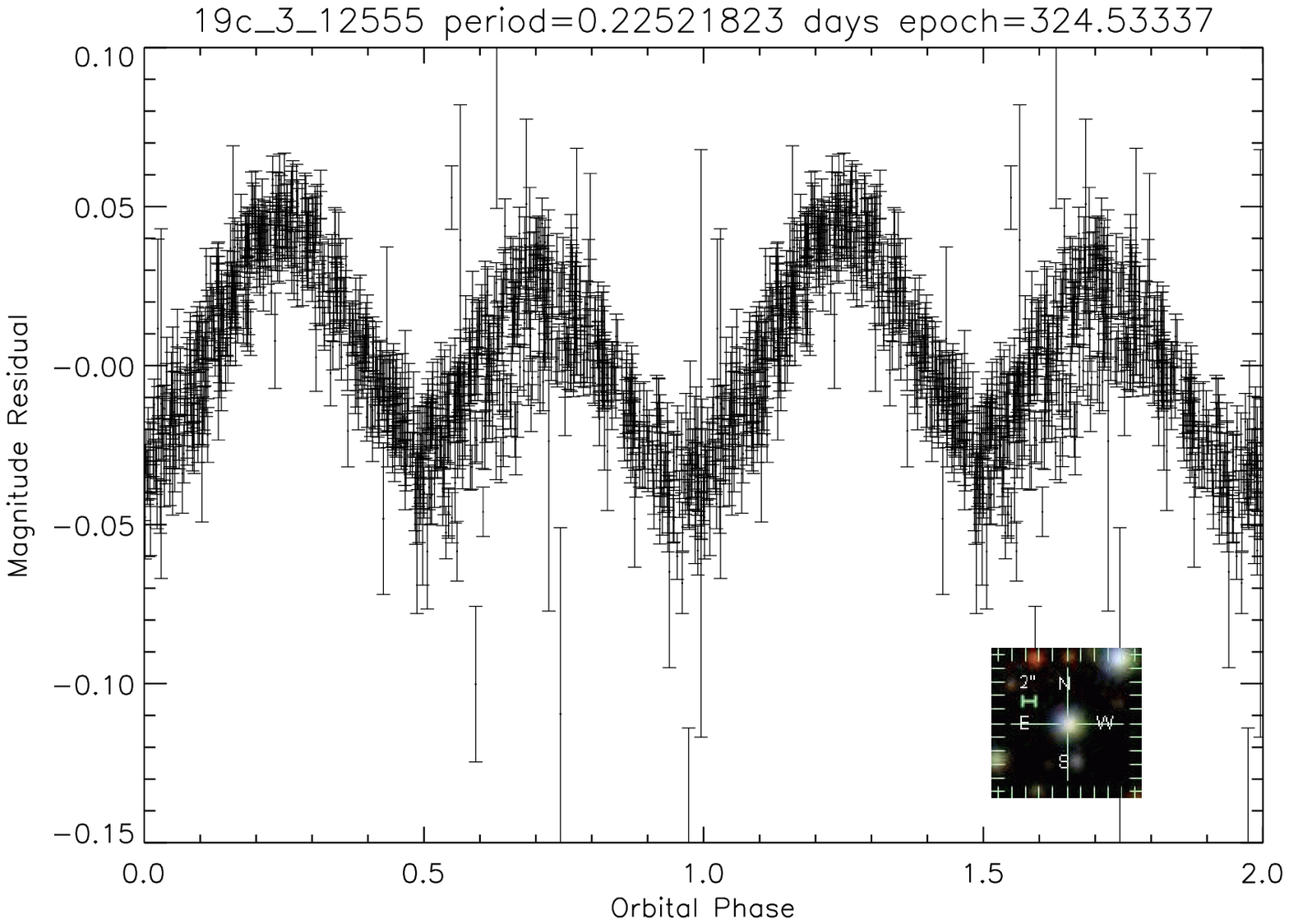} \\
\includegraphics[width=0.5\textwidth]{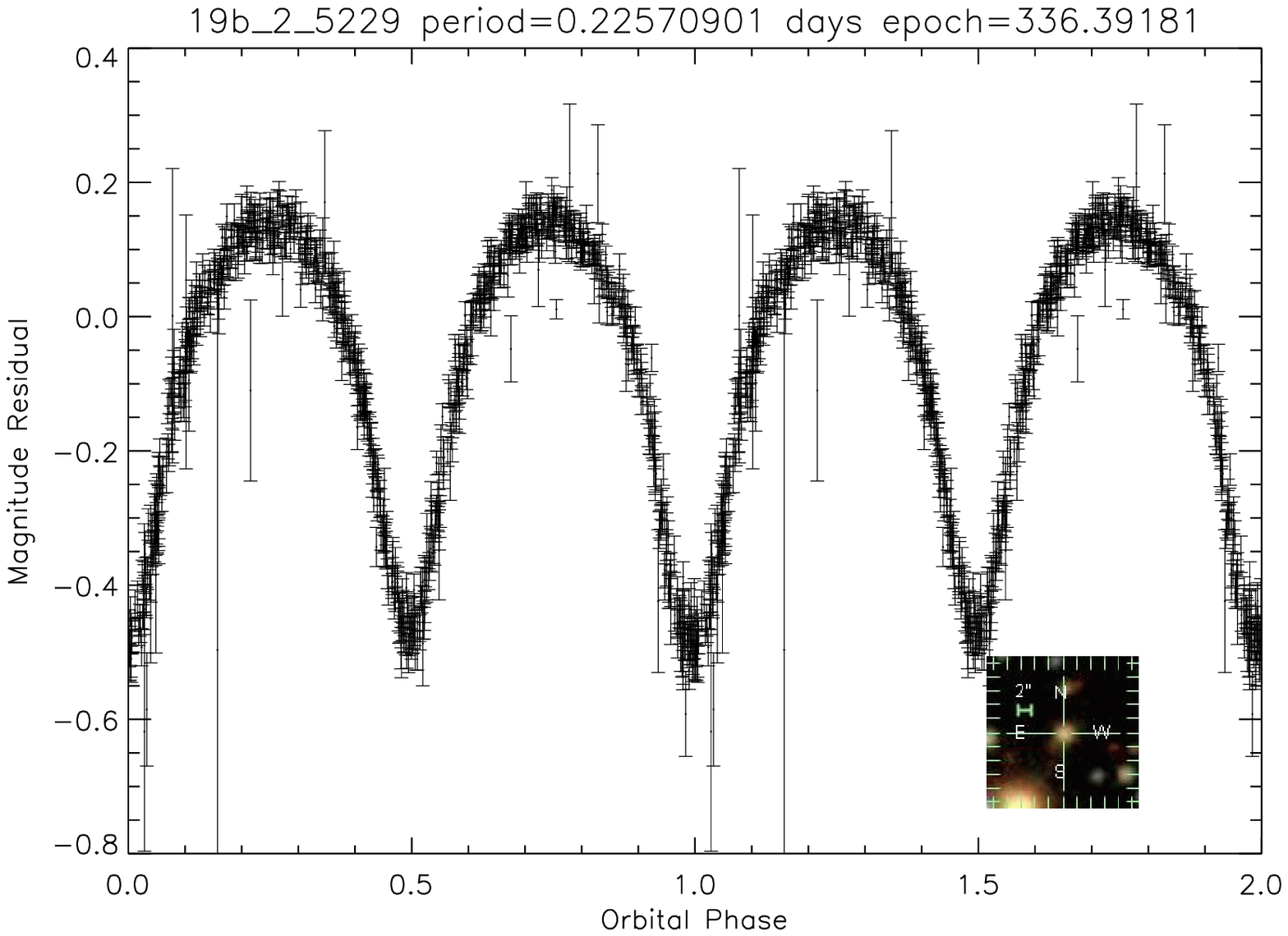} &
\includegraphics[width=0.5\textwidth]{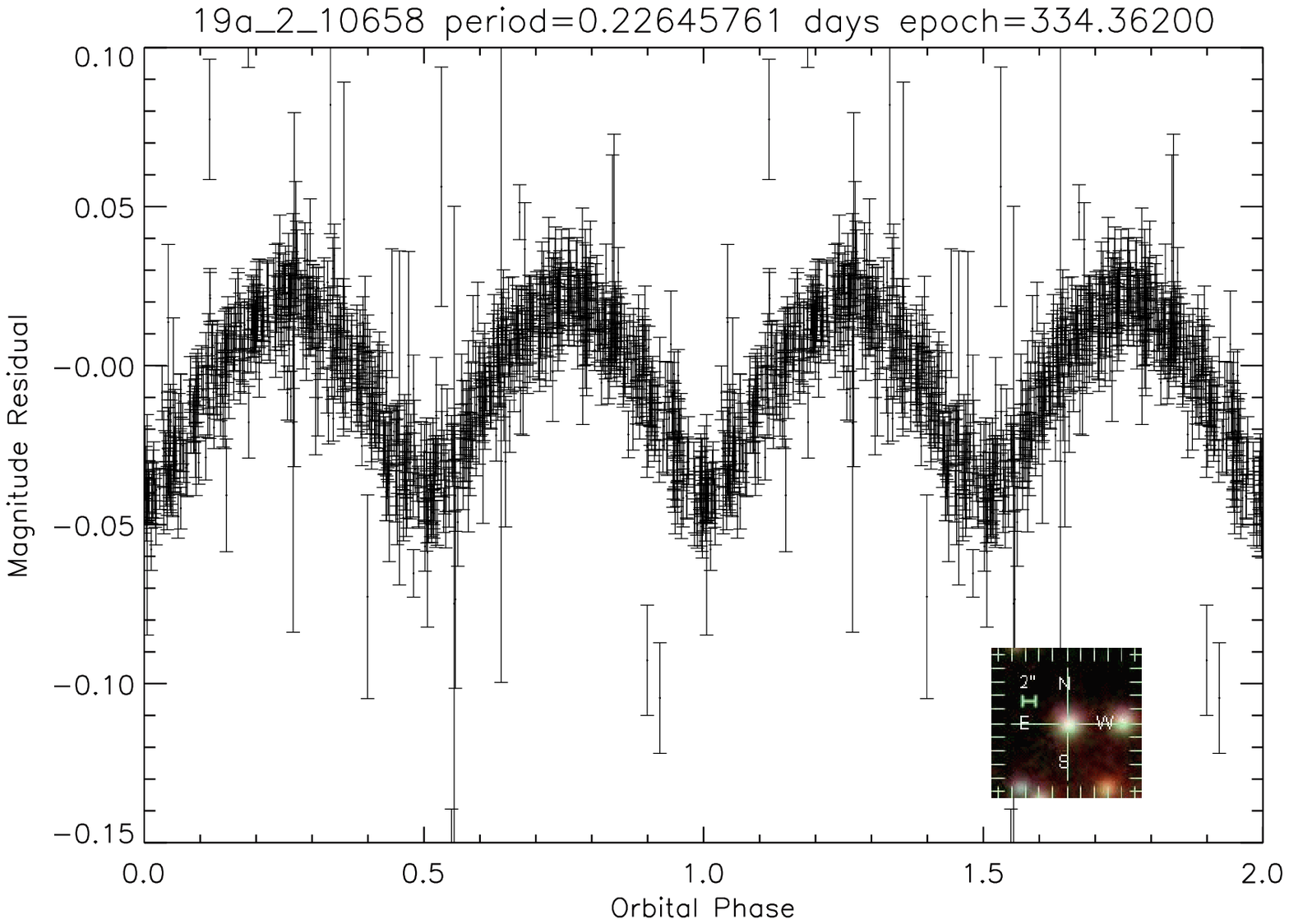} \\
\end{array}$
\end{center}
\caption{Blue sample with $P\leq$0.23 days continued}
\end{figure*}

\begin{figure*}
\begin{center}$
\begin{array}{ll}
\includegraphics[width=0.5\textwidth]{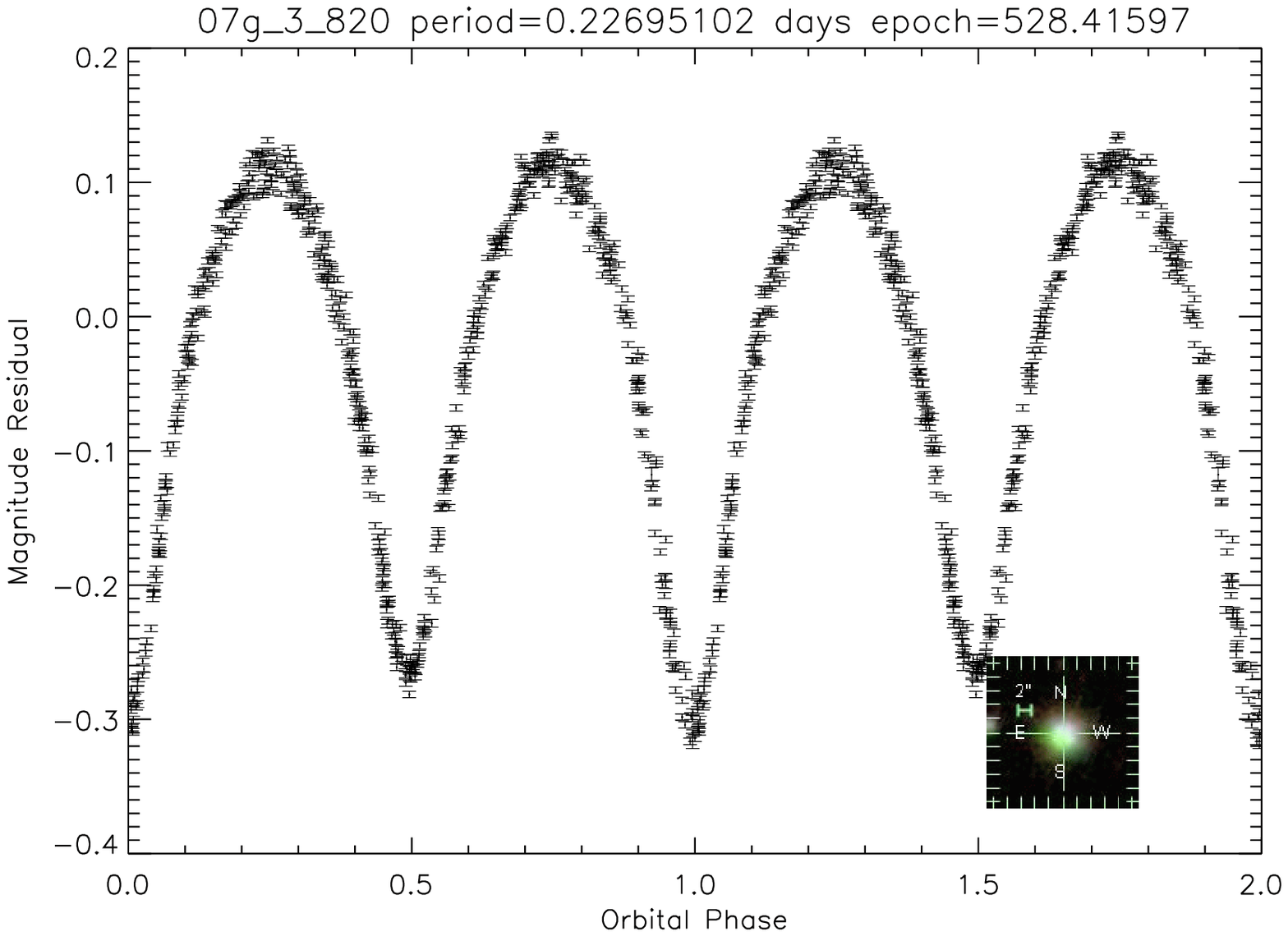} &
\includegraphics[width=0.5\textwidth]{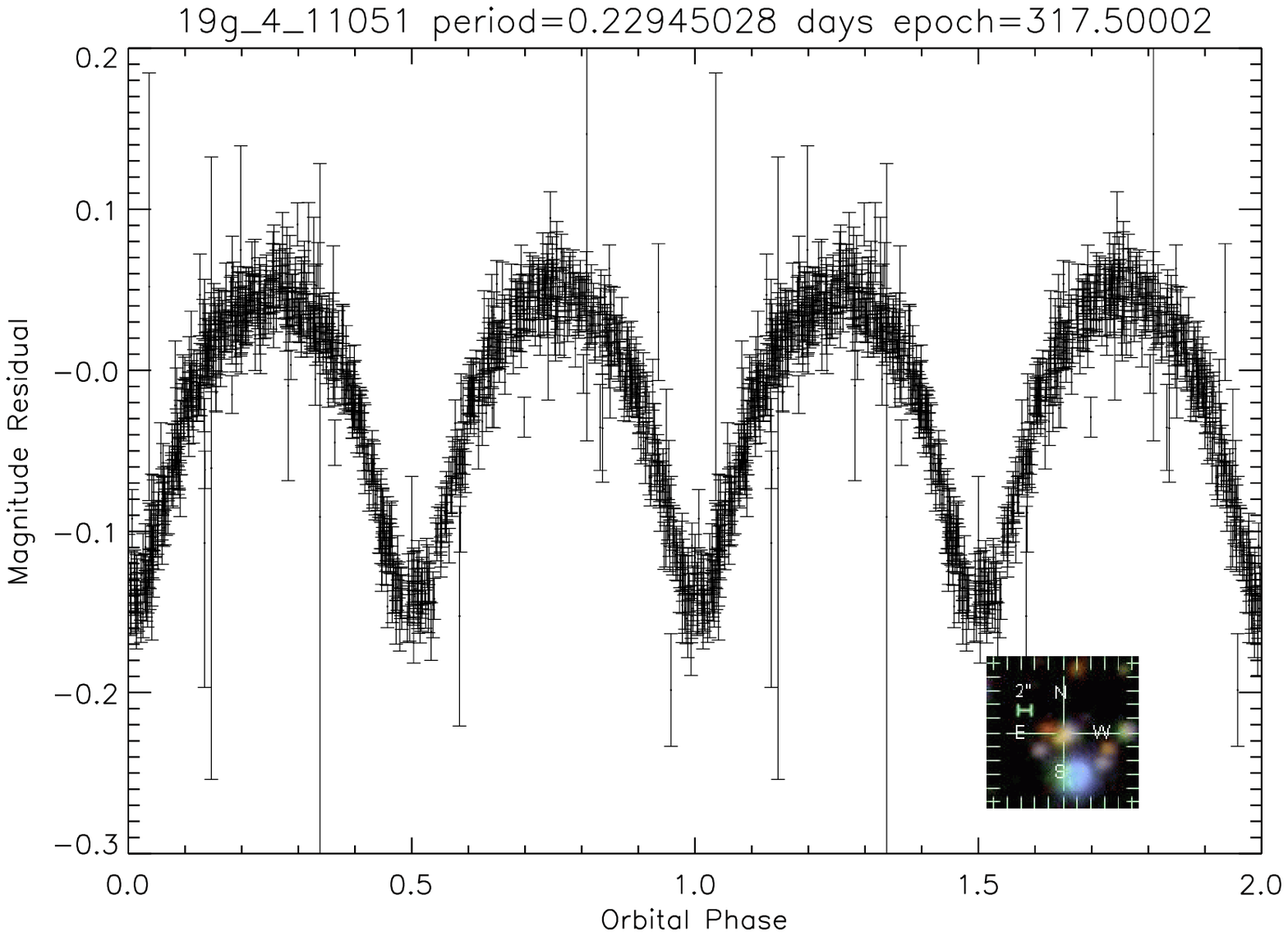} \\
\end{array}$
\end{center}
\caption{Blue sample with $P\leq$0.23 days continued}
\end{figure*}

\clearpage
\begin{table*}[h]
\begin{tabular*}{1.0\textwidth}
   {@{\extracolsep{\fill}}llllllll}
\hline
Name		&Period    &	&RA &	     &	&DEC	&\\
		&(days)	   &\textit{h}   &\textit{m}  &\textit{s}       &\textit{d}   &\textit{m}  &\textit{s}\\	
\hline
19b-3-06008     &0.1121579 &19 &33 &49.3     &36 &54 &01.8\\
07g-3-05744     &0.1512067 &07 &07 &47.6     &12 &59 &42.1\\
17d-3-02440     &0.1571555 &17 &17 &28.7     &04 &07 &30.7\\
19h-3-14922     &0.1798301 &19 &38 &44.1     &36 &30 &14.7\\
19g-3-06701     &0.2090313 &19 &37 &04.0     &36 &36 &24.8\\
07c-4-05645     &0.2107227 &07 &05 &25.9     &13 &23 &14.3\\
19e-3-05704     &0.2196803 &19 &32 &32.8     &36 &32 &11.8\\
19c-2-08140     &0.2272810 &19 &37 &11.3     &36 &22 &39.0\\
19e-4-00861     &0.2284277 &19 &31 &04.5     &36 &30 &01.3\\
19c-1-09997     &0.2466007 &19 &34 &45.7     &36 &22 &47.4\\
03f-1-01082     &0.2501641 &03 &36 &29.1     &38 &58 &10.6\\
03b-3-02411     &0.2625199 &03 &38 &26.0     &39 &38 &16.3\\
19f-1-07389     &0.2698691 &19 &31 &34.6     &36 &08 &37.9\\
07h-4-03156     &0.2703216 &07 &06 &05.8     &13 &04 &10.1\\
19d-4-05861     &0.2721970 &19 &36 &36.6     &36 &48 &12.3\\
07e-2-03887     &0.2733565 &07 &03 &23.5     &12 &34 &39.1\\
19a-4-04542     &0.2936766 &19 &30 &29.7     &36 &48 &35.7\\
07d-2-02291     &0.2013071 &07 &08 &43.1     &12 &51 &50.3\\
07f-1-05360     &0.2056109 &07 &02 &34.0     &12 &42 &25.9\\
07b-4-06173     &0.2123376 &07 &02 &31.3     &13 &22 &60.0\\
07a-1-03517     &0.2143045 &07 &01 &56.1     &12 &45 &21.0\\
07c-3-05395     &0.2148488 &07 &07 &40.8     &13 &09 &25.4\\
07g-3-04935     &0.2148496 &07 &07 &40.8     &13 &09 &25.5\\
03e-4-02972     &0.2193154 &03 &34 &38.5     &39 &22 &22.1\\
03e-4-03812     &0.2218956 &03 &34 &30.8     &39 &24 &48.5\\
19h-1-10160     &0.2233888 &19 &35 &53.7     &36 &08 &47.8\\
19c-3-12555     &0.2252182 &19 &37 &30.9     &36 &46 &55.7\\
19b-2-05229     &0.2257090 &19 &33 &41.2     &36 &24 &31.7\\
19a-2-10658     &0.2264576 &19 &33 &17.5     &36 &17 &17.5\\
07g-3-00820     &0.2269510 &07 &07 &03.1     &12 &59 &58.4\\
19g-4-11051     &0.2294503 &19 &34 &55.9     &36 &39 &56.5\\
\hline  
     \end{tabular*}
\caption{Table showing the sky coordinates for our WTS short-period eclipsing M-dwarf binaries: we list right ascension (RA) and declination (DEC) in the format \textit{hms}(hours, minutes, seconds) and \textit{dms}(degrees, minutes, seconds).}
\label{OnlineTable}
\end{table*}

\end{document}